\documentclass[12pt,draftclsnofoot,onecolumn]{IEEEtran} 

\usepackage{times}
\usepackage[final]{graphicx}
\usepackage{amsmath,amsfonts}
\usepackage{amssymb,amsbsy}
\usepackage{times}

\usepackage{epsf,graphics,epsfig,subfigure} 
\usepackage{cite}

\newcommand\ignore[1]{}

\newtheorem{thm}{Theorem}
\newtheorem{lem}{Lemma}
\newtheorem{prop}{Proposition}
\newtheorem{defn}{Definition}

\newtheorem{rem}{Remark}

\newcommand{\fixed}{ {\sf fixed} }

\newcommand{\chanpow} {\rho_c  }

\newcommand{\trace}{ {\mathrm{Tr}}}
\newcommand{\diag} {{\mathrm{diag}}  }

\newcommand{\opt} { {\sf{opt}} }
\newcommand{\stat} { {\sf{stat}} }

\newcommand{\unstruct}{  {\sf unconst} }
\newcommand{\semi} { {\sf semi} }

\newcommand{\err} { {\sf{err}} }
 
\newcommand{\bH} {{\mathbf{H}  } }

\newcommand{\bQ} { {\mathbf{Q}} }
\newcommand{\bEe}{{\mathit{E}}}    
\newcommand{\bX} {{\mathbf{X}  } }

\newcommand{\bI} {{\mathbf{I}  } }

\newcommand{\bGamma}{ {\mathbf{\Gamma}} }
\newcommand{\bs} {  {\bf s} }

\newcommand{\transnr}{   \frac{ \rho}{ M } }
\newcommand{\perf} { {\sf{perf}} }

\newcommand{\bU}   { {\mathbf{U}}} 

\newcommand{\bfLambda}  {{\mathbf{\Lambda}} } 
\newcommand{\bSigma}  {{\mathbf{\Sigma}} }

\newcommand{\bFa} {  {\bf{F}} }

\newcommand{\bB}   { {\mathbf{B}}} 
\newcommand{\bW}   { {\mathbf{W}}} 
\newcommand{\bb} {{\mathbf{b}}}
\newcommand{\ba} {{\mathbf{a}}}
\newcommand{\bA} {{\mathbf{A}}}

\newcommand{\bu}  { {\mathbf{u}}  }
\newcommand{\by}  { {\mathbf{y}}  }
\newcommand{\bh}  { {\mathbf{h}}  }

\newcommand{\bw}  { {\mathbf{w}  } }
\newcommand{\ud} { {\mathrm{d}}  }  

\newcommand{\bv} { {\mathbf{v}} }

\newcommand{\iid} {  {\sf{iid}} }
\newcommand{\ind} { {\sf{ind}} }

\newcommand{\snr}{{\sf{SNR}} }
\newcommand{\ord}{{\mathcal{O}}}
\newcommand{\littleo}{{\mathnormal{o}}}

\newcommand{\hsp}{\hspace{0.1in} }

\newcommand{\hspp}{\hspace{0.05in} }
\newcommand{\hsppp}{\hspace{0.02in} }
\newcommand{\hspppp} { {\hspace{0.01in}} }

\newcommand{\sinr}{ {\sf{SINR}} } 
\newcommand{\mse} { {\sf{MSE}} }
\newcommand{\mmse}{ {\sf{mmse}} }

\newcommand{\gammartwo} {  \mu_{r, \hsppp 2} }
\newcommand{\gammatone} { {\mathrm{Gap}}_{t} }

\newcommand{\gammatc} { {\mathrm{Gap}}_t^c }
\newcommand{\gammarc} { \mu_{r, \hsppp 2}^{ c} }

\newsavebox{\savepar}

\begin{document}
\title{Low-Complexity Structured Precoding for Spatially Correlated 
MIMO Channels}
\author{\large {\hspace{0.35in}} Vasanthan Raghavan, 
Akbar M.\ Sayeed, Venugopal V.\ Veeravalli$^*$
\thanks{V.\ Raghavan and V.~V.\ Veeravalli are with the Coordinated Science 
Laboratory and the Department of Electrical and Computer Engineering, University 
of Illinois at Urbana-Champaign, Urbana, IL 61801 USA. A.~M.\ Sayeed is with the 
Department of Electrical and Computer Engineering, University of Wisconsin-Madison, 
Madison, WI 53706 USA. Email: {\tt{vasanthan\_raghavan@ieee.org, 
vvv@uiuc.edu.}} $^*$Corresponding author.} 
\newline 
\thanks{ 
This work was partly supported by the NSF under grant \#CCF-0049089 through 
the University of Illinois, and grant \#CCF-0431088 through the University 
of Wisconsin. This paper was presented in part at the 42nd Annual Allerton 
Conference on Communications, Control and Computing, Allerton, IL, 
2006 and will be presented at the IEEE International Symposium on Information 
Theory, Toronto, Canada, 2008.}} 

\maketitle
\vspace{-25mm}
\baselineskip 18pt

\begin{abstract} 
\noindent 
The focus of this paper is on spatial precoding in correlated multi-antenna 
channels, where the number of independent data-streams is adapted to trade-off 
the data-rate with the transmitter complexity. Towards the goal of a low-complexity 
implementation, a {\em structured precoder} is proposed, where the precoder 
matrix evolves fairly slowly at a rate comparable with the statistical evolution 
of the channel. Here, the eigenvectors of the precoder matrix correspond to the 
dominant eigenvectors of the transmit covariance matrix, whereas the power 
allocation across the modes is fixed, known at both the ends, and is of low-complexity. 
A particular case of the proposed scheme (semiunitary precoding), where the spatial 
modes are excited with equal power, is shown to be near-optimal in {\em matched 
channels}. A matched channel is one where the dominant eigenvalues of the 
transmit covariance matrix are well-conditioned and their number equals the 
number of independent data-streams, and the receive covariance matrix is also 
well-conditioned. In {\em mismatched channels}, where the above conditions are 
not met, it is shown that the loss in performance with semiunitary precoding when 
compared with a perfect channel information benchmark is substantial. This 
loss needs to be mitigated via limited feedback techniques that provide partial 
channel information to the transmitter. More importantly, 
we develop matching metrics that capture the degree of matching of a channel to 
the precoder structure {\em continuously}, and allow ordering two matrix channels 
in terms of their mutual information or error probability performance. 
\end{abstract}


\section{Introduction} 
\label{sec1} 
Multiple antenna communications has received significant attention over the 
last decade as a mechanism to increase the rate of information transfer, or 
the reliability of signal reception, or a combination of the two. The focus of 
this work is on point-to-point spatial precoding systems, where the number of 
independent data-streams is constrained to be a subset\footnote{The number of 
data-streams, $M$, is such that $1 \leq M \leq N_t$ with $N_t$ 
denoting the transmit 
antenna dimension. 
Note 
that $M$ is the rank of the input covariance matrix and the number of radio-frequency 
(RF) link chains as well.}, $M$, of the transmit dimension so as to minimize the 
complexity and the cost associated with transmission. Initial works on precoding 
study optimal signaling strategies when perfect channel state information (CSI) is 
available at the transmitter and the receiver. These studies show that a {\em channel 
diagonalizing} input that corresponds to exciting the dominant $M$-dimensional 
eigen-space of the channel, with a power allocation that can be computed via 
waterfilling, is robust under different design 
metrics~\cite{lee_petersen,salz,yang_roy,scaglione_gia_barb,sampath,sampath2,yang_roy2,scaglione_spbgs,palomar_precode}. 

Although perfect CSI provides a benchmark on the performance, it is difficult to 
obtain in practice. More importantly, the system performance is not robust under 
CSI uncertainty. Even a small error in the CSI at the transmitter can lead to a 
dramatic degradation in performance with a scheme that is designed for the 
mismatched CSI~\cite{goldsmith_review,bolsckei,david_corr,vasanth_limfb_precode,vasanth_limfb}. 
Furthermore, even if perfect CSI is available, tight constraints on 
complexity as well as energy 
consumption~\cite{keyes,meindl_davis,rabaey,razavi,pamela} at the RF 
level in the mobile ends may disallow the implementation of optimal 
solutions in practice. This is because Third 
Generation wireless systems and beyond are expected to be multi-carrier in 
nature and the burden of computing the optimal input is magnified by 
the number of sub-carriers and the rate of evolution of the channel realizations. 
Besides this, the structure of the input could change, often dramatically, at the 
rate of evolution of the channel realizations, which also makes it difficult to 
implement. These reasons suggest that a slower rate of adaptation of the input 
signals, that is of low complexity and is more robust to CSI uncertainty, is 
preferred in practice.

In realistic wireless systems, where the channels are spatio-temporally 
correlated, 
the slow rate of statistical evolution implies that it is reasonable to assume 
perfect statistical knowledge of the channel at the transmitter. Since the spatial 
statistics experienced by the individual sub-carriers are 
identical~\cite{akbar_and_venu,venu_capacity,ada_poon}, the burden of computing 
the optimal input with only the statistical information at the transmitter is 
equivalent to that of a narrowband system. Even in this setting, optimal precoding 
has been studied for different spatial correlation 
models~\cite{visotsky,jafar,jorswieck_bfcap,moustakas_stat_phy,shengli_zhou,venu_capacity,tulino_ind,goldsmith_review,bolsckei,harish_varanasi,zhang_palomar,jongren_capside,akhtar}. 
These works show that the eigen-directions of the optimal input covariance matrix 
correspond to a set of the $M$-dominant eigenvectors of the transmit covariance 
matrix and are hence, easily adaptable to changes in statistics. However, computing 
the power allocation across the $M$ modes requires Monte Carlo averaging or 
gradient descent-type 
approaches~\cite{venu_capacity,tulino_ind,goldsmith_review,bolsckei,harish_varanasi}. While 
the computational complexity of the power 
allocation algorithm may be affordable at the base station end, whether it is 
possible or not at the mobile end is questionable. Moreover, there has been no 
systematic study of statistics-based precoding approaches and hence, it is not clear 
as to how far the performance of the statistical scheme is with respect to the 
perfect CSI benchmark. 

It should be noted that all the above works study precoder design with an 
emphasis on obtaining information-theoretic limits on performance. In contrast, 
our focus here is on low-complexity schemes that can be easily implemented and 
easily adapted to changes in channel statistics. In this work, we consider a 
narrowband setup where spatial correlation is modeled by a general 
decomposition~\cite{canonical_bonek,tulino_ind,vasanth_it_rks08} 
that: 1) Is 
based on physical principles, 2) Has been verified by many recent measurement 
campaigns, and 3) Includes as special cases the well-studied 
{\em{i.i.d.\footnote{I.I.D.\ stands for independent and identically 
distributed.} model}}, 
the {\emph{separable correlation model}}~\cite{chuah}, and the {\emph{virtual 
representation}}~\cite{akbar,venu_capacity,akbar_and_venu}.

We propose the notion of {\em structured precoding}, where the power allocation 
across the $M$ spatial modes is fixed and known at both the ends. Two specific 
cases are studied in depth in this work: 1) A statistical {\em semiunitary}\footnote{An 
$N_t \times M$ matrix ${\bf X}$ with $M \leq N_t$ is said to be semiunitary if 
it satisfies ${\bf X}^H {\bf X} = \bI_M$.} precoder, where the eigen-directions of 
the input correspond to the dominant eigenvectors 
of the transmit covariance matrix and the power allocation is uniform, is studied 
theoretically. 2) A precoder, where the eigen-directions are as before, 
and the power is allocated proportionate to the transmit covariance matrix eigenvalues 
below a threshold signal-to-noise ratio ($\snr$) and uniformly above this $\snr$, is 
studied via simulations. Following the philosophy propounded here, 
more complicated schemes, where the power allocation across the modes can be 
computed with low-complexity, possibly as a function of the $\snr$ and the 
statistics, can also be considered. 

Our focus is on two questions: 1) When is the first scheme near-optimal with 
respect to a perfect CSI benchmark?, and 2) What is the ``gap''\footnote{This 
gap can possibly be bridged with a {\em limited feedback} 
scheme~\cite{limfb_honig,david_corr,vasanth_limfb_precode,vasanth_limfb} 
that provides partial channel 
information to the transmitter.} in performance and how does it depend on the 
system and the channel parameters? The performance metric used in this work is 
relative average mutual information loss. We also study relative uncoded error 
probability enhancement and relative mean-squared error ($\mse$) enhancement, 
whenever they can be characterized analytically. 

The answers to the above questions lie in the notion of 
{\em matched} and {\em mismatched channels}, which are introduced in this work. 
A matched channel is one where the channel is effectively matched to the precoding 
scheme with the following two conditioning properties being true: 1) The $M$-dominant 
eigenvalues of the transmit covariance matrix are {\emph{well-conditioned}}\footnote{If 
${\bf \Lambda}_t(1) \geq \cdots \geq {\bf \Lambda}_{t}(M)$ denote the first $M$ 
eigenvalues of the transmit covariance matrix and $\frac{ {\bf \Lambda}_t(1) } 
{ {\bf \Lambda}_t(M) }$ is (or is not) significantly larger than $1$, we loosely 
say that these eigenvalues are ill-(or well-)conditioned.}, whereas the remaining 
$(N_t - M)$ eigenvalues are {\emph{ill-conditioned}} away from the dominant ones, 
and 2) The receive covariance matrix is also {\emph{well-conditioned}}. A mismatched 
channel is one where both the transmit and the receive covariance matrices are 
ill-conditioned, with the additional condition that 
${\sf rank}({\bf H}) \geq M$ with probability $1$. 

We show that matched and mismatched channels correspond to the 
cases where the relative performance of the semiunitary precoder are closest 
and farthest to the perfect CSI precoder, respectively. The degree of 
channel-to-precoder scheme matching can be abstractly measured with {\em matching 
metrics}, that are also introduced in this work. As a by-product of our study, we 
also show that the semiunitary precoder is near-optimal in the {\em relative 
antenna asymptotic setting}\footnote{That is, when $\frac{M}{N_r} \rightarrow 0$ 
or $\infty$ as $\{ M, N_t, N_r\} \rightarrow \infty$.} for any channel. This 
paper generalizes previous work~\cite{vasanth_limfb} on the beamforming case 
($M = 1$), where we studied the performance of the statistical beamforming 
scheme.


\noindent {\bf \em Organization:} 
After elucidating the system model in Section~\ref{sec2}, we benchmark the 
structure of the {\em optimal structured precoder} in the perfect CSI case in 
Section~\ref{sec3}. Using tools from majorization theory, we show that the optimal 
input naturally extends the channel-diagonalizing input from the unconstrained 
case~\cite{lee_petersen,salz,yang_roy,scaglione_gia_barb,sampath,sampath2,yang_roy2,scaglione_spbgs,palomar_precode}. 
In Section~\ref{sec4}, we elaborate on the problem setup of structured precoding. 
In Sections~\ref{sec5}-\ref{sec7}, using tools from random matrix theory and 
eigenvector perturbation theory, we study the asymptotic (in antenna dimensions) 
performance of a {\em statistical semiunitary precoder} that excites the $M$-dominant 
eigenvectors of the transmit covariance matrix. We provide numerical studies to 
illustrate the benefits of the proposed precoding scheme under realistic system 
assumptions in Section~\ref{sec8} 
with a discussion of our results and conclusions in Section~\ref{sec9}. 
Proofs of most of the claims have been relegated to the appendices. 

\noindent {\bf \em Notation:} 
The $M$-dimensional identity matrix is denoted by ${\bI}_M$. The $i,j$-th and 
$i$-th diagonal entries of a matrix $\bX$ are denoted by $\bX(i,j)$ and $\bX(i)$, 
respectively. In more complicated settings (for example, when the matrix $\bX$ 
is represented as a product or sum of many matrices), the above entries are 
denoted by $\bX_{ij}$ and $\bX_i$, respectively. The complex conjugate, conjugate 
transpose, regular transpose and inverse operations are denoted by $(\cdot)^{\star}$, 
$(\cdot)^{H}$, $(\cdot)^{T}$ and $(\cdot)^{-1}$ while the expectation, the trace 
and the determinant operators are given by $\bEe [\cdot]$, $\trace(\cdot)$ and 
$\det(\cdot)$, respectively. The $t$-dimensional complex vector space is denoted 
by ${\mathbb C}^{t}$. The standard big-Oh ($\ord$) and small-oh ($\littleo$) 
notations are used along with the {\em standard ordering for eigenvalues} of an 
$n \times n$-dimensional Hermitian matrix $\bX$: $\lambda_1(\bX) \geq \cdots \geq 
\lambda_n(\bX)$. The largest and the smallest eigenvalues are often denoted also 
by $\lambda_{\max}(\bX)$ and $\lambda_{\min}(\bX)$, respectively. The notation $x^+$ 
stands for $\max(x,0)$.

\section{System Setup} 
\label{sec2} 
We consider a communication model with $N_{t}$ transmit and $N_r$ receive 
antennas, where $M$ ($1 \leq M \leq N_t$) independent data-streams are used in 
signaling. That is, the $M$-dimensional input vector $\bs$ is precoded into an 
$N_t$-dimensional vector via the $N_t \times M$ precoding matrix ${\bf F}$ and 
transmitted over the channel. The discrete-time baseband signal model used is 
\begin{align} 
\label{eq_system_equation} 
\by = {\bf H}  \hsppp {\bf F} \hsppp \bs  + {\bf n}, 
\end{align} 
where $\by$ is the $N_r$-dimensional received vector, ${\bf H}$ is the 
$N_{r} \times N_{t}$-dimensional channel matrix, and ${\bf n}$ is the 
$N_r$-dimensional 
(zero mean, unit variance) additive white Gaussian noise. In practice, the choice 
of $M$ is decided based on a trade-off between complexity, cost and performance 
gain.

\subsection{Channel Model} 
\label{sec_chmod}
The main emphasis of this work is on the impact of spatial correlation. 
We isolate the spatial aspect by assuming a block fading, narrowband model 
for the time-frequency correlation of ${\bf H}$. It is well-known that 
Rayleigh fading (zero mean complex Gaussian) is an accurate model for 
$\bH$ in a non line-of-sight setting and hence, the complete spatial 
statistics are described by the second-order moments of $\{\bH(i,j)\}$. 

The most general, mathematically tractable spatial correlation model is a 
{\emph{canonical decomposition}}\footnote{This model is referred to as the 
``eigen-beam or beamspace model'' in~\cite{canonical_bonek} and is used in capacity 
analysis in~\cite{tulino_ind}.} of the channel along the transmit and the receive 
covariance bases~\cite{canonical_bonek,tulino_ind,vasanth_it_rks08}. In this 
model, we assume that the auto- and the cross-covariance matrices of all 
rows of ${\bf H}$ have the same eigen-basis (denoted by ${\bf U}_t$), and 
the auto- and the cross-covariance matrices of all the columns of ${\bf H}$ 
have the same eigen-basis (denoted by ${\bf U}_r$). Thus, we can decompose 
${\bf H}$ as 
\begin{eqnarray}
\label{canl}
\bH = \bU_r \hsppp \bH_{  \ind } \hsppp  \bU_t^{\sl H},
\end{eqnarray} 
where $\bH_{\ind}$ has independent, but not necessarily identically distributed 
entries, and $\bU_t$ and $\bU_r$ are unitary matrices. The transmit and the 
receive covariance matrices are defined as 
\begin{eqnarray}
\bSigma_t \triangleq \bEe [ \bH^H \bH ] & = & 
\bU_t \hsppp \bEe [ \bH_{\ind}^H  \bH_{\ind} ] \hsppp \bU_t^{\sl H} 
= \bU_t \bfLambda_t \bU_t^{\sl H}, \\ 
\bSigma_r \triangleq \bEe [ \bH \bH^H ] & = & 
\bU_r \hsppp \bEe [ \bH_{\ind}  \bH_{\ind}^H ] \hsppp \bU_r^{\sl H} 
= \bU_r \hsppp \bfLambda_r \hsppp \bU_r^{\sl H}, 
\end{eqnarray} 
where $\bfLambda_t = \bEe [ \bH_{\ind}^H  \bH_{\ind} ]$ and $\bfLambda_r = 
\bEe [ \bH_{\ind}   \bH_{\ind}^H ]$ 
are diagonal. 

Under certain special cases, the model in~(\ref{canl}) reduces 
to some well-known spatial correlation models such as the i.i.d.\ 
model, the separable correlation~\cite{chuah} and the virtual 
representation~\cite{akbar,venu_capacity,akbar_and_venu} 
frameworks. The readers are referred to~\cite{vasanth_limfb_precode} 
for details. The i.i.d.\ model, while being analytically tractable, is 
unrealistic for applications where large antenna spacings or a rich 
scattering environment are not possible. Even though the separable model 
may be an accurate fit under certain channel conditions~\cite{kron_int2}, 
deficiencies acquired by the separability property result in misleading 
estimates of system 
performance~\cite{deficiency_kron1,deficiency_kron2,vasanth_it_rks08}. 
The readers are referred to~\cite{deficiency_kron1,canonical_bonek,zhou} 
for more details on how the canonical, and more specifically the virtual 
model fit measured data better. Given a correlated channel, in this work, 
we will assume without any loss in generality that $M \leq {\sf rank} 
({\bf \Lambda}_t) \leq N_t.$

\subsection{Channel State Information} 
Initial works in the precoding literature have assumed perfect CSI at both the 
transmitter and the receiver. Perfect CSI at the receiver (the coherent case) 
is usually reasonable for systems that adopt a `training followed by signaling' 
model. On the other hand, both the perfect and the no CSI assumptions at the 
transmitter are unrealistic, being too optimistic and too pessimistic, 
respectively. This is so because the perfect CSI condition imposes a huge burden 
on the training or the feedback apparatus on the reverse link while on the other 
hand, the spatial statistics of the channel entries evolve over much slower 
timescales and can be learned at both the ends. In this work, we study the 
coherent case with perfect statistical knowledge at the transmitter. 

\subsection{Transceiver Architecture} 
\label{sec_highlight} 
The transmitted vector ${\bf F}{\bf s}$ (see~(\ref{eq_system_equation})) has a 
power constraint $\rho$. The transmit power constraint can be rewritten as 
\begin{eqnarray} 
\label{power} 
\rho = \bEe \left[ {\bf s}^H \hsppp {\bf F}^H \hsppp {\bf F} \hsppp {\bf s} 
\right] = \trace \left( \bEe \left[ {\bf F} \hsppp {\bf s} \hsppp 
{\bf s}^H \hsppp {\bf F}^H \right] \right) = 
\trace \left( {\bf F} \hsppp \bQ_s \hsppp {\bf F}^H \right), 
\hsp 
\bQ_s \triangleq \bEe \left[ {\bf s} \hsppp {\bf s}^H \right]. 
\end{eqnarray}
By decomposing ${\bf F}$ and $\bQ_s$ using singular value decomposition (SVD) 
and renormalizing, it can be seen that the system equation can be written 
as: 
\begin{eqnarray}
\by & = & {\bf H} \hsppp {\bf F} \hsppp {\bf s} + {\bf n} , \hsp 
{\bf F} = \sqrt{ \frac{\rho}{M} } \hsppp {\bf V}_{ {\bf F} } \hsppp 
{\bfLambda}_{{\bf F}}^{1/2},  \label{general}
\end{eqnarray} 
where ${\bf V}_{{\bf F}}$ is an $N_t \times M$ semiunitary matrix, 
${\bfLambda}_{{\bf F}}$ is an $M \times M$ non-negative definite 
power shaping (allocation)
matrix with $\trace(\bfLambda_{ {\bf F} } ) \leq M$, and $\bs$ is an $M \times 1$ 
vector with i.i.d.\ components that have zero mean and variance one. That is, 
the general precoder can be thought of as a power loading by $\bfLambda_{\bFa}$, 
followed by a rotation with ${\bf V}_{\bFa}$. 

The optimal reception strategy of the input symbols corresponds to non-linear 
maximum likelihood (ML) decoding. However, the exponential complexity of ML 
decoding in both antenna dimensions and coherence length implies that simpler 
receiver architectures are preferred. In this work, we assume a linear minimum 
mean-squared error (${\sf MMSE}$) receiver. With this receiver, the symbol 
corresponding to the $k$-th data-stream is recovered by projecting the received 
signal ${\bf y}$ on to the $N_r \times 1$ vector 
\begin{eqnarray}
{\bf g}_k & = & \sqrt{ \frac{\rho}{M} }  \left( \frac{\rho}{M} 
\bH {\bf F} {\bf F}^H {\bf H}^H + {\bf I}_{N_r}  \right)^{-1} 
{\bf H} {\bf f}_{k},
\end{eqnarray} 
where ${\bf f}_{k}$ is the $k$-th column of ${\bf F}$. That is, the recovered 
symbol is $\widehat{{\bf s}}(k) = 
{\bf g}_k^H {\bf y}$, and the signal-to-interference-noise ratio ($\sinr$) at the 
output of the linear filter ${\bf g}_k$ is 
\begin{eqnarray}
\sinr_k =  
\frac{1} { \left[ \left( {\bI}_M + 
\frac{ \rho  } { M } {\bf F}^{\sl H} {\bf H}^{\sl H} 
{\bf H} {\bf F} \right)^{-1} \right]_{k} } - 1.  
\label{sinr_exp}
\end{eqnarray} 
Also, note that the $\mse$ of the $k$-th data-stream, $\mse_k$, 
is given by $\left[ \left( {\bI}_M + \frac{ \rho  } { M } {\bf F}^{\sl H} 
{\bf H}^{\sl H} {\bf H} {\bf F} \right)^{-1}\right]_{k}$.

\subsection{A Case for Structured Precoding}
\label{case_struct}
Almost all of the current works on precoder design do not assume any specific 
structure on the precoder matrix ${\bf F}$. This is because the main focus of 
these works is on characterizing the fundamental performance limits of precoding. 
That is, to study optimal signaling schemes from a mutual information or 
an error probability viewpoint. 

The structure\footnote{By structure, we mean a set of eigenvectors and 
eigenvalues of ${\bf F}_{\opt}$, that are captured by ${\bf V}_{ {\bf F}_{\opt} }$ 
and ${\bf \Lambda}_{ {\bf F}_{\opt} }$, in~(\ref{general}).} of the optimal precoder, 
${\bf F}_{\opt}$, critically depends on the knowledge of the eigenspace of $\bH$ 
(see Sec.~\ref{sec3}). Even a small inaccuracy in the knowledge of the eigenspace 
of $\bH$ could lead to a precoder with a significantly degraded 
performance~\cite{goldsmith_review,bolsckei,david_corr,vasanth_limfb_precode,vasanth_limfb}. 
While this issue does not arise in the perfect CSI case, it is critical in 
systems with imperfect CSI. In particular, imperfect channel 
knowledge arises in practice due to constraints on the quality and frequency 
of channel or statistical feedback and channel estimation at the receiver. 

Moreover, even if perfect CSI is available at the transmitter, 
the efficient utilization of this information is constrained by 
fundamental limits on energy per bit constraints at the 
computational or processing 
level~\cite{keyes,meindl_davis,rabaey,razavi,pamela}. 
These limits in turn imply that a large number of computations are difficult 
to realize in low-power devices, such as those found at the mobile ends. 
For example, the move towards multi-carrier signaling and the fast rate at 
which channel realizations evolve leads to computational limits on how 
many SVD operations can be afforded. Another key aspect to note is that 
the eigenspace of the optimal input could change dramatically from one 
channel realization to the next, and this poses constraints on the 
adaptivity of the solutions proposed in the literature. In 
fact, RF design constraints imposed by the above limits are often the 
principal stumbling blocks in realizing multi-antenna systems in practice. 
The readers are referred to~\cite{razavi} for a broad array of RF design 
challenges, imposed by computational and complexity constraints. 

All of the above reasons suggest that it may not be possible for ${\bf F}$ 
to be designed at an arbitrarily fast rate. They 
also suggest that ${\bf F}$ cannot have arbitrary structure and one 
cannot learn it with arbitrarily fine precision. The case of statistical 
precoding, where the optimal input is adapted in response to the statistical 
information has thus received significant attention. In this case, computing 
the optimal power allocation across the excited modes requires either Monte 
Carlo averaging or gradient descent-type approaches (see Sec.~\ref{sec4}). 
The affordability of the complexity of these approaches at the mobile end is 
again questionable. 

These reasons motivate us to study {\em structured precoding,} where the 
eigen-modes as well as the power allocation across them are determined via 
low-complexity operations on the channel statistics. The additional structure 
imposed on ${\bf F}$ serves the following purposes: 1) Isolating the impact 
of inaccuracy in the singular vectors and singular values of ${\bf F}$ on 
performance with respect to a genie-aided design, 2) Given that there are 
resource constraints on the reverse link quantization, identifying those 
features of the channel ${\bf H}$ that require an appropriate resource 
allocation so as to optimize system performance, and 3) Obtaining more 
realistic `intermediate' benchmarks for systems in practice. 

We first focus on a specific class of {\emph{semiunitary precoder}}, 
where ${\bf \Lambda}_{\bf F} = \bI_M$. We then consider the more general 
structured precoder case, where ${\bf \Lambda}_{{\bf F}}$ is fixed, but 
is chosen different from the identity matrix.

\section{Perfect CSI Benchmark for Structured Precoding} 
\label{sec3}
Towards the eventual goal of studying a structured statistical precoding 
scheme, we first characterize the optimal perfect CSI benchmark in this 
section. 

\subsection{Unconstrained Precoders}
If only one data-stream is excited 
($M = 1$), the received $\snr$ is given by $\rho \hsppp 
\frac{ |{\bf z}^H {\bf H} {\bf f} |^2} { {\bf z}^H {\bf z} }$, where 
${\bf f}$ is the beamforming vector and ${\bf z}$ is the combining vector. 
It is straightforward to note that the jointly optimal design of ${\bf z}$ 
and ${\bf f}$ can be reduced to a beamformer 
design by using the combining vector $\frac{ {\bf H} {\bf f} } 
{ \sqrt{ {\bf f}^H {\bf H}^H {\bf H} {\bf f} } }$, and that the optimal choices 
${\bf f}_{\opt}$ and ${\bf z}_{\opt}$ are the dominant right singular vector of 
${\bf H}$ and $\frac{ {\bf H} {\bf f}_{\opt}  } { \sqrt{ \lambda_{\max}( {\bf H}^H 
{\bf H} ) } }$, respectively~\cite{paulraj_book}. In this case, the received 
$\snr$ coincides with 
$\rho \hspp \lambda_{\max}( {\bf H}^H {\bf H} )$. 

In contrast to beamforming, the precoding case with $M > 1$ requires a recourse 
to the study of eigenvalues of products of Hermitian matrices. For the (general) 
unconstrained precoding case, the joint precoder-equalizer design turns out to 
have a {\em channel diagonalizing} structure. To state this result, we need 
some additional notation. Let 
an SVD of ${\bf H}$ be given by 
${\bf H} = {\bf U}_{\bf H} {\bf \Lambda}_{\bf H} {\bf V}_{\bf H}^{\sl H}$, where 
${\bf V}_{\bf H} = [ {\bf v}_1 \cdots {\bf v}_{N_t}  ]$. Without any loss in 
generality, we assume that the non-trivial singular values of ${\bf H}$ are 
arranged in the standard order.

\begin{lem}
\label{lem_unstruct} 
The optimal choice of ${\bf V}_{ {\bf F}_{\opt} }$ and 
${\bf \Lambda}_{ {\bf F}_{\opt} }$
in~(\ref{general}) are as follows: ${\bf V}_{ {\bf F}_{\opt} }$ corresponds to 
$[ {\bf v}_1 \cdots {\bf v}_M ]$, and the 
diagonal entries of $\bfLambda_{ {\bf F}_{\opt} }$ are obtained via waterfilling. 
\end{lem}
\begin{proof} 
The optimality of the channel diagonalizing structure has been proved 
in~\cite{lee_petersen,salz,yang_roy,scaglione_gia_barb}, with the design metric 
being the average $\mse$ of the data-streams. Other design metrics where the 
channel diagonalizing structure is optimal include weighted $\mse$ of the 
data-streams~\cite{sampath,sampath2}, determinant of the $\mse$ 
matrix~\cite{yang_roy2}, and a peak-power 
constraint metric~\cite{scaglione_spbgs}. 
A unified convex programming framework for precoder optimization is proposed 
in~\cite{palomar_precode} by studying two broad classes of functions: 
Schur-concave\footnote{The definitions of Schur-concave and Schur-convex 
functions are provided in Appendix~\ref{app_majorize}.} and Schur-convex 
functions. In~\cite{palomar_precode}, the authors show that most of the above 
design criteria can be formulated as either a Schur-concave or Schur-convex 
function of the $\mse$ and the channel diagonalizing structure is optimal in 
either case. 
\end{proof}

\subsection{Semiunitary Precoders} 
When the precoders are constrained to be structured, it is intuitive (but not 
obvious) to expect a channel diagonalizing structure to be optimal. The following 
series of propositions elucidate the optimality of this structure in the semiunitary 
case with certain restrictions on the objective function. The more general 
structured case will be considered thereafter. The readers are referred to 
App.~\ref{app_majorize} for many relevant definitions and results from 
majorization theory. Following the introduction from App.~\ref{app_majorize}, we 
are prepared for the following. 

\subsubsection{Precoders that Optimize Schur-concave Objective Functions}
\begin{prop} 
\label{prop_opt_precoder}
Let $f : {\mathbb{R}}^M \mapsto {\mathbb{R}}$ be a Schur-concave function over 
its domain. Also, let $f(\cdot)$ be monotonically increasing in its arguments. 
That is, let the univariate function $f( \cdots , x_k , \cdots ): {\mathbb{R}} 
\mapsto {\mathbb{R}}$ be monotonically increasing for all $k$. If $\mse = 
[ \mse_1 \hspp \cdots \hspp \mse_M]$, then the optimal choice of semiunitary 
precoder ${\bf F}_{\opt}$ that minimizes $f(\mse)$ is given by 
\begin{eqnarray} 
\label{fopt1}
{\bf F}_{\opt} = 
[ \bv_1 \hsppp \cdots \hsppp \bv_M]. 
\end{eqnarray} 
\end{prop} 
\begin{proof}
See Appendix~\ref{pf_prop_opt_precoder}. 
\end{proof} 

The utility of the above proposition can be gauged from the fact that a large 
class of useful functions satisfy the Schur-concavity property. For example, 
from Remark~\ref{maj_rem} in App.~\ref{app_majorize}, we see that any weighted 
arithmetic or geometric mean of $\{ \mse_k \}$ (with weights chosen appropriately) 
is Schur-concave. The same remark illustrates the limitations of this partitioning 
because the mutual information function {\emph{cannot}} (in general) be expressed 
as a Schur-concave (or a Schur-convex) function of $\mse$. 

In the special case of Gaussian inputs, the objective function 
$f(\cdot)$ to be maximized is 
\begin{eqnarray} 
f(\cdot) = \log \det \left( \bI_M + \frac{\rho}{M} {\bf F}^H 
{\bf H}^H {\bf H} {\bf F} \right) = 
-\log \det \left( {\bf E} \right), 
\end{eqnarray}
where ${\bf E}$ is the mean-squared error matrix defined as 
\begin{eqnarray} 
\bEe[ ({\bf s} - \widehat{ {\bf s} } ) ({\bf s} - \widehat{ {\bf s} } ) ^H] 
\triangleq  \left( \bI_M + \frac{\rho}{M} {\bf F}^H {\bf H}^H {\bf H} {\bf F} 
\right)^{-1}. 
\end{eqnarray}
It can be shown that maximizing the mutual information with the Gaussian input 
(or alternately, minimizing the determinant of ${\bf E}$) can be easily accommodated 
in the framework of Prop.~\ref{prop_opt_precoder};~see~\cite{palomar_precode} for 
details. Alternately, an easy consequence of 
Lemma~\ref{lem_poincare}~(see App.~\ref{app_majorize}) is the fact that a channel 
diagonalizing structure maximizes mutual information and this has been established 
in~\cite{david_heath_multimode}. Also note that if $M = N_t$, any choice of ${\bf F}$ 
unitary leads to the same value of $f(\cdot)$. Extending the proof 
of~\cite{david_heath_multimode} to the case of a non-Gaussian input requires 
closed-form expressions for the mutual information, which are (in general) 
difficult to obtain.

\subsubsection{Precoders that Minimize the Average Error Probability} 
\label{ssec1}
Besides mutual information, uncoded error probability is another important 
metric that describes the performance of a communication system. We now show 
how the machinery of majorization theory can be used to study the error probability. 
We state the most general form of this study in the following proposition, with 
its particularization to the error probability case illustrated thereafter. 
\begin{prop}
\label{prop_schur_cvx}
Let $h: {\mathbb{R}} \mapsto {\mathbb{R}}$ be a continuous, increasing, and 
convex function of its argument. 
The optimal choice of ${\bf F}$ that minimizes 
$\sum_{k=1}^M h (\mse_k)$ is given by 
\begin{eqnarray} 
\label{fopt2}
{\bf F}_{\opt} = 
\left[ \bv_1 \hsppp \cdots \hsppp \bv_M \right] \hsppp \bGamma , 
\end{eqnarray} 
where $\bGamma$ is an appropriately chosen unitary matrix (see 
App.~\ref{pf_prop_opt_precoder} for details on construction). 
\end{prop}
\begin{proof}
See Appendix~\ref{pf_prop_opt_precoder}. 
\end{proof}

If 
$h(\cdot)$ is as in 
Prop.~\ref{prop_schur_cvx}, and $g : {\mathbb{R}}^M \mapsto {\mathbb{R}}$ is 
defined as 
\begin{eqnarray}
g(\mse) \triangleq \sum_{k=1}^M h(\mse_k), 
\end{eqnarray}
then it is important to note from Lemma~\ref{lem_schur_cvx} in App.~\ref{app_majorize} 
that $g(\cdot)$ is a Schur-convex function of 
$\mse$. Thus, in general, Prop.~\ref{prop_schur_cvx} is neither a consequence of 
nor implies Prop.~\ref{prop_opt_precoder}. 

We now show how Prop.~\ref{prop_schur_cvx} is useful in the error probability 
setting. Let $P_{\err}$ denote the probability that at least one of the $M$ 
data-streams is in error. Then, 
\begin{eqnarray}
P_{\err} = 1 - \prod_{k=1}^M (1 - P_{k}), 
\end{eqnarray} 
where $P_{k}$ is the probability that the $k$-th data-stream is in error. 
If some fixed constellation is used for signaling across all the data-streams, 
we can write $P_k$ as 
\begin{eqnarray}
P_k = \alpha {\cal Q} \left( \beta \hsppp  \big( \sinr_k \big)^{1/2} \right), 
\end{eqnarray}
where $\sinr_k$ is the received $\sinr$ of the $k$-th data-stream after linear 
processing~\cite{proakis}, $\alpha$ and $\beta$ are constants dependent only on 
the type of the constellation, and ${\cal Q}(\cdot)$ is the ${\cal Q}$-function 
associated with a standard Gaussian random variable. Assuming that the error 
probability of the weakest data-stream is sufficiently small (which is reasonable 
for most design problems), we have $P_{\err} \approx \sum_{k = 1}^M P_{k}$. 
Alternately, one could consider a metric that measures the average error 
probability of the individual data-streams: $\frac{1}{M} \hsppp \sum_{k = 1}^M P_{k}$. 
Thus, in either case, we are interested in studying the optimal choice of precoder 
${\bf F}$ that minimizes $\sum_{k = 1}^M P_{k}$. 

It is straightforward to note that $P_{k}(\cdot)$ is a continuous and increasing 
function of $\mse$. Besides, it is shown in~\cite{palomar_precode} that $P_{k}(\cdot)$ 
is a convex function\footnote{In particular, it is shown in~\cite[App.~H]{palomar_precode} 
that if the corresponding bit error rate values satisfy ${\sf BER} < 0.02$, this 
is true independent of the input constellation. Moreover, in the case of BPSK 
and QPSK constellations, $P_{k}(\cdot)$ is convex over the entire domain of 
$\mse$. Note that, as stated in~\cite{palomar_precode}, 
the assumption of ${\sf BER} < 0.02$ is mild in a practical 
scenario since the uncoded ${\sf BER}$ is usually much smaller than $0.02$.} 
of $\mse$ as long as the argument is sufficiently small. We are thus justified in 
assuming that $P_k(\cdot)$ is convex, continuous and increasing in $\mse$. Then, 
Prop.~\ref{prop_schur_cvx} shows that $P_{\err}$ is minimized by ${\bf F}_{\opt}$ 
as in (\ref{fopt2}). 

\subsubsection{Precoders that Optimize Schur-convex Objective Functions} 
\label{ssec2}
It is natural to probe the optimality of ${\bf F}_{\opt}$ in (\ref{fopt2}) if 
instead of the average error probability, we considered the error probability 
corresponding to the weakest data-stream. For this, we now need the counterpart 
of Prop.~\ref{prop_opt_precoder} which is as follows. 
\begin{prop} 
\label{prop_opt_scv}
Let $f : {\mathbb{R}}^M \mapsto {\mathbb{R}}$ be a Schur-convex function over 
its domain. Also, let $f(\cdot)$ be monotonically increasing in its arguments. 
The optimal choice of semiunitary precoder ${\bf F}_{\opt}$ that minimizes 
$f(\mse)$ is given by 
\begin{eqnarray} 
\label{fopt}
{\bf F}_{\opt} = 
[ \bv_1 \hsppp \cdots \hsppp \bv_M] \hsppp 
\bGamma, 
\end{eqnarray} 
where $\bGamma$ is the same unitary matrix as defined in Prop.~\ref{prop_schur_cvx}. 
\end{prop} 
\begin{proof}
The proof follows along the same lines as Prop.~\ref{prop_schur_cvx}. 
No details are provided. 
\end{proof}

To answer the question that led towards the above proposition, note from 
Lemma~\ref{lem_used} in App.~\ref{app_majorize} that 
$\max_k P_k$ is a Schur-convex 
function of $\mse$. Thus from Prop.~\ref{prop_opt_scv}, the optimal precoder is 
as in (\ref{fopt}). Further, note that the matrix ${\bf \Gamma}$ in the description 
of ${\bf F}_{\opt}$ in (\ref{fopt2}) and (\ref{fopt}) can be ignored since ${\bf s}$ 
is i.i.d.\ and therefore, so is ${\bf \Gamma} {\bf s}$. 

\subsection{General Structured Precoders}
We now generalize our results to the general structured case. 
\begin{prop}
\label{prop_gem} 
Let the structure of the precoder be ${\bf F} = {\bf V}_{\bf F} \hsppp 
{\bf \Lambda}_{\fixed}^{1/2}$, where ${\bf \Lambda}_{\fixed}$ is some fixed 
matrix of rank $M$ with $\trace({\bf \Lambda}_{\fixed}) \leq M$, albeit chosen 
{\em arbitrarily}. That is, in the ensuing optimization ${\bf \Lambda}_{\fixed}$ 
is fixed and we only optimize over ${\bf V}_{\bf F}$. As before, the structure 
of the optimal ${\bf V}_{ \bf F}$ depends on the nature of the objective function. 
\begin{itemize} 
\item 
Schur-concave objective functions (and in particular, the mutual information with 
Gaussian input) are optimized by ${\bf F}$ of the form: 
\begin{eqnarray}
\label{foptx}
{\bf F}_{\opt} = 
[ \bv_1 \hsppp \cdots \hsppp \bv_M] \hspp  {\bf \Lambda}_{\fixed}^{1/2}.
\end{eqnarray} 
\item 
Schur-convex objective functions (and in particular, the average uncoded error 
probability) are optimized by ${\bf F}$ of the form: 
\begin{eqnarray} 
\label{foptx1}
{\bf F}_{\opt} = 
[ \bv_1 \hsppp \cdots \hsppp \bv_M] 
\hspp {\bf \Lambda}_{\fixed}^{1/2} \hspp {\bf \Gamma} 
\end{eqnarray} 
for an appropriately chosen unitary matrix ${\bf \Gamma}$. 
\end{itemize}
\ignore{ 
Let the structure of the precoder be ${\bf F} = {\bf V}_{\bf F} \hsppp 
{\bf \Lambda}_{\fixed}^{1/2}$ where ${\bf \Lambda}_{\fixed}$ is a fixed matrix. 
In the first two cases studied in the previous section, the optimal precoder is 
\begin{eqnarray} 
\label{foptx}
{\bf F}_{\opt} = \exp(j \theta) \hsppp [ \bv_1 \hsppp \cdots \hsppp \bv_M] \hspp  
{\bf \Lambda}_{\fixed}^{1/2}  
\end{eqnarray} 
whereas in the next two cases, we have 
\begin{eqnarray} 
\label{foptx1}
{\bf F}_{\opt} = \exp(j \theta) \hsppp [ \bv_1 \hsppp \cdots \hsppp \bv_M] \hspp 
{\bf \Lambda}_{\fixed}^{1/2} \hspp {\bf \Gamma} 
\end{eqnarray} 
for an appropriately chosen unitary matrix ${\bf \Gamma}$ and $\theta \in 
{\mathbb{R}}$. 
}
\end{prop}
\begin{proof}
We follow the same proof techniques of 
Prop.~\ref{prop_opt_precoder}-\ref{prop_opt_scv}. See Appendix~\ref{pf_prop_opt_precoder} 
for details. 
\end{proof}
Thus, even in the more general structured precoding case, the channel 
diagonalizing structure is optimal. 

\section{Statistical Precoding: Preliminaries} 
\label{sec4} 
We now assume that instantaneous channel information is not available at the 
transmitter, but channel statistics are known. 

\subsection{Notations} 
\label{nots}
While much of the notations 
required in the rest of the paper have been established in 
Sec.~\ref{sec_chmod}, we find it convenient to restate some of them 
that are often used in the ensuing sections. 
We assume that ${\bf H}$ is described by either the separable model or the 
more general non-separable model of~(\ref{canl}). Let the variance of 
${\bf H}_{\ind}(i,j)$ be denoted by $\sigma_{ij}^2$. The eigenvalues of the 
transmit covariance matrix are denoted by $\{ {\bf \Lambda}_t(k) \}$ in the 
separable case while in the non-separable case, they 
are denoted by $\gamma_{t,k} \triangleq \sum_{i=1}^{N_r} \sigma_{ik}^2$. 
In either case, we assume 
that the columns of $\bH_{\ind}$ are arranged such that the transmit 
eigenvalues are in decreasing order. The channel power of ${\bf H}$, $\rho_c$, 
is given by $\rho_c = \sum_{i=1}^{N_r} {\bf \Lambda}_r(i) = 
\sum_{i=1}^{N_t} {\bf \Lambda}_t(i)$. The normalized channel power is 
$\gamma_r \triangleq \frac{\rho_c}{N_r}$. 

In the separable case, let ${\widetilde{\bfLambda} }_t$ denote the 
principal $M \times M$ sub-matrix of $\bfLambda_t$ and 
${\bf \widetilde{H}}_{\iid}$ denote the $N_r \times M$ principal sub-matrix 
of ${\bf H}_{\iid}$. That is, 
\begin{eqnarray}
{\bf H}_{\iid} = \left[ \begin{array}{cc} 
\underbrace { {\bf \widetilde{H}}_{\iid} }_{ N_r \times M }  
&   \underbrace{ \times }  _{N_r \times (N_t - M)}
\end{array} \right]. 
\end{eqnarray} 
Without any explicit reference to $k$, we will often denote by 
${\widehat{\bfLambda} }_t$, the $(M-1) \times (M-1)$ matrix obtained from 
${\widetilde{\bfLambda} }_t$ by removing the $k$-th row and $k$-th column 
and by ${\bf \widehat{H}}_{\iid}$, the matrix obtained from 
${\bf \widetilde{H}}_{\iid}$ by removing the $k$-th column alone. 
In the non-separable case, let $\widetilde{\bH}_{\ind}$ denote the 
$N_r \times M$-dimensional principal sub-matrix of ${\bH}_{\ind}$.

\subsection{Unconstrained Precoders} 
\begin{lem} 
\label{lem_unstructured_opt}
The optimal precoder ${\bf F}_{\stat, \hsppp \opt}$ is of the form 
${\bf V}_{\stat} \hsppp {\bf \Lambda}_{\stat}^{1/2}$, 
where ${\bf V}_{\stat}$ is a set of $M$ dominant eigenvectors of the transmit 
covariance matrix ${\bf \Sigma}_t$ and ${\bf \Lambda}_{\stat}$ is the unique 
solution to the following constrained optimization: 
\begin{eqnarray}
{\bf \Lambda}_{\stat} = \arg \max _{ {\bf \Lambda} \in {\cal L} } 
\bEe_{\bH }  \left[ \log \det \left( \bI_{N_r} + \frac{\rho}{M} \hspp 
\widetilde{\bH}_{\ind} \hsppp {\bf \Lambda} \hsppp {\widetilde{\bH}}_{\ind}^H \right) 
\right] \label{copt}
\end{eqnarray}
with ${\cal L} = \{  {\bf \Lambda} \}$ denoting the 
convex set of all diagonal $M \times M$ non-negative definite matrices such that 
$\trace( {\bf \Lambda} ) \leq M$. 
\endproof 
\end{lem}
The optimality of the dominant eigenvectors of ${\bf \Sigma}_t$ is not surprising 
(see~\cite{visotsky,jafar,jorswieck_bfcap,moustakas_stat_phy,venu_capacity,tulino_ind,goldsmith_review,bolsckei} 
and references therein for problems of a similar nature). The optimization 
in~(\ref{copt}) is standard: Maximizing a concave function over a convex set. 
A gradient descent-type approach for this is provided 
in~\cite{zhang_palomar} and a Monte Carlo approach is provided 
in~\cite{venu_capacity,tulino_ind,harish_varanasi}. 

\subsection{Structured Statistical Precoders} 
As explained in~Sec.~\ref{case_struct}, the complexity of solving for 
${\bf \Lambda}_{\stat}$ in~(\ref{copt}) may be 
unaffordable in many practical scenarios. 
We therefore pursue two statistics-based precoders: ${\bf F}_{\semi}$ and 
${\bf F}_{\fixed}$, with ${\bf F}_{\semi} = {\bf V}_{\stat}$ and 
${\bf F}_{\fixed} = {\bf V}_{\stat} \hsppp {\bf \Lambda}_{\fixed}^{1/2}$. 
The choice of ${\bf \Lambda}_{\fixed}$ that is of interest here is: 
\begin{eqnarray}
\label{lam_fixed}
{\bf \Lambda}_{\fixed}(k) = \left\{   \begin{array}{cc}
M \cdot \frac{{\bf \Lambda}_t(k) } { \sum_{j=1}^M {\bf \Lambda}_t(j)  } 
& {\rm if} \hspp \rho < \snr_{\sf T}, \\ 
1 & {\rm if} \hspp \rho \geq \snr_{\sf T}. 
\end{array}
\right.
\end{eqnarray}
The threshold $\snr$ ($\snr_{\sf T}$) is such that 
\begin{eqnarray}
\snr_{\sf T} = \alpha \frac{M}{ {\bf \Lambda}_t(M) } 
\end{eqnarray}
for an appropriate choice of $\alpha, \alpha > 1$. This choice is motivated by 
our recent work~\cite{vasanth_isit07_heath} on transient-$\snr$ (the $\snr$ at 
which exciting $M$ modes is information theoretically optimal) design.

For a given channel realization, let $I_{\stat, \hsppp \semi}(\rho)$ and 
$P_{\err, \hsppp \stat, \hsppp \semi}(\rho)$ denote the mutual information and 
error probability achievable with ${\bf F}_{\semi}$, while 
$I_{\stat, \hsppp \fixed}(\rho)$ and $P_{\err, \hsppp \stat, \hsppp \fixed}(\rho)$ 
denote the corresponding quantities with ${\bf F}_{\fixed}$, all at an $\snr$ of 
$\rho$. Similarly, denote the corresponding quantities with the three perfect CSI 
precoders described in Lemma~\ref{lem_unstruct}, (\ref{fopt1}) and (\ref{foptx}) 
by: $I_{\perf, \hsppp \unstruct}(\rho)$, $I_{\perf, \hsppp \semi}(\rho)$, 
$I_{\perf, \hsppp \fixed}(\rho)$, and $P_{\err, \hsppp \perf, \hsppp \unstruct}(\rho)$, 
$P_{\err, \hsppp \perf, \hsppp \semi}(\rho)$, 
$P_{\err, \hsppp \perf, \hsppp \fixed}(\rho)$, respectively. It is important to note 
the distinction between these quantities. While $I_{\stat, \hsppp \bullet}(\rho)$ and 
$P_{\err, \hsppp \stat, \hsppp \bullet}(\rho)$ are functions of the channel 
realization ${\bf H}$, the precoder structure itself is independent of ${\bf H}$, 
but only dependent on the channel statistics. On the other hand, 
$I_{\perf, \hsppp \bullet}(\rho)$ and $P_{\err, \hsppp \perf, \hsppp \bullet}(\rho)$ 
in addition to being dependent on the channel realization also correspond to 
precoders whose structure is dependent on ${\bf H}$ and chosen optimally. 

\subsection{Average Relative Difference Metrics} 
Towards the goal of studying the proposed scheme(s), we develop 
{\em universal} metrics that 
capture the performance gap between the proposed precoder(s) and an ideal benchmark. 
We first motivate the choice of our metric in an abstract context. 

Let `scheme $1$' and `scheme $2$' denote two signaling schemes with 
$I_{ {\sf scheme}, \hsppp 1 }(\rho)$ and $I_{ {\sf scheme}, \hsppp 2 }(\rho)$ 
denoting the mutual information of the two schemes at an $\snr, \rho$. Our goal 
is to quantify\footnote{In our setting, `scheme $1$' corresponds to a perfect CSI 
precoder and `scheme $2$' to a structured statistical precoder.} 
whether scheme $1$ is better 
than scheme $2$ or not, and if so, by how much. For any signaling scheme, 
the average mutual information is a function of $\rho$ as well as the statistical 
description of the channel. Irrespective of the spatial correlation, the average 
mutual information of any scheme tends to zero as $\rho \rightarrow 0$ and tends to 
infinity as $\rho \rightarrow \infty$. For this reason, the difference in average 
mutual information between the two schemes can converge to zero as $\rho \rightarrow 0$ 
at a rate different from that of either scheme, and could blow up to infinity as 
$\rho \rightarrow \infty$. Thus, the difference in average mutual information is 
{\em not} a good measure for comparing the two schemes. 

An efficient comparison of the two schemes is possible by using either of the 
following set of {\em average relative difference metrics}: 
\begin{eqnarray} 
\Delta I_{  {\rm scheme } \hsppp 1, \hspp {\rm scheme} \hsppp 2 } 
& \triangleq & \frac{ \bEe_{\bH} \left[ I_{ {\sf scheme}, \hsppp 1 }(\rho) - 
I_{ {\sf scheme}, \hsppp 2 }(\rho) \right] } 
{ \bEe_{\bH}[ I_{ {\sf scheme}, \hsppp 2 }(\rho) ] }, 
\label{kl2} \\ 
\widetilde{\Delta I} _{ {\rm scheme } \hsppp 1, \hspp {\rm scheme} \hsppp 2  } 
& \triangleq & \bEe_{\bH} \left[ \frac{I_{ {\sf scheme}, \hsppp 1 }(\rho) - 
I_{ {\sf scheme}, \hsppp 2 }(\rho)} 
{I_{ {\sf scheme}, \hsppp 2 }(\rho)} \right]. 
\label{kl1} 
\end{eqnarray} 
Note that the choice of scheme $2$ in the denominator of~(\ref{kl2}) and~(\ref{kl1}) 
is the scheme that performs relatively poorly. Thus, $\Delta I_{\bullet}$ and 
$\widetilde{\Delta I} _{\bullet}$ correspond to a worst-case measure of relative 
performance. The metrics are more meaningful (than the difference metric) in 
studying the relative gap (or closeness) between the 
schemes\footnote{\label{fn_delta_comp}Empirical studies indicate that the correlation 
coefficient between 
$\frac{I_{ {\sf scheme}, \hsppp 1 }(\rho)} {I_{ {\sf scheme}, \hsppp 2 }(\rho)}$ 
and $I_{ {\sf scheme}, \hsppp 2 }(\rho)$ is negative. While this claim seems 
plausible given the reciprocal role of $I_{ {\sf scheme}, \hsppp 2 }(\rho)$ in 
the two terms, we do not have a concrete mathematical proof of this claim. If this 
claim were to be true, we would have $\Delta I _{\bullet} \leq 
\widetilde{ \Delta I}_{\bullet}$. In any case, it should be clear that 
$\Delta I_{\bullet}$ and $\widetilde{\Delta I}_{\bullet}$ are related to each 
other by an $\ord(1)$ factor. In Sec.~\ref{sec5} and~\ref{sec6}, we will characterize 
either coefficient depending on its tractability.}, 
{\emph{independent}} of the $\snr$. While we 
have used the case of average mutual information to motivate the need for a relative 
difference metric, the same argument is applicable in the error probability case. In 
fact, the need for such a metric is more critical in the error probability case since 
the error probabilities of the schemes that are being compared (and hence, the 
difference between them) are small. 

\subsection{Problem Setup} 
The main goal of this paper is to quantify, as a function of the statistics and 
antenna dimensions, 
\begin{eqnarray} 
\label{lk3}
\Delta I_{\semi} & \triangleq  
& \frac{ \bEe_{\bH} \left[ I_{\perf, \hsppp \unstruct}(\rho) - 
I_{\stat, \hsppp \semi}(\rho) \right]  } 
{   \bEe_{\bH} \left[ I_{\stat, \hsppp \semi}(\rho) \right]} 
\end{eqnarray}
in the case of mutual information, and 
\begin{eqnarray} 
\label{lk4}
\Delta P_{\semi} & \triangleq  & \bEe_{\bH} \left[ 
\frac{ P_{\err, \hsppp \stat, \hsppp \semi}(\rho) - 
P_{\err, \hsppp \perf, \hsppp \unstruct}(\rho) } 
{   P_{\err, \hsppp \perf, \hsppp \unstruct}(\rho) } \right] 
\end{eqnarray}
in the case of error probability. In addition, we are also interested in the 
corresponding quantities for ${\bf F}_{\sf fixed}$ in~(\ref{lam_fixed}): 
$\Delta I_{\sf fixed}$ and $\Delta P_{\sf fixed}$. 

While closed-form expressions for the above metrics seem difficult to obtain 
across all $\snr$ regimes, the following simplifying assumptions render these 
metrics theoretically tractable. 
\begin{itemize}
\item 
{\bf{\emph{Asymptotics of Antenna Dimension(s):}}} 
Any performance metric computation in the spatially 
correlated, finite antenna setting suffers from fundamental difficulties 
associated with a lack of knowledge of 
the joint probability density function of singular values of the channel matrix. 
However, under many settings, in the asymptotics of antenna dimension(s), the 
density function of eigenvalues converges (in an appropriate sense) to a certain 
deterministic density function. Many recent works on multi-antenna 
channels~(see~\cite{venu_capacity,tulino_ind,goldsmith_review,bolsckei} and references 
therein) exploit this fundamental property in the characterization of various 
information theoretic quantities of interest. 

\ignore{ 
The characterization of the various average relative difference metrics in this 
work also critically depends on invoking the assumption of antenna asymptotics. 
In this context, recent results have shown that the convergence of finite antenna 
results to asymptotic estimates is usually fast. In particular, numerical as well 
as theoretical results show that this convergence is on the order of inverse of 
antenna dimension~\cite{vasanth_it_weak06}. Thus the asymptotic results, that will 
be stated later in this section, are also useful in the finite antenna setting 
(see Sec.~\ref{sec8}). 
}

In this work, we find it useful to separate our study 
into two cases: 1) An easily tractable 
case of {\emph{relative receive antenna asymptotics}}, where $\frac{M}{N_r} 
\rightarrow 0$, and 2) A more difficult case of {\emph{proportional growth of 
antenna dimensions}}, where both $\{ M, N_r \} \rightarrow \infty$ with 
$\frac{M}{N_r} \rightarrow \gamma$ and $\gamma \in (0,\infty)$ is a constant. The 
first case includes the following sub-cases in a unified way: a) $N_t$ and 
$M$ are finite and $N_r \rightarrow \infty$, b) $\big\{ M, N_r \big\} \rightarrow 
\infty$ with $\frac{M}{N_r} \rightarrow 0$, and c) via a relabeling of indices 
the case where $\frac{M}{N_r} \rightarrow \infty$ with either $N_r$ finite or 
$N_r \rightarrow \infty$. 

\item 
{\bf{\emph{Signaling Constellation:}}} 
In the error probability case, it will be shown in Sec.~\ref{sec6} that the 
relative difference metric can be written in terms of the $\sinr$ of the individual 
data-streams. Since exact closed-form expressions are known for the $\sinr$s 
(see~(\ref{sinr_exp})) of a linear MMSE receiver, independent of the signaling 
constellation, there is no need to constrain the inputs to be of any particular 
type. On the other hand, in the case of mutual information, when Gaussian inputs 
are used for signaling, the average mutual information is given by the well-known 
$\log \det(\cdot)$ formula. 
However, in the non-Gaussian 
case, closed-form expressions are difficult to obtain for mutual information. Thus, 
we will restrict our attention to average relative mutual information loss in the 
Gaussian case. In the non-Gaussian case, the relative $\mse$ enhancement 
is a good indicator\footnote{The mutual information is related to the 
$\mse$ of the optimal MMSE receiver through the relationship established 
in~\cite{guo_mmse}, and not the $\mse$ of the linear MMSE receiver. Despite this 
difficulty, the $\mse$ enhancement with a linear MMSE receiver is a good indicator 
of mutual information loss in the non-Gaussian case~\cite{guo_mmse}.} of the mutual 
information loss. Besides this, the $\mse$ enhancement serves as a soft 
decision metric when the processed received data is fed through more complex, non-linear 
receiver architectures such as a turbo- or LDPC-decoder.

\item 
{\bf{\emph{High-$\snr$ Regime:}}} 
Computing {\emph{universal}} upper bounds for the metrics in~(\ref{lk3}) 
and~(\ref{lk4}), and the corresponding quantities for ${\bf F}_{\sf fixed}$, 
that are {\emph{tight}} across the entire $\snr$ range seems to be a 
difficult proposition. However, when the $\snr$ is reasonably high (more 
precisely, $\rho \geq \alpha \frac{M}{ {\bf \Lambda}_t(M) }$ for some 
suitable $\alpha > 1$), we will see that considerable 
simplifications and hence, closed-form characterizations are possible. 
In this $\snr$ regime, the semiunitary precoder coincides with the precoder 
in~(\ref{lam_fixed}) as does the performance of another commonly-used 
low-complexity receiver, the zeroforcing receiver. 
\end{itemize}

\section{Mutual Information Loss with Semiunitary Precoding} 
\label{sec5} 
In this section, we focus on the (average) relative loss in mutual 
information with ${\bf F}_{\sf semi}$, assuming Gaussian inputs. The 
difference $\Delta I_{\semi}$ (see~(\ref{lk3})) can be written as 
\begin{eqnarray}
\Delta I_{\semi} & = & \underbrace{\frac{ \bEe_{\bH} 
\left[ I_{\perf, \hspp \unstruct}(\rho) - 
I_{\perf,\hspp \semi}(\rho) \right] }
{ \bEe_{\bH} 
\left[ I_{\stat, \hspp \semi}(\rho) \right]  } }_{\Delta I_1}+ 
\underbrace{ 
\frac{ \bEe_{\bH} \left[ I_{\perf, \hspp \semi}(\rho) - 
I_{\stat,\hspp \semi}(\rho) \right] }
{ \bEe_{\bH} \left[ I_{\stat, \hspp \semi}(\rho) \right]  } }_{\Delta I _2}. 
\end{eqnarray}
Since the argument within the expectation of the numerator of $\Delta I_1$ is 
not explicitly dependent on the spatial correlation model, it is straightforward 
to obtain a bound for $\Delta I_1$. 
\begin{prop}
\label{Delta_I1}
If $\rho$ is such that $\rho \geq \alpha 
\bEe_{\bH} \left[ \frac{M} { { \bf \Lambda}_{\bH}(M)  } \right]$ for some 
$\alpha > 1$, $\Delta I_1$ is bounded as 
\begin{eqnarray}
\label{closs1}
\Delta I_1 \leq \frac{2 M} 
{ \alpha^2 \bEe_{\bH} \left[ I_{\stat, \hspp \semi}(\rho) \right]  } 
\cdot \frac{\bEe_{\bH} \left[ \left ( 
\frac{1}{  {\bf \Lambda}_{\bH}(M) } \right)^2  \right] }
{ \left( \bEe_{\bH} 
\left[ \frac{1}{ {\bf \Lambda}_{\bH}(M) } \right] \right)^2 }. 
\end{eqnarray}
\end{prop}
{\vspace{0.1in}}
\begin{proof}
See Appendix~\ref{term1}.
\end{proof} 

Intuitively, as $\alpha$ and hence the $\snr$ increases, the waterfilling power 
allocation of the optimal precoding scheme converges to uniform power 
allocation across the $M$ modes (see~\cite{goldsmith_review,bolsckei,venu_capacity} etc.) 
and thus, $\Delta I_1$ decreases. The bound provided in~(\ref{closs1}) is not 
tight since we have not characterized the exact probability ${\rm Pr}(n_{\bH} < M)$ 
(in App.~\ref{term1}) that determines $\Delta I_1$. But the above 
bound is sufficient to capture the performance loss with uniform power allocation. 

Characterization of $\Delta I_2$, which is explicitly dependent on the spatial 
correlation model, is non-trivial. In the following series of theorems, we provide 
bounds for different correlation models and regimes. We first consider 
the relative antenna asymptotic case.

\subsection{Separable Model} 

\begin{thm} 
\label{thm_loss_cap} 
Let the channel ${\bf H}$ be described by the separable model. From the 
remark in Footnote~\ref{fn_delta_comp}, $\Delta I_2$ is well-approximated 
by its more tractable version, $\widetilde{\Delta I}_2:$ 
\begin{eqnarray}
\widetilde{\Delta I}_2 \triangleq 
\bEe _{{\bf H} } \left[ \frac{ I_{\perf, \hspp \semi}(\rho) 
 - I_{\stat , \hspp \semi } (\rho) } 
 { I_{\stat, \hspp \semi}(\rho) }  \right].
\end{eqnarray} 
For any 
fixed value of $\rho$, $\widetilde{ \Delta I}_2$ is bounded as 
\begin{eqnarray}
\label{x1}
\widetilde{ \Delta I}_2  \leq \frac{2 \kappa_1}{\gamma_r} \cdot 
\frac{ \sqrt{ \sum_{i=1}^{N_r} ( {\bf \Lambda}_r(i) )^2  } }
{N_r} \cdot \frac{1}{M}
\sum_{k = 1}^M \frac{1} { \log \left( 1  + \transnr \hsppp 
\bfLambda_t(k) \right) }, 
\end{eqnarray} 
where $\kappa_1$ is a constant determined from an application of 
Lemma~\ref{yinbailemma} (in App.~\ref{app_majorize}). 
\end{thm}
{\vspace{0.1in}}
\begin{proof}
See Appendix~\ref{app_thm2}.  
\end{proof}

\subsection{Canonical Model} 
\begin{thm} 
\label{thm_loss_cap1} 
Consider the canonical case with $\frac{N_t}{N_r} \rightarrow 0$. Using the 
generalized asymptotic eigenvalue characterization in Lemma~\ref{yinbailemma} 
(in App.~\ref{app_majorize}) and following the approach of 
Theorem~\ref{thm_loss_cap}, we have 
\begin{eqnarray}
\label{x2}
\Delta I_2  \leq  2\kappa_2 \cdot \sqrt{ \frac{N_t}{N_r} } \cdot 
\frac{N_r}{M} \sum_{k=1}^M \left[ \frac{1}{ 
\gamma_{t,k} \log \left( 1 + \frac{\rho }{M} \hsppp \gamma_{t,k} \right)
 } \right]
\end{eqnarray}
{\vspace{0.1in}} 
for some constant $\kappa_2$ determined from Lemma~\ref{yinbailemma}. 
The proof is not provided. 
\endproof 
\end{thm}


\subsection{Special Case: Beamforming} 
We now pay attention to the beamforming case ($M = 1$), 
the low-complexity of which makes it an attractive signaling choice in many 
wireless standards. While the $\snr$ regime where beamforming is capacity-optimal 
has been established in prior 
work~\cite{goldsmith_review,bolsckei,venu_capacity,vasanth_isit07_heath}, the performance 
gap between statistical and perfect CSI beamforming is less clear. Using tools 
from eigenvector perturbation theory, introduced in~\cite{vasanth_limfb}, we 
establish the following results. 

First, note that the term $\Delta I_1$ is redundant in the beamforming case. 
Let $I_{\perf}(\rho)$ and $I_{\stat}(\rho)$ denote the mutual information 
achievable by beamforming with perfect CSI and statistical information alone, 
respectively. Define the loss term 
\begin{eqnarray} 
\Delta I_{\sf bf} \triangleq \frac{ {\bEe}_{\bH} 
\left[ I_{\perf}(\rho) - I_{\stat}(\rho) \right]} 
{ \bEe_{\bH} \left[ I_{\stat}(\rho) \right] }. 
\end{eqnarray}
The following discussion complements recent work on the performance gap with 
the separable model~\cite{mckay}, that have been established by exploiting some 
recent advances in random matrix theory. Unlike~\cite{mckay} which is 
based on exact random matrix theory results and is applicable only for 
${\bEe} \left[ I_{\perf}(\rho) - I_{\stat}(\rho) \right]$ in the separable 
case, we generalize the results to the canonical modeling framework, but 
do not consider fine refinement of constants in the following results for 
the sake of brevity. 

\begin{prop} 
\label{thm_tp_bf} 
There exists a constant $\kappa_3$ such that $\Delta I_{\sf bf}$ is given by 
{\vspace{-0.05in}}
\begin{eqnarray} 
\label{eqn_kron_bf} 
\Delta I_{\sf bf} 
& \leq & \frac{ \log \left( 1 + \rho \hsppp \kappa_3 \cdot 
\sqrt{ \frac{ N_t \log(N_r) }{N_r}  } \right) } 
{\bEe_{\bH}\left[ I_{\stat}(\rho) \right] }. 
\end{eqnarray} 
The constant $\kappa_3$ is model- (separable or canonical) and 
regime- (proportional growth or relative asymptotics) dependent. 
Simple bounds for $\kappa_3$ are as follows: 
1) ${\bf \Lambda}_t(1) \left( 
1 + \kappa_{3,\hspppp 1} \frac{ \sqrt{N_t N_r}  }{ \rho_c } \right)$ 
for the separable 
and relative asymptotics case, 2) $
\gamma_{t,1} + 
\kappa_{3, \hspppp 2} \sqrt{N_tN_r}$ for the canonical and relative 
asymptotics case, 3) $\frac{\kappa_{3,\hspppp 3} \hsppp N_r} {\rho_c} 
\cdot {\bf \Lambda}_t(1)$ in the proportional growth setting for the 
separable case, and 4) $\kappa_{3, \hspppp 4}\hsppp N_r$ for the 
canonical case. The constants $\kappa_{3,\hspppp i}, \hspp i = 1, \cdots, 4$ 
are independent of $N_t, N_r, {\bf \Sigma}_t$ and ${\bf \Sigma}_r$. 
\end{prop}
\begin{proof}
See Appendix~\ref{app_tp_bf}. 
\end{proof}

\subsection{Proportional Growth of Antenna Dimensions: Separable Case} 
\begin{thm}
\label{thm_proportional2}
Let $\bH$ be characterized by the separable model. Let 
$\{M, N_r \} \rightarrow \infty$ with $\frac{M}{N_r} \rightarrow \gamma$ 
and $\gamma \in (0, \infty)$. Let the following conditions hold: 
1) $\frac{ {\bf \Lambda}_t(1) } 
{ {\bf \Lambda}_t(M) } = \ord(1)$, 2) $\frac{ {\bf \Lambda}_r(1) } 
{ {\bf \Lambda}_t(M) } = \ord(1)$, 3) $\frac{ {\bf \Lambda}_r(M) } 
{ {\bf \Lambda}_t(M) } = \ord(1)$, 4) $\frac{ \sum_{k=1}^M {\bf \Lambda}_t(k) } 
{\rho_c} = b_1 = \ord(1)$, and 
5) $\frac{ \sum_{k=1}^M {\bf \Lambda}_r(k) }{\rho_c} = b_2 = \ord(1)$. 
If $\rho \geq \alpha \frac{M} { {\bf \Lambda}_t(M) }$ for some $\alpha > 1$, 
$\Delta I_2$ is bounded as 
\begin{eqnarray}
\Delta I_2 & \leq & \frac{ \log(e/M) + \kappa_4}
{ \log(\rho/e) + \frac{1}{M} 
\sum_{k=1}^M \log \left( \frac{\bfLambda_t(k) \bfLambda_r(k)} {\rho_c}
\right)  } \label{upperbound_thm2} \\ 
\kappa_4 & = & \kappa_4' + 
\min \left( \bEe_{\bH} \left[ \log \left( 
\frac{\lambda_{\max}(\bH_{\iid}^H \bfLambda_r \bH_{\iid} )} {G_{M,\hsppp \bfLambda_r}} 
\right)  \right], 
\bEe_{\bH} \left[ \log \left( 
\frac{\lambda_{\max}(\bH_{\iid} \bfLambda_t \bH_{\iid}^H )} {G_{M,\hsppp \bfLambda_t}} 
\right)  \right] \right)
\end{eqnarray} 
where $\kappa_4'$ depends only on the constants in the statement of the 
theorem, and $G_{M,\hsppp \bfLambda_{\bullet}}$ are the geometric means of 
eigenvalues, defined as 
\begin{eqnarray} 
G_{M,\hsppp \bfLambda_r} \triangleq 
\left( \prod_{k = 1}^M \bfLambda_r(k) \right)^{1/M}, 
{\hspace{0.2in}} 
G_{M,\hsppp \bfLambda_t} \triangleq 
\left( \prod_{k = 1}^M \bfLambda_t(k) \right)^{1/M}. 
\end{eqnarray}
\end{thm}
{\vspace{0.1in}}
\begin{proof} 
See Appendix~\ref{app_thm_proportional2}. 
\end{proof}

\subsection{Discussion}
\label{match_4}
It is of interest to understand the structure of the 
scheme that is optimal from a mutual information viewpoint for a given 
channel. While many advances have been made along this direction (in particular, 
regarding the eigenvectors of the optimal 
input)~\cite{visotsky,jafar,jorswieck_bfcap,moustakas_stat_phy,shengli_zhou,venu_capacity,tulino_ind,goldsmith_review,bolsckei,harish_varanasi,zhang_palomar,jongren_capside,akhtar}, 
a complete understanding is rendered difficult by the lack of a comprehensive 
random matrix theory for correlated channels. 
Theorems~\ref{thm_loss_cap}-\ref{thm_loss_cap1} provide an alternative approach, 
where we characterize the structure of ${\bf H}$ that is `best' or `worst' 
for a given precoding scheme. 

Let us now freeze ${\bf \Lambda}_r$ to be a fixed matrix so as to develop an 
understanding of the structure of ${\bf \Lambda}_t$ that minimizes performance 
loss. Given that a constraint $\sum_{i=1}^{N_t} {\bf \Lambda}_t(i) = \rho_c$ 
has to be met, it can be checked that performance loss in (\ref{x1}), 
(\ref{x2}) and (\ref{upperbound_thm2}) is minimized by the following 
choice: $\bfLambda_t(1) = \cdots = \bfLambda_t(M) = \frac{\chanpow}{M}$ 
and $\bfLambda_t(M+1) = \cdots = \bfLambda_t(N_t) = 0$. On the other 
extreme, the worst choice of ${\bf \Lambda}_t$ that maximizes the performance 
loss is of the form: ${\bf \Lambda}_t(1) \approx \rho_c$ and ${\bf \Lambda}_t(i) 
\approx 0, i \geq 2$, but with the added constraint that 
${\sf rank}({\bf \Lambda}_t ) \geq M$. It is important to note that the 
largest gap\footnote{In fact, if 
${\sf rank}({\bf \Lambda}_t) = 1$, the statistical precoder achieves the 
same throughput as the optimal precoder.} is {\em not} achieved when 
${\sf rank}({\bf \Lambda}_t) = 1$. Motivated by Theorem~\ref{thm_proportional2}, 
we define a {\em matching metric for the transmitter side}: 
\begin{eqnarray} 
{\cal M}_t \triangleq \prod_{i=1}^M {\bf \Lambda}_t(i), 
\label{matchtxside}
\end{eqnarray}
that captures the closeness of a given channel from the best and worst channels 
(characterized above). As ${\cal M}_t$ increases, the channel becomes more 
matched on the transmitter side and the performance loss decreases and {\em vice 
versa}. 

Capturing the impact of ${\bf \Lambda}_r$ on performance loss is 
difficult since ${\bf \Lambda}_r$ is hidden in the first-order analysis of 
Theorems~\ref{thm_loss_cap1} and~\ref{thm_proportional2}. Nevertheless, 
(\ref{x1}) shows that a {\em matching metric for the receiver side} 
can be defined as 
\begin{eqnarray} 
{\cal M}_r \triangleq \sum_{i=1}^{N_r} \left( {\bf \Lambda}_r(i)  \right)^2. 
\end{eqnarray} 
Again, with a constraint $\sum_{i=1}^{N_r} {\bf \Lambda}_r(i) = \rho_c$ to be 
met, it can be seen that ${\cal M}_r$ is minimized by ${\bf \Lambda}_r = 
\frac{\rho_c}{N_r} \hspp {\bf I}_{N_r}$ and maximized by 
${\bf \Lambda}_r(1) \approx \rho_c$ and ${\bf \Lambda}_r(i) \approx 0, i \geq 2$, 
but with the added constraint that ${\sf rank}({\bf \Lambda}_r ) \geq M$. 
It can be seen that the performance loss is not maximized 
when ${\sf rank}({\bf \Lambda}_r) < M$. 

A channel that is matched on both the transmitter and the receiver sides is 
referred to as a {\em matched channel} and is optimal for the given precoder 
structure (fixed choice of $M$). 
The structure of the matched channel can be summarized as: 1) The 
rank of ${\bf \Lambda}_t$ is $M$ with the dominant transmit eigenvalues 
being well-conditioned, and 2) ${\bf \Lambda}_r$ is also well-conditioned. 
A channel that is ill-conditioned on both the transmit and the receive sides 
such that ${\sf rank}({\bf H}) \geq M$ (with probability $1$) is said to be 
a {\em mismatched channel}. 

An interesting consequence of the study in Theorems~\ref{thm_loss_cap} 
and~\ref{thm_loss_cap1} is that channel hardening, that occurs as $N_r$ increases, 
results in the vanishing of $\Delta I_{\semi}$. That is, {\em statistical 
information is as good as perfect CSI in the receive antenna asymptotics.} 
This behavior is 
peculiar of this asymptotic regime and will also be observed in the error 
probability case. The high-$\snr$ characterization for signaling with $M$ 
spatial modes ($\rho \geq \alpha \frac{M}{ {\bf \Lambda}_t(M) }$ for some 
$\alpha > 1$) has also been identified in prior work~\cite{vasanth_isit07_heath}.

\section{Error Probability Enhancement with Semiunitary Precoding} 
\label{sec6} 
In this section, we study the (average) relative error probability enhancement, 
$\Delta P_{\semi}$, with semiunitary precoding in the high-$\snr$ regime. 
Towards this goal, we first note that $\Delta P_{\semi}$ in~(\ref{lk4}) 
can be written\footnote{Note that $\Delta P_{\semi}$ is independent of 
how error probability is defined: Averaged across data-streams or at least 
one data-stream in error.} as 
\begin{eqnarray}
\Delta P_{\semi} & = & 
\bEe_{\bH} \left[ \frac{ \sum_{k=1}^M 
P_{k, \hsppp \stat, \hsppp \semi}(\rho) 
- P_{k, \hsppp \perf, \hsppp \unstruct}(\rho) 
 } { \sum_{k=1}^M P_{k, \hsppp \perf, \hsppp \unstruct} (\rho) } 
\right] \\ 
& \stackrel{(a)}{\leq } & \bEe_{\bH} \left[ \frac{1}{M} \cdot 
\sum_{k = 1}^M 
\frac{ P_{k,\hsppp \stat, \hsppp \semi}(\rho) - 
P_{k,\hsppp \perf, \hsppp \unstruct}(\rho)  } 
{  P_{k, \hsppp \perf, \hsppp \unstruct} (\rho) } \right], 
\end{eqnarray} 
where (a) follows from Lemma~\ref{lemma_usable}. 

\begin{prop} 
\label{prop_prob_err} 
The loss term, $\Delta P_{\semi}$, can be bounded as 
\begin{eqnarray}
\Delta P_{\semi} \leq \bEe_{\bH} \left[ 
\frac{1}{M} \cdot \sum_{k = 1}^M 
\frac{ \exp\Big(\frac{ \beta^2 \hsppp \Delta \sinr_k }{2} \Big) 
\sqrt{1 + \frac{ \Delta \sinr_k }{ \sinr_{k, \hsppp \stat, \hsppp \semi} } 
} } 
{1 - \Big(\frac{1} {\beta^2 \hsppp \sinr_{k,\hsppp \perf, \hsppp \unstruct}
} \Big) } - 1 \right],  
\label{deltap_semi}
\end{eqnarray} 
where 
\begin{eqnarray}
\Delta \sinr_k & \triangleq &
\sinr_{k,\hsppp \perf, \hsppp \unstruct} - 
\sinr_{k,\hsppp \stat, \hsppp \semi} \nonumber 
\\ &= &   1 + \frac{ {\bf \Lambda}_{\sf wf}(k) 
\lambda_k (\bfLambda_t \bH_{\iid}^H \bfLambda_r \bH_{\iid}) }{\rho_c}
- \frac{\det \left( {\bI}_M + \frac{\rho}{M \hsppp \rho_c} \cdot 
{\widetilde{\bfLambda} }_t^{1/2} \hsppp {\bf \widetilde{H}}_{\iid}^{\sl H} \hsppp 
\bfLambda_r \hsppp {\bf \widetilde{H}}_{\iid} \hsppp {\widetilde{\bfLambda} }_t^{1/2}
\right) } 
{\det \left( {\bI}_{M-1} + \frac{\rho}{M \hsppp \rho_c} \cdot 
{\widehat{\bfLambda} }_t^{1/2} 
\hsppp {\bf \widehat{H}}_{\iid}^{\sl H} \hsppp \bfLambda_r  \hsppp 
{\bf \widehat{H}}_{\iid} \hsppp {\widehat{\bfLambda} }_t^{1/2} 
\right) }. \nonumber 
\end{eqnarray}
{\vspace{0.1in}}
See notations established in Sec.~\ref{nots}. 
\end{prop} 
\begin{proof}
See Appendix~\ref{app_perr_semi}. 
\end{proof}

As in Sec.~\ref{sec5}, we consider the separable and canonical models 
for the relative antenna asymptotic case separately.

\subsection{Separable Model} 
\begin{thm}
\label{thm_prob_loss} 
In the separable case, 
if $\rho \geq \alpha \frac{M}{ {\bf \Lambda}_t(M)}$ for some $\alpha > 1$, 
$\Delta P_{\semi}$ can be bounded as 
\begin{eqnarray}
\Delta P_{\sf semi} & \leq & \frac{1}{\beta^2 M} \sum_{k=1}^M \frac{1} 
{ \frac{ \rho {\bf \Lambda}_t(k)}{M} - 1 } + 
\beta^2 \left( 1 + \frac{M}{\alpha} \right) \nonumber \\ 
& & {\hspace{0.02in}} 
+  \frac{ \beta^2 \rho \sum_{k=1}^M {\bf \Lambda}_t(k) }{M}  \left( 
\frac{1}{\alpha} + \frac{1}{\alpha^2} \cdot 
\frac{ \bEe \left[ \left( \frac{1} { {\bf \Lambda}_{\bH}(M) } \right)^2  \right] } 
{  \left( \bEe \left[ \frac{1}{ {\bf \Lambda}_{\bH}(M) } \right]  \right)^2 } 
+ \frac{1}{\gamma_r} \hspp \ord \left( \frac{ \sqrt{N_t} + \sqrt{M} } 
{\sqrt{N_r}}   \right) \right). 
\label{y1}
\end{eqnarray} 
Thus the dominant term of $\Delta P_{\semi}$ in the relative antenna asymptotics 
and large $\alpha$ is of the form: 
$\frac{1}{\beta^2 \rho } \cdot \sum_{k=1}^M \frac{1}{\bfLambda_t(k)} + 
\beta^2 \frac{\sum_{k=1}^M {\bf \Lambda}_t(k)}{ {\bf \Lambda}_t(M) }$. 
\end{thm}
{\vspace{0.1in}}
\begin{proof}
See Appendix~\ref{app_thm1}. 
\end{proof}

\subsection{Canonical Model} 
We characterize $\Delta P_2$, the performance gap between the statistical and 
perfect CSI semiunitary precoders, alone for the sake of simplicity. Along the 
development of Theorem~\ref{thm_prob_loss}, it is straightforward to extend 
this result to $\Delta P_{\semi}$. 
\begin{thm}
\label{thm_prob_loss_can} 
Let $\rho \geq \alpha \frac{M}{\gamma_{t,M} } = \alpha 
\frac{M} { \sum_{i} \sigma_{iM}^2 }$. The dominant term of 
$\Delta P_{2}$ is bounded as 
\begin{eqnarray}
\Delta P_{2} & \leq & \frac{ \beta^2  \rho}{2 \alpha} \cdot 
\frac{ \sum_{k=1}^M \gamma_{t,k} }{M} + 
\frac{1}{\beta^2 \rho} \sum_{k=1}^M \frac{1}{ \gamma_{t,k} } + 
\frac{ \beta^2 \rho }{2 \gamma_r } \cdot 
\frac{ \sum_{k=1}^M \gamma_{t,k} } {M} \cdot 
\ord \left( \frac{ \sqrt{M} + \sqrt{N_t} } 
{\sqrt{N_r}} \right) \\  
& = & \frac{ \beta^2 }{2} \cdot 
\frac{ \sum_{i=1}^{N_r} \sum_{k=1}^M \sigma_{ik}^2  }  
{ \sum_{i=1}^{N_r} \sigma_{iM}^2 } + 
\frac{1 }{\beta^2 \alpha M} \cdot \sum_{i} \sigma_{iM}^2 \cdot 
\sum_{k=1}^M \frac{ 1 } { \sum_i \sigma_{ik}^2 }. 
\label{y2}
\end{eqnarray} 
\end{thm}
{\vspace{0.1in}}
\begin{proof}
The proof follows along the same lines as Theorem~\ref{thm_prob_loss} 
by applying the second part of Lemma~\ref{yinbailemma} (see 
App.~\ref{app_majorize}). No explicit proof is provided. 
\end{proof} 


\subsection{Special Case: Beamforming}
In the beamforming setting, our earlier work~\cite{vasanth_limfb,vasanth_isit06} 
leverages advances in eigenvector perturbation theory to provide bounds on 
$\Delta P_{\sf bf}$, the gap in performance between statistical and perfect CSI 
beamforming. These results are summarized in the following lemmas. 
\begin{lem}
\label{thm_rate1} 
Let $\bH$ be described by the separable model. 
Assume that ${\bf \Lambda}_t(1) > {\bf \Lambda}_t(2) \left( 1 + \frac{2 } 
{\gamma_r \hsppp N_r^{\eta}} \right)$ for some $\eta > 0$. There exists a constant 
$K_1$ such that 
\begin{eqnarray}
\label{eqn_show}
\Delta P_{\sf bf} \leq  
K_1 \cdot \frac{ \sqrt{ \gammartwo} } { \gammatone \hsppp \gamma_r  } 
\cdot \sqrt{ \frac{N_t \log(N_r)} {N_r } }, 
\end{eqnarray}  
where $\gammartwo$ corresponds to the second moment of the receive eigen-modes 
and $\gammatone$ corresponds to the separation between the transmit eigen-modes, 
and are defined as 
\begin{eqnarray} 
\gammartwo \triangleq 
\frac{\sum_{k=1}^{N_r} \left( {\bf \Lambda}_r(k) \right)^2 } {N_r}, 
\hsp  \hsp 
\gammatone  \triangleq 
1 - \frac{{ \bf \Lambda}_t(2) } { {\bf \Lambda}_t(1)}. 
\end{eqnarray}
\endproof 
\end{lem} 

\begin{lem}
\label{thm_rate2}
Let $\bH$ be described by the canonical model. If $\frac{ \gamma_{t,1} } 
{ N_r } > \frac{\gamma_{t,2} }{ N_r } + 
\frac{2}{N_r^{\eta}}$ for some $\eta > 0$, there exists a constant $K_2$ such that 
\begin{eqnarray} 
\Delta P_{\sf bf} 
\leq K_2 \cdot \left( \gammatc \cdot \gammarc \right)^{1/2} 
\hsppp \sqrt{ \frac{N_t \log(N_r) } {N_r} }, 
\end{eqnarray} 
where $\gammatc$ and $\gammarc$ are defined as 
\begin{eqnarray}
\gammatc \triangleq  \frac{1}{N_t-1} \sum_{k=2}^{N_t} 
\frac{N_r^2} { \left( \gamma_{t,1} - \gamma_{t,k} \right)^2 }, 
\hsp \hsp  
\gammarc  \triangleq  
\max_{j > 1} \frac{ \sum_{i} \sigma_{ij}^2 \sigma_{i1}^2 } {N_r}. 
\end{eqnarray}
\endproof 
\end{lem}

Thus in the asymptotics of $N_r$ relative to $N_t$, 
even channel statistical information is sufficient for near-perfect CSI 
performance. Further, given a fixed $N_t$ and $N_r$, ill-conditioning of 
$\bSigma_t$ and well-conditioning of $\bSigma_r$ reduces 
$\Delta P_{\sf bf}$. 
We also provided evidence in~\cite{vasanth_limfb,vasanth_isit06} that, of 
these two factors, the conditioning 
of $\bSigma_t$ is more critical than that of $\bSigma_r$. 
Theorems~\ref{thm_prob_loss}-\ref{thm_prob_loss_can} provide a multi-mode 
generalization of these results.

\subsection{Discussion}
As in the mutual information case, we are interested in channels that 
minimize and maximize the performance loss $\Delta P_{\sf semi}$. 
From~(\ref{y1}) and~(\ref{y2}), it is observed that the choice of 
${\bf \Lambda}_t$ that minimizes performance loss is such that: 
1) It minimizes $\frac{ {\bf \Lambda}_t(k) } {  {\bf \Lambda}_t(M) }, \hspp 
1 \leq k \leq M$, and 2) It also minimizes $\sum_{k=1}^M 
\frac{1}{ {\bf \Lambda}_t(k) }$. Both of 
these constraints are met by a channel 
that maximizes ${\cal M}_t$ (as defined in~(\ref{matchtxside}) for 
the mutual information case). 
That is, a channel that is matched on the transmitter side from a 
mutual information viewpoint is also matched on the transmitter side 
from an error probability viewpoint. However, it is difficult to make similar 
conclusions about matching on the receiver side. 

\ignore{ 
Similarly, a 

Our first-order analysis provides no insights on the impact of conditioning of 
$\bSigma_r$ on performance loss. But numerical results (as well as analysis in 
the proportional growth case that will soon follow) point to a trend that the 
constant $C_1$ in Lemma~\ref{yinbailemma} and (\ref{eqnc1}) [See 
Appendix~\ref{app_thm1}] which is independent of $\bSigma_t, \hsppp 
\bSigma_r, \hsppp N_t, \hsppp N_r$ can be made dependent on $\bSigma_r$ and 
that the well-conditioning of $\bSigma_r$ reduces this covariance-dependent 
$C_1$. Nevertheless, it is to be noted that $C_1$ and hence $\bSigma_r$ play a 
second-order role (See equations (\ref{eqnc1}) and (\ref{eqnciter}) in 
Appendix~\ref{app_thm1}) in determining 
performance loss, in agreement with our earlier result on statistical 
beamforming~\cite{vasanth_limfb}. 
}

On the other hand, 
note that as the constellation size increases, $\beta$ decreases. Thus, for 
any fixed $\rho$, the first dominant term of $\Delta P_{\sf semi}$ in~(\ref{y1}) 
and~(\ref{y2}) 
increases as the constellation size increases, whereas the second term decreases. 
The tension between the two dominant terms determines the optimal choice of 
constellation to use at a fixed $\snr$ over a given channel. In the extreme case 
of asymptotically high $\snr$, the first term vanishes and $\Delta P_{\semi}$ is 
minimized with the largest constellation available in the signaling set. The 
optimality of a larger constellation at high-$\snr$ from an error probability 
viewpoint is to be intuitively expected. Further, as in the mutual information 
case, channel hardening results in vanishing $\Delta P_{\semi}$ as $N_r$ 
increases. 
In the more realistic case of proportional growth of antenna dimensions, 
it is difficult to establish that $\Delta \sinr_k \rightarrow 0$ as 
$\rho \rightarrow \infty$. We postpone the study of this case to future work.

\section{MSE Enhancement with Statistical Precoding} 
\label{sec7}
We finally consider the (average) relative $\mse$ enhancement. 
Define 
$\Delta \mse$ as 
\begin{eqnarray}
\Delta \mse \triangleq \frac{1}{M}
\bEe_{\bH} \left[ \sum_{k=1}^M 
\frac{\mse_{k, \hsppp \stat, \hsppp \semi} - 
\mse_{k, \hsppp \perf, \hsppp \unstruct} }
{ \mse_{k, \hsppp \perf, \hsppp \unstruct}} \right]. 
\end{eqnarray}
The following proposition establishes the trend of $\Delta \mse$ under 
certain settings. 
\begin{prop}
In the receive antenna asymptotics case, if $\rho \geq \alpha 
\frac{M}{ {\bf\Lambda}_t(M)  }$, $\Delta \mse$ is bounded as 
\begin{eqnarray} 
\frac{ \Delta \mse} {1 + \frac{M}{\alpha}}
& \leq &  \frac{M}{\alpha} + \frac{M}{\gamma_r} \cdot 
\ord \left( \frac{\sqrt{M} + \sqrt{N_t} }  {\sqrt{N_r}} \right)+ 
\frac{1}{M} \sum_{k=1}^M \frac{ {\bf \Lambda}_t(k) 
\left(  {\bf \Lambda}_{\sf wf}(k) - \frac{\rho}{M} \right) } 
{ 1 + \frac{ \rho {\bf \Lambda}_t(k) } {M} } . 
\label{rec_asy}
\end{eqnarray}
{\vspace{0.1in}}
As $\snr$ increases, the dominant term of $\Delta \mse$ is 
\begin{eqnarray} 
\Delta \mse \leq \frac{M}{\gamma_r} \cdot 
\ord \left( \frac{\sqrt{M} + \sqrt{N_t} }  {\sqrt{N_r}} \right). 
\end{eqnarray}
{\vspace{-0.1in}}
\end{prop}
\begin{proof} 
Note that $\mse_{k ,\hsppp \bullet}$ is defined as 
$\mse_{k ,\hsppp \bullet} = \frac{1}{1 + \sinr_{k, \hsppp \bullet}}$ 
and hence, we have 
\begin{eqnarray}
\Delta \mse = \sum_{k=1}^M \bEe_{\bH} \left[ \frac{\Delta \sinr_k} 
{1 + \sinr_{k, \hsppp \stat, \hsppp \semi}} \right]. 
\end{eqnarray} 
Following~(\ref{t1y}) and~(\ref{t2}) in Appendix~\ref{app_thm1}, (\ref{rec_asy}) 
follows immediately in the receive antenna asymptotics case. 
\end{proof}

While we expect $\Delta \mse \rightarrow 0$ in the proportional growth 
case also, we do not have a mathematical proof of this fact. This will be 
addressed in future work. 

\ignore{
This expectation arises from 
Theorem~\ref{thm_proportional2} which shows that $\Delta I_2 \rightarrow 0$ 
and the connection between mutual information and $\mmse$ in~(\ref{ami_general}). 
In fact, in this setting a straightforward computation following on~(\ref{s1}) shows 
that 
\begin{eqnarray}
\Delta \mse = \sum_{k=1}^M \bEe_{\bH} \left[ 
\frac{\lambda_k(\bfLambda_t \bH_{\iid}^H \bfLambda_r \bH_{\iid} ) }
{\bfLambda_t(k) \left[ \bh_k^H \bfLambda_r \bh_k - 
\bh_k^H \bfLambda_r {\bf u} ({\bf u}^H \bfLambda_r {\bf u})^{-1} 
{\bf u}^H \bfLambda_r \bh_k \right]} - 1 \right]
\nonumber 
\end{eqnarray} 
where $\bh_k$ is the $k$-th column of $\bH_{\iid}$ and ${\bf u}$ is the 
$N_r \times (M-1)$ principal sub-matrix of the left singular vector matrix of 
$\widehat{\bH}_{\iid}$. The fact that we are required to compute the expectation 
of a complicated function of a random matrix renders further analysis very 
difficult. 
}

\section{Numerical Studies}
\label{sec8} 
In this section, we illustrate the results established in this paper via 
some numerical studies. We consider $4 \times 4$ channels for our study 
where $M = 2$ data-streams are excited with: 1) Gaussian inputs for the 
mutual information case, and 2) QPSK inputs for the error probability case. 
In all the cases, the channel power is normalized to $N_t N_r = 16$.

\begin{figure}[htb!]
\centering 
\includegraphics[height=9cm,width=10cm]{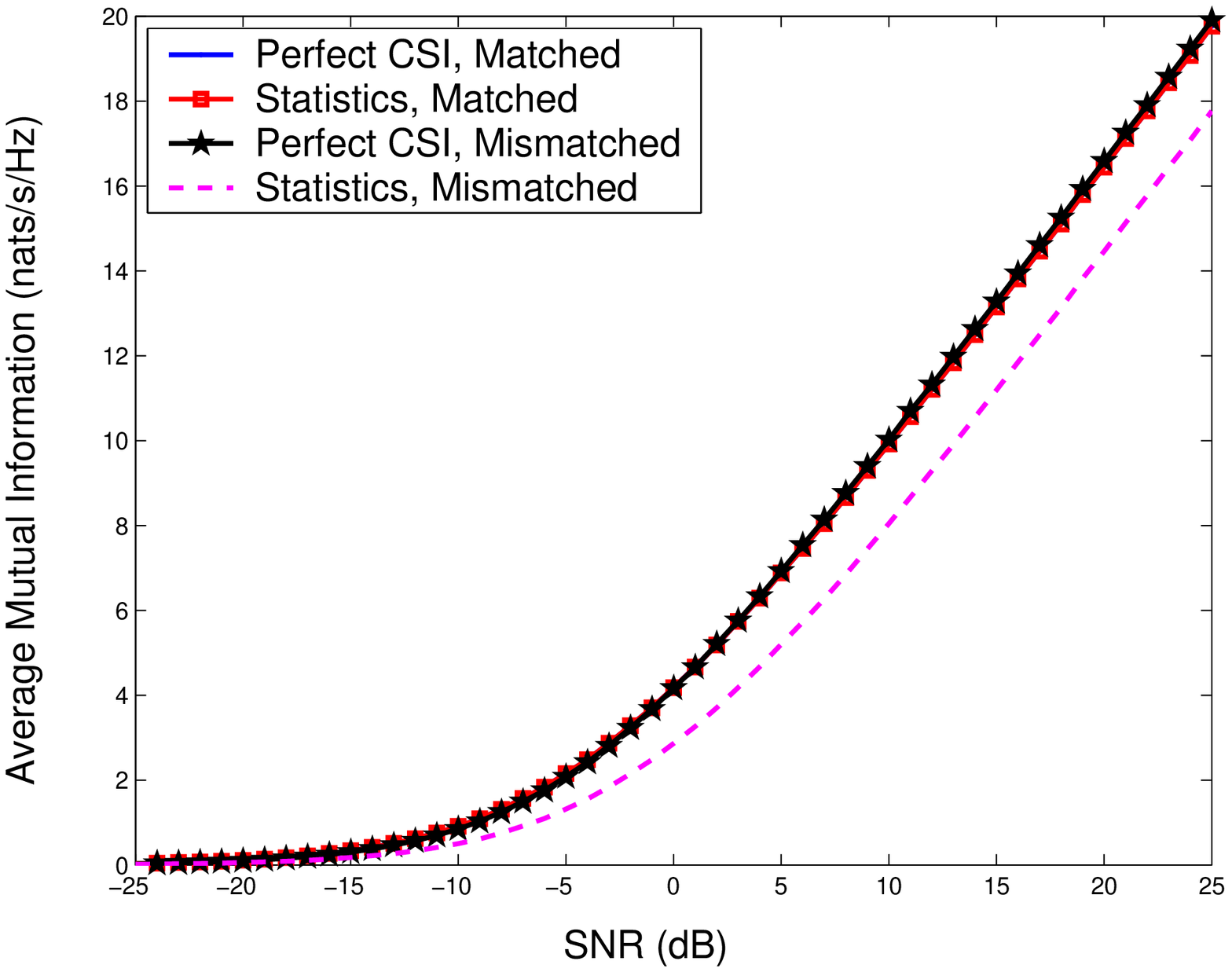}
\caption{\label{fig2} Mutual information of the perfect CSI and the 
statistical semiunitary precoders over {\em matched} and {\em mismatched} channels.} 
\end{figure}

\ignore{ 
\begin{figure}[htb!]
\begin{center}
\begin{tabular}{cc}
\includegraphics[height=2.5in,width=3in]{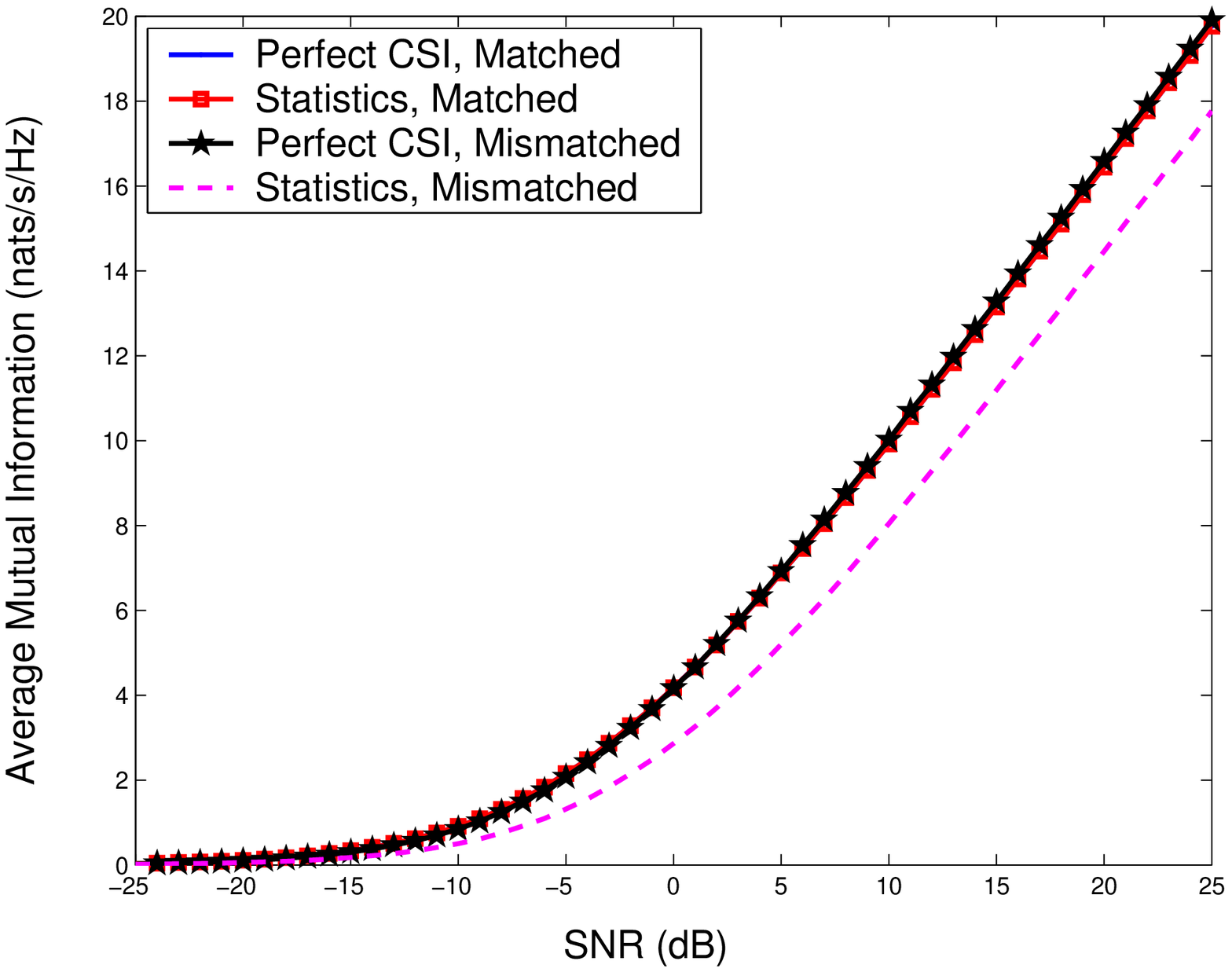} & 
\includegraphics[height=2.5in,width=3in]{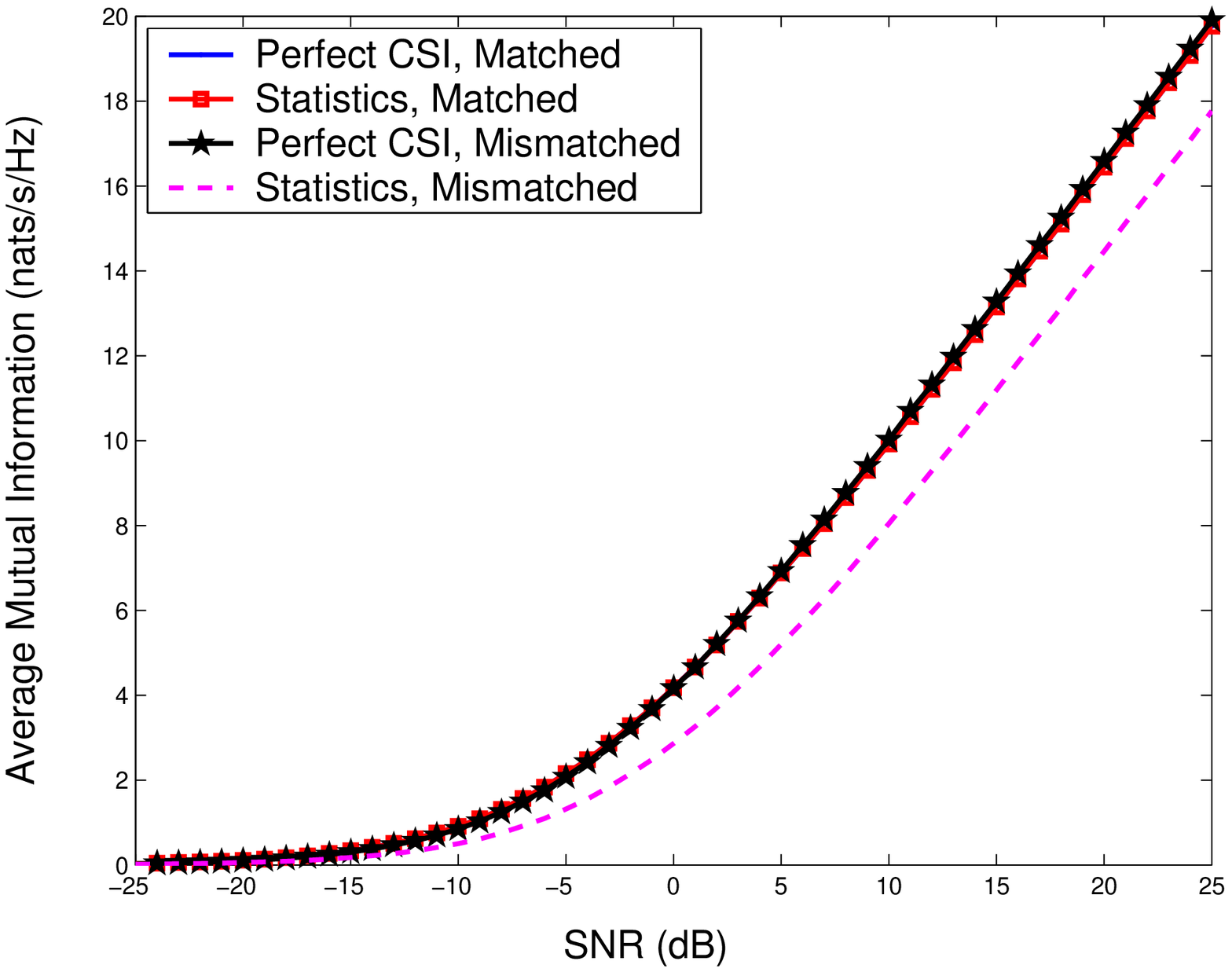}
\\ (a) & (b) 
\end{tabular} 
\caption{\label{fig2} (a) Mutual information and (b) Error probability 
performance of the perfect CSI and the statistical {\em semiunitary} 
precoders over {\em matched} and {\em mismatched} channels.} 
\end{center}
\end{figure}
}

\begin{itemize}
\item 
{\bf {\em Matched vs.\ Mismatched Channels:}} The first study illustrates 
the performance of statistical semiunitary precoding over matched and 
mismatched channels. We consider a $4 \times 4$ {\em matched} channel with 
normalized separable model, where $\diag({\bf \Lambda}_t) = 
[8 \hspp 8 \hspp 0\hspp 0]$. The {\em mismatched} channel is characterized 
by $\diag({\bf \Lambda}_t) = [4 \hspp 4 \hspp 4 \hspp 4]$. In both the cases, 
${\bf \Lambda}_r = 4 \hsppp {\bf I}_4$. Fig.~\ref{fig2} shows the average 
mutual information with perfect CSI and statistical semiunitary precoding 
in the two channels. 

As explained before, the mutual information in the four cases are given by: 
\begin{eqnarray}
I_{{\sf matched}, \hsppp {\sf perf}}(\rho) = 
I_{{\sf matched}, \hsppp {\sf stat}}(\rho) & = &
\bEe \left[ \sum_{i= 1}^M \log \left( 1 + \frac{\rho}{M} \hsppp \frac{N_t}{M} 
\hsppp \lambda_i( \widetilde{\bf H}_{\iid}^H {\widetilde{\bf H}}_{\iid}  )  \right)  
\right] \label{e1}
\\ I_{ {\sf mismatched}, \hsppp {\sf perf} }(\rho) & = & 
\bEe \left[ \sum_{i = 1}^M \log \left( 1 + \frac{\rho}{M} \hsppp 
\lambda_i( {\bf H}_{\iid}^H {\bf H}_{\iid} )  \right)  \right] 
\label{e2} 
\\ 
I_{ {\sf mismatched}, \hsppp {\sf stat}}(\rho)  & = &
\bEe \left[ \sum_{i= 1}^M \log \left( 1 + \frac{\rho}{M}
\hsppp \lambda_i( \widetilde{\bf H}_{\iid}^H {\widetilde{\bf H}}_{\iid}  )  \right)  
\right], \label{e3} 
\end{eqnarray}
where $\widetilde{{\bf H}}_{\iid}$ and ${\bf H}_{\iid}$ are 
$N_r \times M$ and $N_r \times N_t$ i.i.d.\ matrices. As can be seen 
from~(\ref{e1}), (\ref{e3}) and Fig.~\ref{fig2}, the performance of the mismatched 
statistical precoder is $10 \log_{10} \left( \frac{N_t}{M}  \right) \approx 3$ dB 
away from both the matched precoders. It is also surprising that the 
matched precoders have nearly the same performance as the mismatched 
(i.i.d.\ channel) optimal precoder. This seems to be related to the choice of 
$N_t, N_r$, $M$ and eigen-properties of i.i.d.\ random matrices. 

\begin{figure}[htb!]
\begin{center}
\begin{tabular}{cc}
\includegraphics[height=2.5in,width=3in]{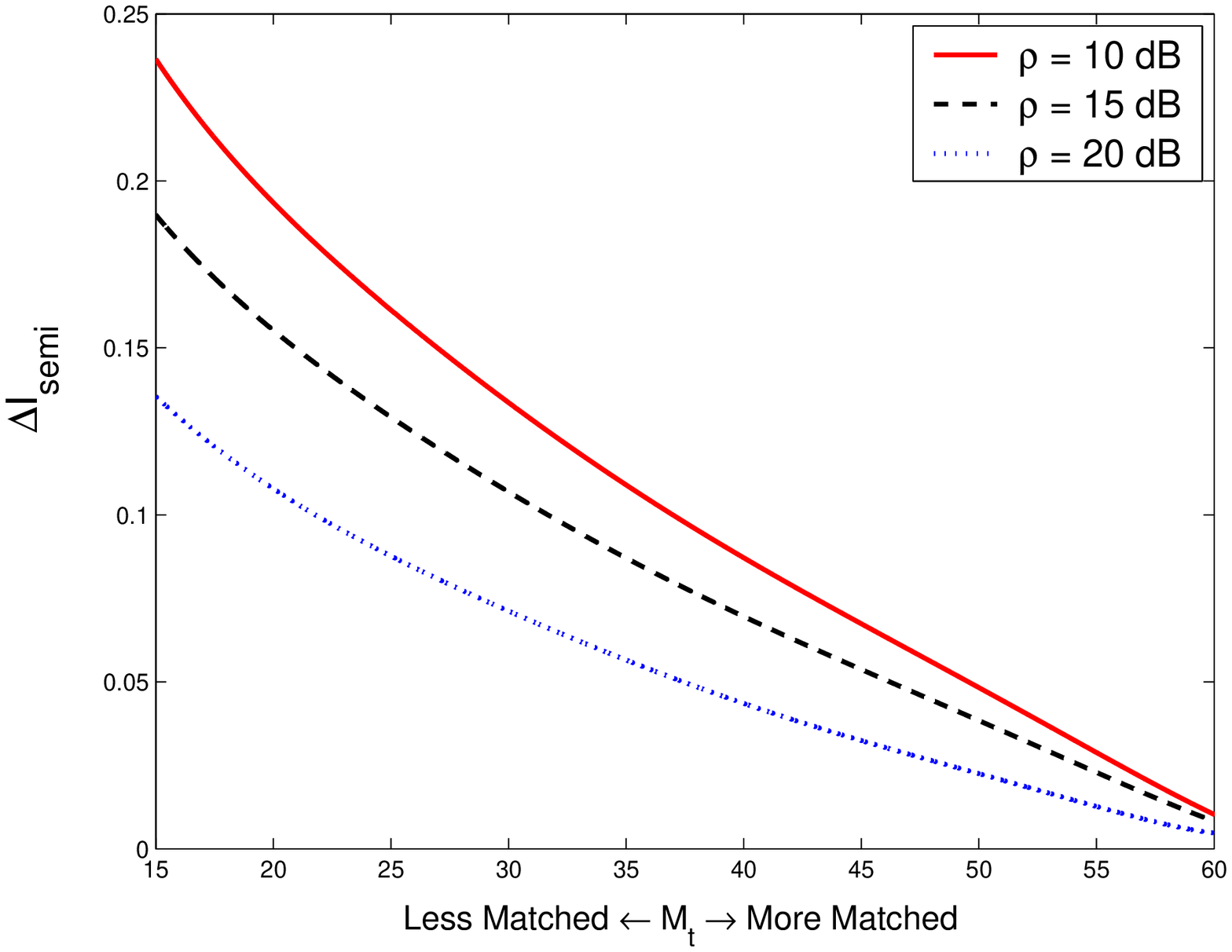} & 
\includegraphics[height=2.5in,width=3in]{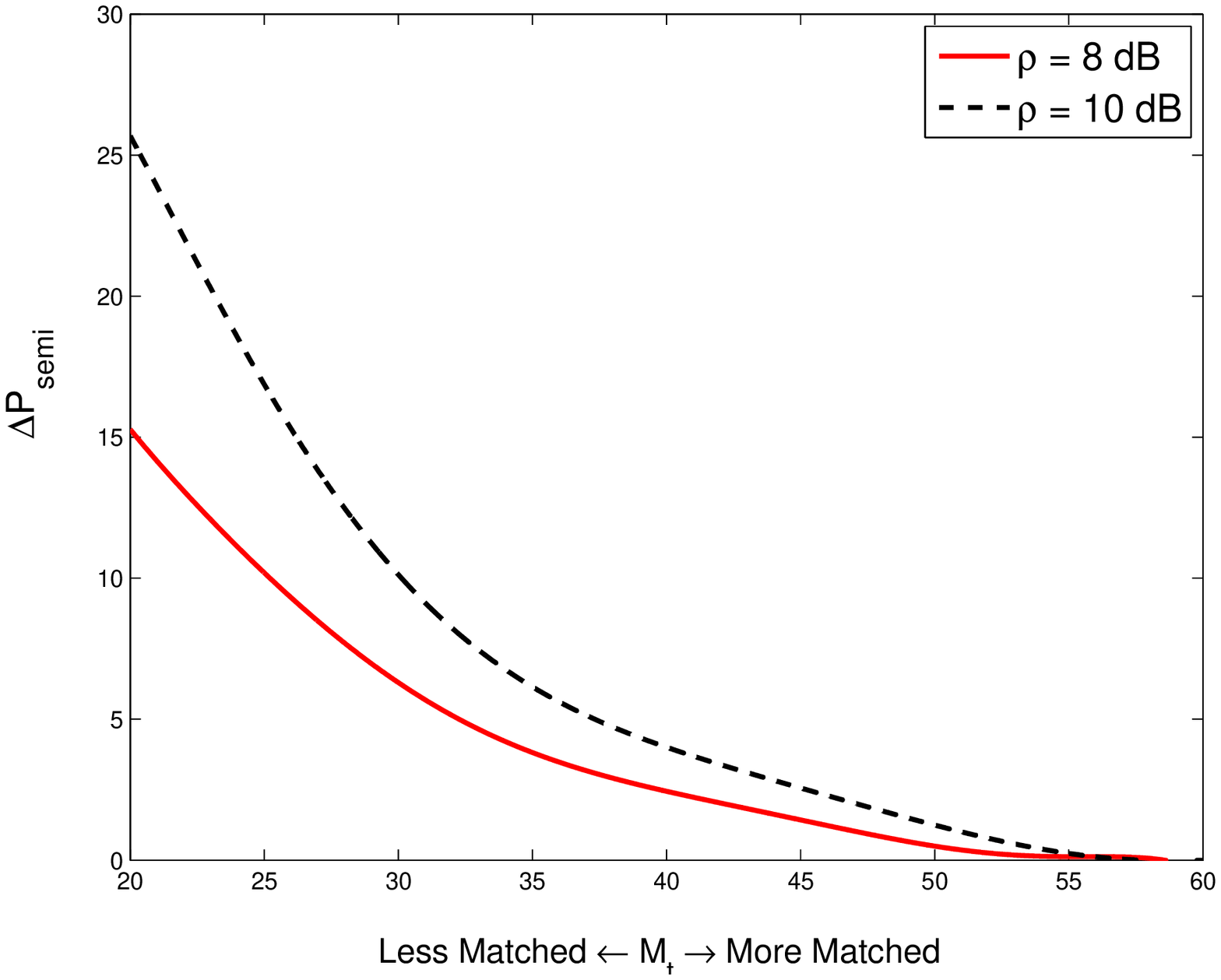}
\\ (a) & (b) 
\end{tabular} 
\caption{\label{fig3} Gap in performance between statistical and perfect CSI 
semiunitary precoding as a function of the matching metric, ${\cal M}_t$: 
(a) Mutual information and (b) Error probability.} 
\end{center}
\end{figure}

\item 
{\bf \em Performance Gap as a Function of Matching Metric:} 
The second study focuses on the gap in performance between the perfect CSI 
and the statistical precoders, as a function of the degree of matching of the 
channel to the precoder structure. We consider $4 \times 4$ channels 
with $M = 2$, and freeze ${\bf U}_t$, ${\bf U}_r$ to some arbitrary choice 
in our study. We also freeze ${\bf \Lambda}_r$ to $4 \hsppp {\bf I}_4$ so as 
to focus on the impact of matching on the transmitter side. Note that the 
matching metric (defined in Sec.~\ref{match_4}), ${\cal M}_t = 
\prod_{k=1}^M {\bf \Lambda}_t(k)$, takes values in the range $(0, 64]$ in our 
setting. A family of $\sim {\hspace{-0.05in}} 1700$ channels (each characterized 
uniquely by ${\bf \Lambda}_t(k), {\hspace{0.05in}} k = 1, \cdots , N_t$) is 
generated such that $\sum_{k=1}^{N_t} {\bf \Lambda}_t(k) = \rho_c = 16$ 
and ${\cal M}_t$ takes values over its range. The channels become more matched 
(on the transmitter side) to the precoder structure as ${\cal M}_t$ increases. 

While much of our study in the preceding sections is based on asymptotic random 
matrix theory, Fig.~\ref{fig3} illustrates that the notion of matched channels 
developed in this work is useful in characterizing performance, even in 
practically relevant regimes like $4 \times 4$ channels. Fig~\ref{fig3}(a) 
illustrates that $\Delta I_{\semi}$ decreases as the channel becomes more 
matched on the transmitter side for three choices of $\rho$, whereas 
Fig~\ref{fig3}(b) illustrates the same trend for $\Delta P_{\semi}$. Note that 
for a given channel as $\rho$ increases, $\Delta I_{\semi}$ decreases whereas 
$\Delta P_{\semi}$ increases. This is because of the contrasting behaviors of 
$I_{\stat, \hsppp \semi}(\rho)$ and $P_{\err, \hsppp \perf, \hsppp \unstruct}(\rho)$ 
as $\rho$ increases. 

It is important to note the following. In general, there exists no 
ordering relationship between any two matrix channels~\cite{olkin}. 
Nevertheless, Fig.~\ref{fig3} shows that the relative (mutual 
information or error probability) performance of two channels can be 
compared by using ${\cal M}_t$ and ${\cal M}_r$. A channel that is 
more matched leads to a smaller value of $\Delta I_{\bullet}$, as 
well as $\Delta P_{\bullet}$ for any fixed $\snr$. 
 
\begin{figure}[htb!]
\begin{center}
\begin{tabular}{cc}
\includegraphics[height=2.5in,width=3in]{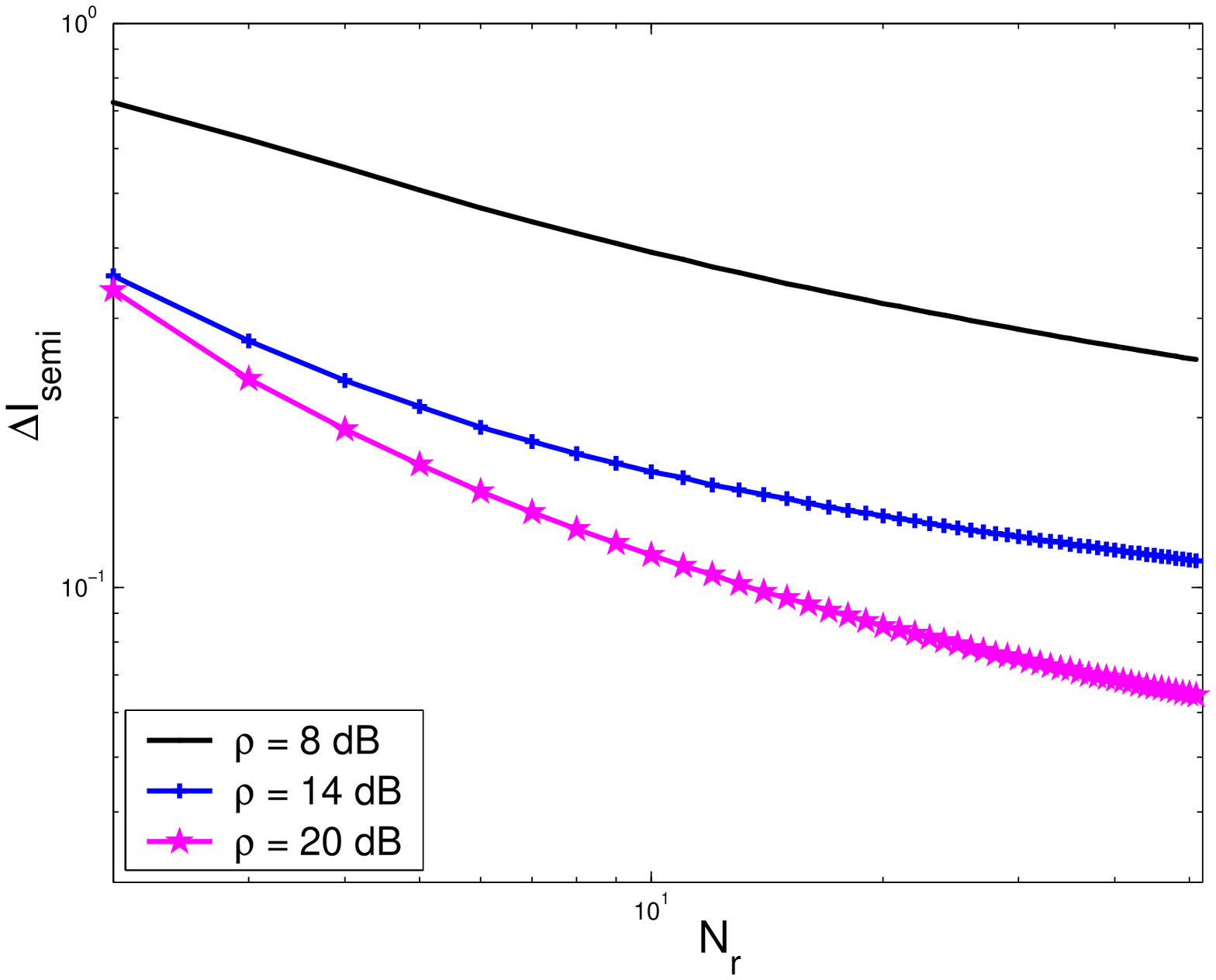} & 
\includegraphics[height=2.5in,width=3in]{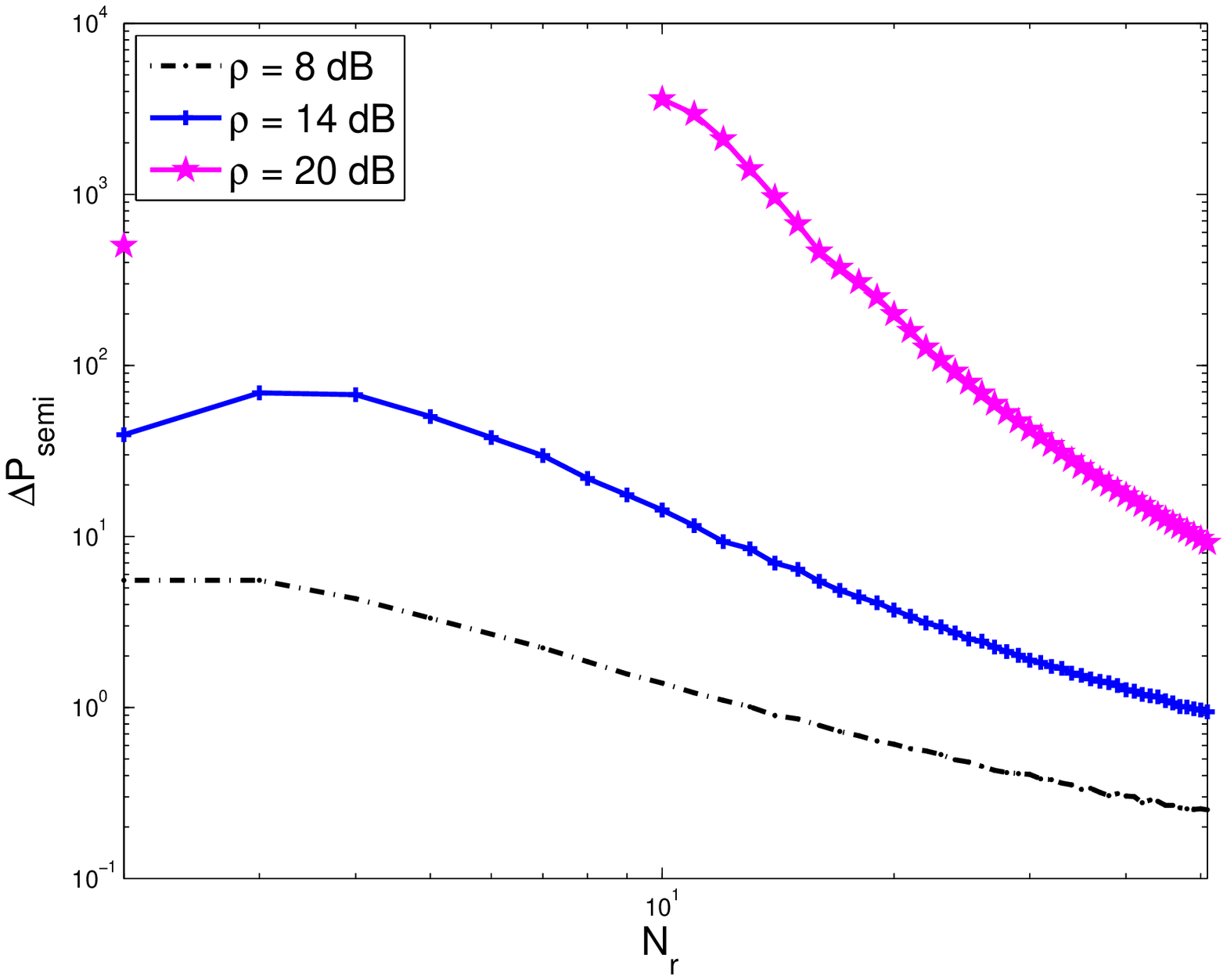}
\\ (a) & (b) 
\end{tabular} 
\caption{\label{fig4} Asymptotic optimality of the statistical semiunitary 
precoder for fixed $N_t = 4$, $M = 2$ as $N_r$ increases: (a) Mutual 
information and (b) Error probability.} 
\end{center}
\end{figure}

\item 
{\bf \em {Asymptotic Optimality:}} 
The third study illustrates the asymptotic optimality of statistical 
precoding. Fig~\ref{fig4} plots $\Delta I_{\semi}$ and $\Delta P_{\semi}$ as a 
function of $N_r$ with $N_t$ and $M$ fixed at $N_t = 4$ and $M = 2$. 
The channels have separable correlation with ${\bf \Lambda}_t = {\bf I}_4$ 
whereas ${\bf \Lambda}_r = \frac{4}{N_r} \hsppp {\bf I}_{N_r}$ and hence, 
$\rho_c = 4$ for all the channels. As can be seen from the study in the 
previous sections as well as the figures, channel hardening, where the eigenvectors 
of ${\bf H}^H {\bf H}$ converge to the eigenvectors of ${\bf \Sigma}_t = 
\bEe[ {\bf H}^H {\bf H}]$ as $\frac{N_t }{N_r} \rightarrow 0$ ensures that 
even channel statistical information is as good as perfect CSI with respect 
to performance. 

\begin{figure}[htb!]
\begin{center}
\begin{tabular}{cc}
\includegraphics[height=2.5in,width=3in]{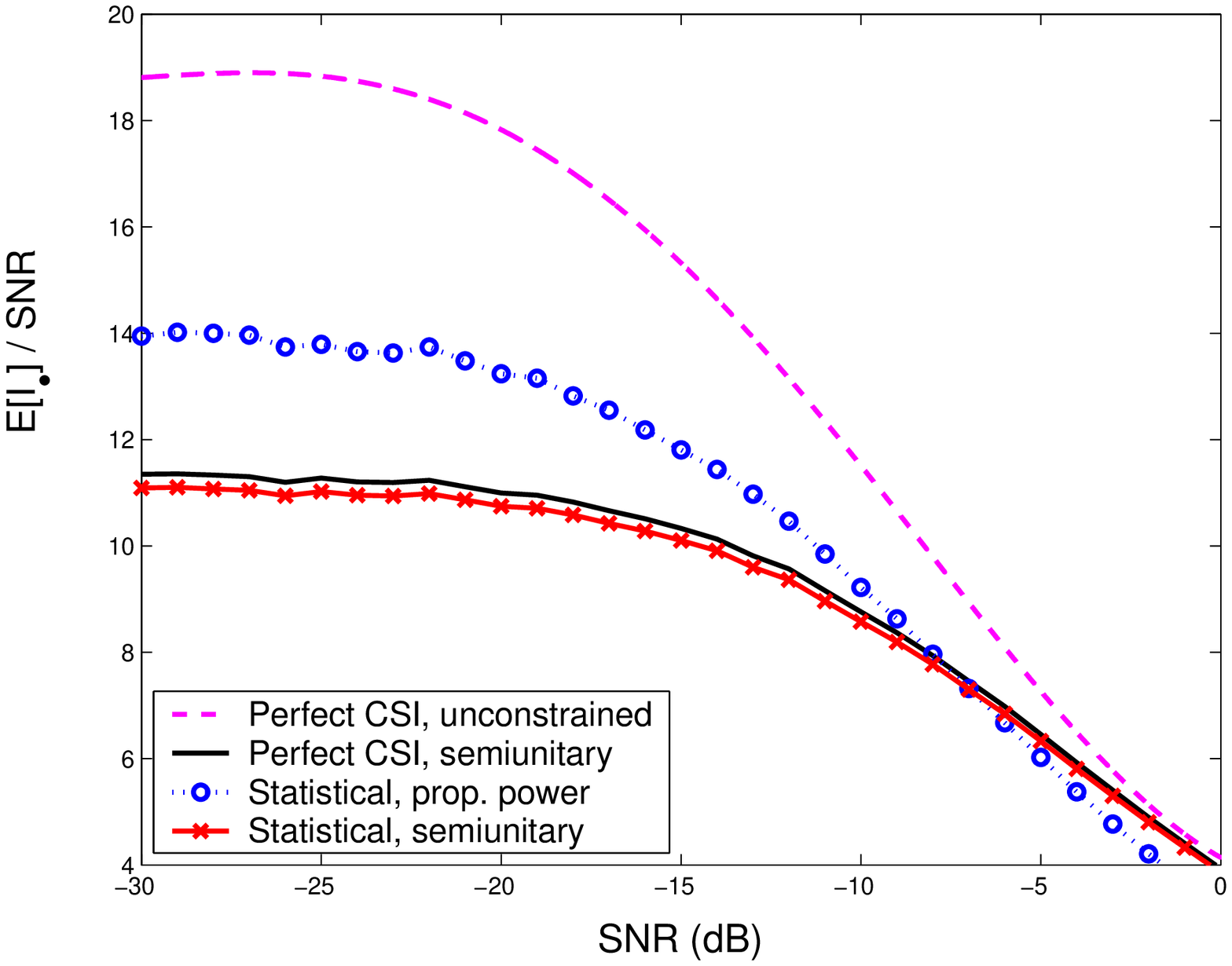} & 
\includegraphics[height=2.5in,width=3in]{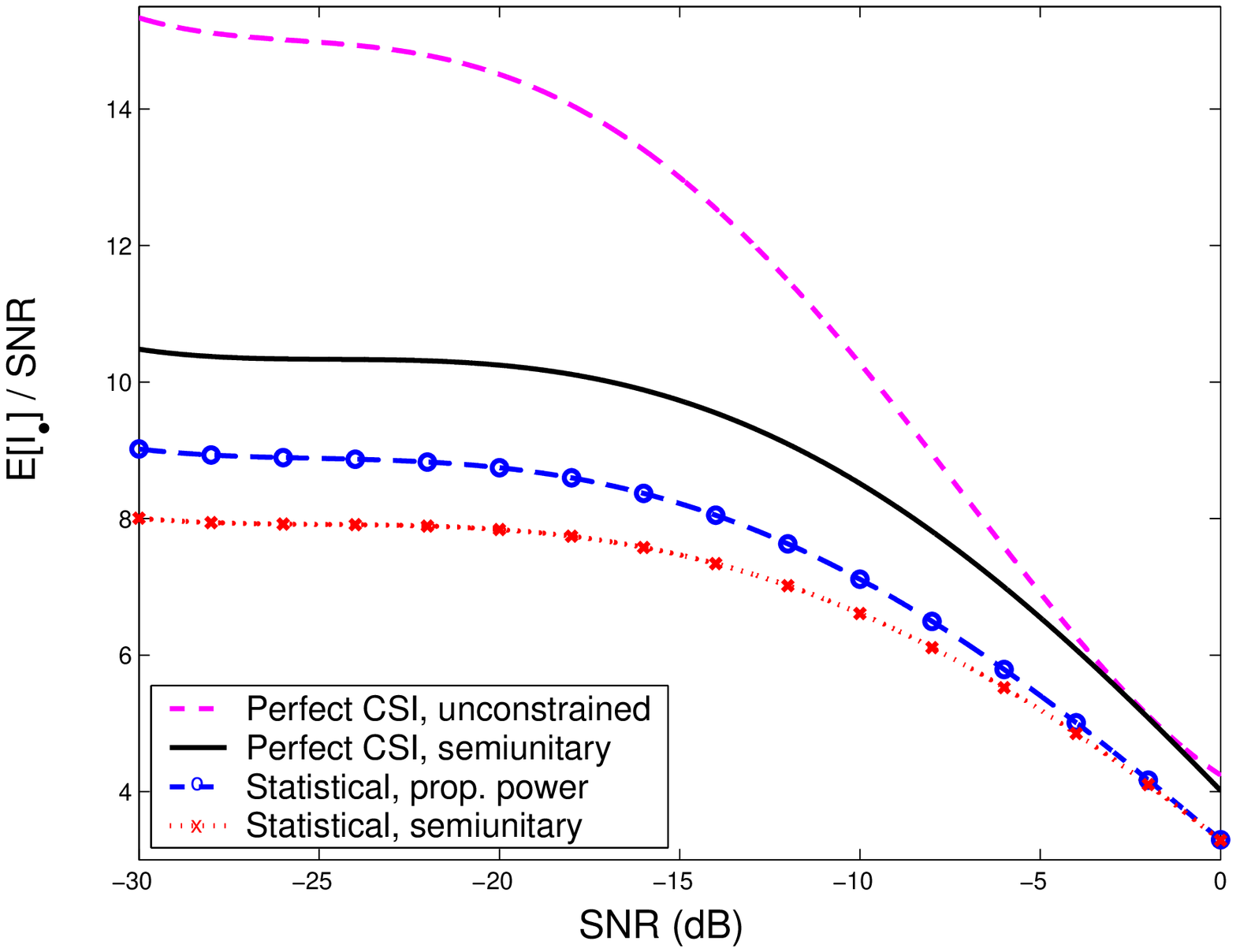} 
\\ (a) & (b) 
\end{tabular} 
\caption{\label{fig5} Low- and medium-$\snr$ mutual information performance 
of the statistical precoder in~(\ref{lam_fixed}) when compared with the 
semiunitary precoder for a) separable and b) non-separable (canonical) models.} 
\end{center}
\end{figure}

\item 
{\bf \em Low- and Medium-$\snr$ Regimes:} 
The last study of this section studies the mutual information performance of 
a statistical precoder in~(\ref{lam_fixed}) when compared with a semiunitary 
precoder in the low- and the medium-$\snr$ regimes. In the high-$\snr$ regime, 
the optimal perfect CSI precoder excites the $M$ modes uniformly with equal 
power. However, in the low-$\snr$ regime, the perfect CSI precoder allocates 
power to the transmit eigen-modes non-uniformly. The precoder structure 
in~(\ref{lam_fixed}) excites the $M = 2$ modes with power proportional to the 
transmit eigenvalues and hence, performs better than the semiunitary precoder. 
Fig~\ref{fig5}(a) shows the performance of the statistical precoder in a 
channel with separable correlation, while Fig.~\ref{fig5}(b) corresponds to a 
channel with non-separable correlation. In the separable case, the transmit 
and the receive eigenvalues are given by $\diag({\bf \Lambda}_t) = 
[ 9.80 \hspp 5.66 \hspp 0.45 \hspp 0.09]$ and $\diag( {\bf \Lambda}_r ) = 
[8.58 \hspp 4.20 \hspp 1.98 \hspp 1.24]$ whereas in the canonical case the 
variance matrix, ${\bf M} = (\sigma_{ij}^2)$, is given by 
\begin{eqnarray}
{\bf M} = \left[ 
\begin{array}{cccc}
    1.66 &    0.31 &   1.71 &    0.31 \\ 
    2.24 &    0.18 &   0.15 &    0.54 \\
    1.97 &    1.46 &   0.70 &    0.28 \\ 
    1.65 &    1.65 &   0.49 &    0.71
\end{array} \right]. 
\end{eqnarray}
It is interesting to note that the 
perfect CSI semiunitary precoder may either perform better or worse than that 
of the precoder in~(\ref{lam_fixed}). Future work will look at this aspect more 
carefully. 

\end{itemize}

\section{Concluding Remarks} 
\label{sec9} 
The main focus of this work is on precoding for spatially correlated 
multi-antenna channels that are often encountered in practice. Motivated and 
inspired by many recent wireless standardization efforts, we proposed 
low-complexity {\em structured precoding} techniques in this paper. Here, the 
eigen-modes of the precoder are chosen to be the dominant eigenvectors of the 
transmit covariance matrix, whereas the power allocation across the excited 
modes are obtained via certain simple, low-complexity methods. A special case 
of structured precoder is a semiunitary precoder, where the spatial modes are 
excited with uniform power. 

In this work, we first established the structure of the optimal perfect CSI 
structured precoder and showed that it naturally extends the channel diagonalizing 
architecture of the perfect CSI unconstrained precoder. We motivated the need 
for a relative difference metric that captures the impact of lack of perfect 
CSI on the precoder performance, independent of the operating $\snr$. We then 
analytically characterized the average relative mutual information loss (as 
well as the average relative uncoded error probability enhancement) of the 
statistical semiunitary precoder using tools from random matrix and eigenvector 
perturbation theories. 

Our results show that given a precoder architecture (that is, fixed antenna 
dimensions and precoder rank), the relative difference metrics are minimized by 
a channel that is matched to it. A matched channel is one that has: 1) The same 
number of dominant transmit eigen-modes as the precoder rank, and 2) The dominant 
transmit as well as the receive eigen-modes that are well-conditioned. Our 
theoretical study also characterizes {\em matching metrics} that enable the 
comparison of two channels with respect to performance loss captured by the 
relative difference metrics. In particular, as the channel becomes more 
matched to the precoder structure and the matching metrics change accordingly 
{\em continuously}, the performance loss decreases monotonically and {\em vice versa}. 
Numerical studies are provided to illustrate our results. 

Our work is a first attempt to analytically study the performance of 
low-complexity statistical precoding with respect to a perfect CSI benchmark. 
Much of this study has been rendered possible due to substantial advances 
in capturing the eigen-properties of random matrices with independent entries. 
Nevertheless, there exist many directions along which this work can be 
developed. We now list a few of these directions. 

This work is limited to the high-$\snr$, large antenna asymptotic regime 
where a comprehensive random matrix theory is available to capture precoder 
performance~\cite{hachem}. Even in this regime, it may be possible 
as in~\cite{mckay} to refine the constants in the bounds for the relative loss 
terms and obtain further insights on the impact of 
spatial correlation on performance loss. Besides that, in the case of proportional 
growth of antenna dimensions with a non-separable correlation model, both mutual 
information as well as error probability have not been characterized completely 
in this work. Lack of availability of closed-form mutual information expressions 
for non-Gaussian inputs limits the development of this work. The notion of 
precoder-channel matching introduced in this work can be developed further to aid 
in the design of low-complexity, structured and adaptive signaling schemes. 
In the case of mismatched channels, the construction of limited feedback schemes 
to bridge the gap in performance has been undertaken 
in~\cite{vasanth_limfb_precode,bhaskar_rao2,vincent_lau}. 
The question of trade-offs between spatial versus spatio-temporal 
precoding~\cite{vasanth_to_code} and extensions to more general Ricean 
fading~\cite{taricco_ricean}, multi-user~\cite{nihar_fb}, 
wideband~\cite{heath_wideband} systems are also of interest. 

\appendix 
\subsection{Key Mathematical Results} 
\label{app_majorize} 
We now introduce some key mathematical results that will be needed in the ensuing 
proofs. 

\noindent {\bf \em Majorization Theory:} 
We start with a few results from majorization theory~\cite{olkin}. 
\begin{defn} 
Let $\ba$ and $\bb$ be two vectors in ${\mathbb{R}}^m$ in non-increasing 
order\footnote{The non-increasing order for vectors results in ambiguity in a 
majorization relationship. To resolve this, in this section, we will 
assume that any two comparable vectors are always in the non-increasing order.}, 
i.e., $\ba(1) \geq \cdots \geq \ba(m)$ and $\bb(1) \geq \cdots \geq \bb(m)$. 
Then $\ba$ is {\em majorized by} $\bb$ (denoted by $\ba \prec \bb$) if 
\begin{eqnarray} 
\sum_{i=1}^k \ba(i) \leq \sum_{i=1}^k \bb(i), \hspp 1 \leq k \leq m 
\end{eqnarray}
with equality if $k = m$. 
\endproof 
\end{defn} 

\begin{rem}
\label{rem1}
For example, if $m = 3$, any positive vector $\ba$ such that $\sum_{i=1}^3 \ba(i) = 1$ 
satisfies the following majorization relationship: 
\begin{eqnarray}
\ba_{\sf low} \prec \ba \prec \ba_{\sf high}
\end{eqnarray}
where $\ba_{\sf low} = \left[\frac{1}{3} \hsp \frac{1}{3} \hsp \frac{1}{3} \right]$ 
and $\ba_{\sf high} = \left[1 \hsp 0 \hsp 0 \right]$. Another example of a majorization 
relationship is provided by an $m \times m$ Hermitian matrix $\bX$, with 
$m$-dimensional vectors ${\bf e}$ and ${\bf d}$ denoting the eigenvalues and 
diagonal entries of $\bX$, respectively. We have ${\bf d} \prec {\bf e}$. 
From the definition, it can also be easily checked that if $\ba \prec \bb$, then 
$- \ba \prec -\bb$. 
\end{rem} 

\begin{lem} 
\label{lem_ustochastic}
A matrix $\bQ$ 
is said to be unitary-stochastic if there exists a unitary matrix $\bGamma$ 
such that $\bQ(i,j) = |\bGamma(i,j)|^2$~\cite[Sec.~2B.5,~p.~23]{olkin}. By 
definition, a unitary-stochastic matrix is doubly stochastic. If $\bu \prec \bv$, 
there exists a unitary-stochastic matrix $\bQ$ such that $\bu = \bv \bQ$. 
\endproof 
\end{lem} 

\begin{defn}
Let $\ba$ and $\bb$ be two vectors in ${\mathbb{R}}^m$ in non-increasing 
order. Then $\ba$ is {\em weakly submajorized by} $\bb$ (denoted by $\ba \prec _w 
\bb$) if 
\begin{eqnarray} 
\label{esup}
\sum_{i=1}^k \ba(i) \leq \sum_{i=1}^k \bb(i), \hspp 1 \leq k \leq m. 
\end{eqnarray}
If the inequality is in the opposite direction in~(\ref{esup}), then 
$\ba$ is {\em weakly 
supermajorized by} $\bb$ and is denoted by $\ba \prec^w \bb$. Note that if 
$\ba \prec_w \bb$, then $\bb \prec^w \ba$ and {\em vice versa}. 
\endproof 
\end{defn}

\begin{lem}
\label{lem_submajorize}
A vector $\ba$ is submajorized by $\bb$ if and only if $\sum g( \ba(i) ) \leq 
\sum g ( \bb(i) )$ for all continuous, increasing convex functions 
$g : {\mathbb{R}} \mapsto {\mathbb{R}}$. For supermajorization, replace $g(\cdot)$ 
by all continuous, decreasing convex functions. If $g(\cdot)$ is decreasing, 
convex and $\ba \prec^w \bb$, we have 
\begin{eqnarray}
\left[ g(\ba(1) ) \hspp \cdots \hspp g(\ba(m))  \right] \hspp \prec_w \hspp 
\left[ g(\bb(1) ) \hspp \cdots \hspp g(\bb(m))  \right]. 
\end{eqnarray}
\end{lem}
\begin{proof}
See~\cite[p.\ 10]{olkin} for the first statement. For the second, 
see~\cite[p.\ 116]{olkin}. 
\end{proof}

\begin{defn} 
A function $f: {\cal A} \mapsto {\mathbb{R}}$ with ${\cal A} \subset {\mathbb{R}}^m$ 
is said to be {\em Schur-concave on ${\cal A}$} if $\{ \ba, \bb \} \in {\cal A}$ and 
$\ba \prec \bb$ implies that $f(\ba) \geq f(\bb)$. If however, $f(\ba) \leq f(\bb)$ 
for all such $\ba$ and $\bb$, $f(\cdot)$ is said to be {\em Schur-convex on ${\cal A}$.} 
If a function is Schur-concave (or -convex) over ${\mathbb{R}}^m$, we just say 
that it is Schur-concave (or -convex). Note that $f(\cdot)$ is Schur-concave 
if and only if $-f(\cdot)$ is Schur-convex. 
\endproof 
\end{defn} 

\begin{rem} 
\label{maj_rem} 
An example of Schur-convex and Schur-concave functions is as follows. Let 
${\bf x} = \left[ x_1 \hsppp \cdots \hsppp  x_m \right]$ with $x_i \geq x_{i+1}$. 
Consider the weighted arithmetic mean of $\{x_i \}$ given by $f({\bf x}) = 
\sum_{i=1}^m w_i x_i$. The function $f(\cdot)$ is Schur-convex if $w_i \geq 0$ and 
$w_1 \hsppp \leq \hsppp \cdots \hsppp \leq \hsppp w_m$. If $w_i \geq 0$, but are in 
the reverse order, then $f(\cdot)$ is Schur-concave. See~\cite[Lemma 4]{palomar_precode} 
for proof of this claim. It is important to note that the sets of Schur-concave and 
Schur-convex functions neither partition nor cover the space of all functions, nor 
are they disjoint. 
\end{rem}

\begin{lem}
\label{lem_schur_cvx}
Let $f : {\mathbb{R}} \mapsto {\mathbb{R}}$ be a continuous convex 
function. Then, $\sum_{i=1}^m f(x_i)$ is Schur-convex. That is, if $\bu$ and 
$\bv$ are two $m \times 1$ vectors such that $\bu \prec \bv$, then, 
$\sum_{i = 1}^m f(\bu(i)) \leq \sum_{i=1}^m f(\bv(i))$. 
Let $\phi : {\mathbb{R}}^m \mapsto {\mathbb{R}}$ be Schur-convex and 
the univariate function $\phi( \cdots , x_i , \cdots ): {\mathbb{R}} 
\mapsto {\mathbb{R}}$ be monotonically decreasing for all $i$. If 
$\ba \prec^w \bb$, we have $\phi(\ba) \leq \phi(\bb)$. 
\end{lem} 
\begin{proof}
See~\cite[p.\ 11]{olkin} for the first statement and~\cite[p.\ 59]{olkin} for 
the second. 
\end{proof}

\begin{lem}
\label{lem_used}
Let $f : {\mathbb{R}} \mapsto {\mathbb{R}}$ be a continuous convex function. 
Then, $\max_{i = 1, \hspp \cdots , \hspp m} f(x_i)$ is continuous and Schur-convex. 
\end{lem}
\begin{proof} 
A composition of an increasing, Schur-convex function with a convex function 
results in a Schur-convex function~\cite[p.\ 63]{olkin}. The proof follows 
by noting that $\max_i x_i$ is a function that is increasing in its arguments 
and is Schur-convex. 
\end{proof} 

\begin{lem}
\label{lemma_usable} 
Let $\{ x_i , i = 1, \cdots , K \}$ and $\{ y_i , i = 1, \cdots , K \}$ 
be two $K$-tuples such that $\{ x_i, y_i \} \geq 0$ for all $i$. Then, 
\begin{eqnarray}
\sum_{i = 1}^K x_i \leq \frac{1}{K} \left( \sum_{i = 1}^K \frac{x_i}{y_i} \right) 
\left( \sum_{i=1}^K y_i \right). 
\end{eqnarray}
\end{lem}
{\vspace{0.1in}}
\begin{proof} 
We prove the lemma by induction. Consider the case $K = 2$. Without loss of 
generality, let $x_1 \leq x_2$ and $y_1 \geq y_2$. We therefore have 
$\frac{x_1}{y_1} \leq \frac{x_2}{y_2}$ which implies that 
\begin{eqnarray}
x_1 + x_2  \leq \frac{x_1}{y_1} \hspp y_2 + \frac{x_2}{y_2} \hspp y_1. 
\end{eqnarray} 
Adding $x_1 + x_2$ on both sides and rearranging, we see that the statement is 
true for $K = 2$. Let the statement be true for $K = n-1$ for any ordering where 
$x_1 \leq \cdots \leq x_{n-1}$ and $y_1 \geq \cdots \geq y_{n-1}$. We will show 
that the statement is true for the $K = n$ case, where we augment the $(n-1)$-tuples 
with $x_n$ and $y_n$. Without loss of generality, we can assume that $x_1 \leq \cdots 
\leq x_n$ and $y_1 \geq \cdots \geq y_n$ after possible rearrangement and relabeling 
of indices. We have 
\begin{eqnarray}
\sum_{i=1}^{n-1} x_i + x_n & \stackrel{(a)}{\leq} & \frac{1}{n-1} 
\left( \sum_{i=1}^{n-1} \frac{x_i}{y_i} \right) 
\left( \sum_{i=1}^{n-1} y_i \right)  + \frac{ x_n}{y_n} \hspp y_n 
\\ 
& \stackrel{(b)}{=} &  \frac{1}{n} 
\left( \sum_{i=1}^{n-1} \frac{x_i}{y_i} \right) \left( \sum_{i=1}^{n-1} y_i \right) 
+ \frac{1}{n} \frac{x_n}{y_n} \hspp y_n   \nonumber \\ 
& &  {\hspace{0.1in}} + \underbrace{ 
\frac{1}{n(n-1)} \left( \sum_{i=1}^{n-1} \frac{x_i}{y_i} \right) 
\left( \sum_{i=1}^{n-1} y_i \right) + 
\frac{n-1}{n} \cdot\frac{x_n}{y_n} \hspp y_n }_{A} 
\\ 
nA & = &  \left( \sum_{i=1}^{n-1} \frac{x_i}{y_i} - \frac{(n-1)x_n}{y_n} 
\right) 
\cdot \left( \frac{ \sum_{i=1}^{n-1} y_i } {n-1} - y_n \right) + 
y_n  \left( \sum_{i=1}^{n-1} \frac{x_i}{y_i} \right)
+  \frac{x_n}{y_n} \left( \sum_{i=1}^{n-1}y_i \right), 
\nonumber 
\end{eqnarray}
where (a) follows from the induction hypothesis and (b) by breaking the sum into 
two pieces. The statement holds for $K = n$ upon rearrangement after using the 
increasing and decreasing ordering assumption of $x_i$ and $y_i$, respectively. 
\end{proof}

\noindent {\bf{\emph{Matrix Theory:}}} 
The {\emph{Poincare separation theorem}} connects the eigenvalues of semiunitary 
transformations with those of the transformed 
matrix~\cite[Cor. 4.3.16, p.\ 190]{horn_and_johnson}. 
\begin{lem}
\label{lem_poincare}
Let $\bA$ be an $n \times n$ Hermitian matrix. Let $r$ be such that 
$1 \leq r \leq n$ and let $\bw_1 , \cdots , \bw_r$ be a set of orthonormal 
vectors in ${\mathbb{C}}^n$. Define $\bB = \bW^{\sl H} \bA \bW$ where $\bW = 
\left[ \bw_1 \hspp \cdots \hspp \bw_r \right]$. Let 
the eigenvalues of $\bA$ and $\bB$ be arranged in non-increasing order. 
Then, we have $\lambda_k(\bB) \leq \lambda_{k}(\bA)$ for all $k = 1, 
\cdots, r$. 
\endproof 
\end{lem}

The following lemma provides bounds for eigenvalues of sums and products of Hermitian 
matrices~\cite{horn_and_johnson}. 
\begin{lem}
\label{lem_eig_bds}
If ${\bf A}$ and ${\bf B}$ are $n \times n$ Hermitian matrices, then 
\begin{eqnarray}
\lambda_k({\bf A}) \lambda_{\min}({\bf B}) & \leq \lambda_k({\bf A B}) \leq & 
\lambda_k({\bf A}) \lambda_{\max } ({\bf B}),  
\hsp k = 1, \hspp \cdots , \hspp n, 
\\ 
\lambda_k({\bf A}) + \lambda_{\min}({\bf B}) & 
\leq \lambda_k({\bf A}+ {\bf B}) \leq & 
\lambda_k({\bf A}) + \lambda_{\max}({\bf B}), \hsp 
k = 1, \hspp \cdots , \hspp n. 
\end{eqnarray}
We also have 
\begin{eqnarray}
\label{eqn_olkin}
\sum_{k=1}^n \lambda_k( {\bf A B} ) \leq \sum_{k=1}^n \lambda_k({\bf A}) \
\lambda_k({\bf B}). 
\end{eqnarray}
\endproof 
\end{lem} 

The following lemma~\cite{silvester} helps in computing the determinant of 
partitioned matrices. 
\begin{lem} 
\label{lem_partition}
If ${\bf X}, {\bf Y}, {\bf Z}$ and ${\bf W}$ are $n \times n$ matrices and ${\bf W}$ 
is invertible, we have 
\begin{eqnarray} 
\det \left[ \begin{array}{cc}
{\bf X} & {\bf Y} \\ 
{\bf Z} & {\bf W} 
\end{array}  \right] = \det({\bf X} - {\bf Y} {\bf W}^{-1} {\bf Z}) 
\cdot \det({\bf W}). 
\end{eqnarray}
\endproof 
\end{lem} 

\noindent {\bf{\emph{Random Matrix Theory:}}} 
We now characterize the eigenvalues of certain families of random matrices. 
\begin{lem}
\label{yinbailemma}
Let ${\bf X}$ be a $p \times n$ complex random matrix with i.i.d.\ entries 
of mean zero, common variance $1$ and a finite fourth moment. Consider two 
cases: 1) $p$ is finite and $n \rightarrow \infty$, and 2) $\{p,n \} 
\rightarrow \infty$ with $p/n \rightarrow 0$. In either case, in the 
asymptotics of $n$, the empirical eigenvalue distribution of 
$\frac{ {\bf X} {\bf X}^{\sl H} - n {\bI}_p } { 2 \sqrt{np}}$ converges 
pointwise with probability $1$ to the semi-circular law $F(x)$ where, 
\begin{eqnarray}
F(x) = \left \{ \begin{array}{cc} 0 & {\rm if} \hspp x < -1, \\
\int_{y = -1} ^{ x } \frac{2}{\pi} \sqrt{1-y^2} \hsppp \ud y 
& {\rm if} \hspp -1 \leq x \leq 1, \\ 
1 & {\rm if} \hspp x > 1.  
\end{array} \right.
\end{eqnarray}
In particular, with probability one, we have 
\begin{eqnarray}
1 - 2 \sqrt{\frac{p}{n}}  \leq 
\liminf_n \frac{ \lambda_{\min}({\bf X} {\bf X}^{\sl H})}{n} \leq 
\limsup_n \frac{ \lambda_{\max}({\bf X} {\bf X}^{\sl H})}{n} \leq 
1 +  2 \sqrt{\frac{p}{n}}. 
\end{eqnarray} 
Let $\bfLambda$ be an $n \times n$ positive definite diagonal matrix. 
Under the same assumptions on ${\bf X}, p, n$ as above, 
there exists a finite constant $\gamma_1 > 0$ (dependent on $p$ and $n$ only 
through ${\bf \Lambda}$) such that, with probability $1$ 
\begin{eqnarray}
\frac{  \sum_{i} \bfLambda(i) }{n} - \gamma_1 \sqrt{\frac{p}{n}} 
\leq 
\liminf_n \frac{ \lambda_{\min}({\bf X} \bfLambda {\bf X}^{\sl H})}{n} \leq 
\limsup_n \frac{ \lambda_{\max}({\bf X} \bfLambda {\bf X}^{\sl H})}{n} \leq 
\frac{  \sum_{i} \bfLambda(i) }{n} +  \gamma_1 \sqrt{\frac{p}{n}}. 
\nonumber 
\end{eqnarray}
On the other hand, let ${\bf X}$ be a $p \times n$ complex random matrix 
with independent entries from a fixed probability space such that 
${\bf X}(i,j)$ is zero mean, has variance $\sigma_{ij}^2$ and 
\begin{eqnarray}
\sup_{n,p} \max_{ij} \bEe[ |{\bf X}(i,j) | ^4 ] \leq \gamma_2 < 
\infty. 
\end{eqnarray}
Also, without loss of generality, assume that $\big \{ 
\sum_{j = 1}^n \sigma_{ij}^2 \big\}$ are arranged in decreasing order. 
Then there exists a finite constant $\gamma_3 > 0$ (independent of $p, n$) 
such that, for all $i$  
\begin{eqnarray}
\frac{  \sum_{j = 1}^n \sigma_{ij}^2 }{n} - \gamma_3 \sqrt{\frac{p}{n}} 
\leq 
\liminf_n \frac{ \lambda_{i}({\bf X} {\bf X}^{\sl H})}{n} \leq 
\limsup_n \frac{ \lambda_{i}({\bf X} {\bf X}^{\sl H})}{n} \leq 
\frac{  \sum_{j = 1}^n \sigma_{ij}^2 }{n} +  \gamma_3 \sqrt{\frac{p}{n}} 
\end{eqnarray}
with probability $1$. 
\end{lem}
\begin{proof} 
We provide an elementary proof of the claim when $p$ is finite, 
$n \rightarrow \infty$ and ${\bf X}(i,j)$ are standard, complex Gaussian. 
Define the set $A_n \triangleq 
\Big\{ \omega : \frac{ \lambda_{\max}( {\bf X}(\omega) {\bf \Lambda} 
{\bf X}(\omega)^H  ) }{n} > 1 + \epsilon_1 + \epsilon_2 \Big\}$. If we can show 
that $\sum_n {\rm Pr} \left( A_n \right) < \infty$, it follows from the 
Borel-Cantelli lemma~\cite{durrett} that ${\rm Pr} \left( \limsup A_n  \right) = 0$. 
By choosing $\epsilon_1$ and $\epsilon_2$ appropriately (as a function of $n$), we 
can establish strict bounds on the eigenvalues.

Breaking ${\bf X} {\bf \Lambda} {\bf X}^H$ into a diagonal component and an 
off-diagonal component and using Lemma~\ref{lem_eig_bds}, it follows via a 
union bound that 
\begin{eqnarray}
{\rm Pr} \left( A_n \right) \leq p {\rm Pr} \left( \frac{ \sum_{i=1}^n 
\left( |{\bf X}(1,i)|^2 - 1 \right) {\bf \Lambda}(i) }{n } > \epsilon_1 
\right) + p^2 {\rm Pr} \left( 
\frac{ |\sum_{i=1}^n {\bf X}(1,i) {\bf \Lambda}(i) {\bf X}(2,i)^{\star} | }  
{n} > \epsilon_2 \right). \nonumber  
\end{eqnarray}
Using a Chernoff-type bound~\cite{durrett}, we have the following: 
\begin{eqnarray}
{\rm Pr}(A_n) \leq p \exp \left( -\frac{\epsilon_1^2 n^2}
{ 2 \sum_{i=1}^n ({\bf \Lambda}(i))^2  }  \right) + 2p^2 \exp 
\left( - \frac{ \epsilon_2^2 n^2 c } { \sum_{i=1}^n ({\bf \Lambda}(i))^2 } 
\right) 
\end{eqnarray} 
for some $c > 0$. The smallest value of $\epsilon_1$ and $\epsilon_2$ that 
can still result in ${\rm Pr} \left( \limsup A_n  \right) = 0$ is such that 
\begin{eqnarray}
\epsilon_1 = \ord(\epsilon_2) = \sqrt{ 
\frac{ \sum_{i=1}^n ({\bf \Lambda}(i) )^2  }{n}  } 
\cdot \frac{1} { n^{1/2 - \eta} }, {\hspace{0.1in}} \eta > 0. 
\end{eqnarray}
Letting $\eta \downarrow 0$, we have 
\begin{eqnarray}
\limsup \frac{ \lambda_{\max} ( {\bf X} {\bf \Lambda} {\bf X}^H ) } {n } 
\leq \frac{ \sum_{i=1}^n {\bf \Lambda}(i) }{n} + \gamma_4 
\sqrt{ \frac{ \sum_{i=1}^n {\bf \Lambda}(i)^2  }{n}  } 
\cdot \frac{1} { \sqrt{n} }, 
\end{eqnarray}
where $\gamma_4 > 0$ is a constant independent of $p$ and $n$. The 
expression for 
$\lambda_{\min}(\cdot)$ is symmetric with that of $\lambda_{\max}(\cdot)$ and 
can be obtained similarly. The extension to the case where ${\bf X}$ has only 
independent entries (not necessarily complex Gaussian) 
also proceeds via the same logic.

Since $p \rightarrow \infty$ in Case 2), the above technique is not useful in 
establishing the claim of the lemma. 
Here, the result follows 
from~\cite{yin_bai},~\cite[Theorem 2.9, p.\ 623]{bai_methodologies}. 
The generalizations with ${\bf \Lambda}$ and independent entries follow via the 
same proof technique as in~\cite{yin_bai} and hence no proofs are provided. 
The readers are referred to~\cite{bai_methodologies} for a brief summary of the 
general technique. 
\end{proof}

\subsection{Proofs of Prop.~\ref{prop_opt_precoder}-\ref{prop_gem}}
\label{pf_prop_opt_precoder} 

\noindent {\bf{\emph{Proof of Prop.~\ref{prop_opt_precoder}:}}} 
Let ${\bf F}$ be a fixed $N_t \times M$ semiunitary precoder and define 
\begin{eqnarray} 
{\bf B} \triangleq \left( \bI_M + \frac{\rho}{M} 
{\bf F}^H {\bf H}^H {\bf H} {\bf F} \right)^{-1}. 
\end{eqnarray}
From~(\ref{sinr_exp}), note that the vector $\mse$ is the vector of 
diagonal entries of ${\bf B}$. Following Lemma~\ref{lem_poincare}, we have 
$\lambda_k 
\left( {\bf F}^H {\bf H}^H {\bf H} {\bf F} \right) \leq \lambda_k \left( 
{\bf H}^H {\bf H} \right)$ for $k = 1, \cdots , M$. That is, the eigenvalues of 
${\bf B}$ satisfy 
\begin{eqnarray} 
\label{lbnd1} 
\lambda_k({\bf B}) \geq \frac{1}{1 + 
\frac{\rho}{M} \lambda_{M-k+1}({\bf H}^H {\bf H})}, 
\hspp k = 1, \cdots, M. 
\end{eqnarray}
Denote by ${\bf \lambda}_{ {\bf B} }$ the vector of eigenvalues of ${\bf B}$. 
The Schur-concavity of $f(\cdot)$ and the fact that the diagonal entries of a 
Hermitian matrix are majorized by its eigenvalues when used with ${\bf B}$ 
results in $f \left( \mse \right) \geq f \left(  {\bf \lambda}_{ {\bf B} } \right)$. 
The monotonicity of $f(\cdot)$ when combined with (\ref{lbnd1}) implies that 
\begin{eqnarray}
f \left( \mse \right) \geq f \left(  \left[ \cdots, \hspp 
\frac{1}{1 + \frac{\rho}{M} \lambda_{M-k+1}({\bf H}^H {\bf H})}, \hspp \cdots \right] 
\right). 
\label{neweqn}
\end{eqnarray}
Note that the lower bound in~(\ref{neweqn}) is independent of the choice of 
${\bf F}$, and hence, also serves as a universal lower bound. Furthermore, the 
choice of ${\bf F}$ in (\ref{fopt1}) meets the lower bound and is hence optimal. 
\endproof

\noindent {\bf{\emph{Proof of Prop.~\ref{prop_schur_cvx}:}}} 
Let ${\bf F}$ be a fixed semiunitary matrix. Define the $M \times 1$ vectors 
${\bf d}$ and ${\bf e}$ with ${\bf d}(k) \triangleq \bB(k)$, where 
$\bB = \left( \bI_M + \frac{\rho}{M} {\bf F}^H {\bf H}^H {\bf H} {\bf F} \right)^{-1}$ 
and ${\bf e}(k) \triangleq \frac{1}{M} \sum_{i=1}^M 
\frac{1}{1 + \frac{\rho}{M} \lambda_i \left({\bf F}^H {\bf H}^H {\bf H} {\bf F} 
\right)}$, respectively. Note that ${\bf e}(k)$ is equal for all $k$ and 
hence, from Remark~\ref{rem1} we have ${\bf d} \succ {\bf e}$. From 
Lemma~\ref{lem_schur_cvx}, we have that $\sum_{k=1}^M h(\cdot)$ is Schur-convex. 
Hence, 
\begin{eqnarray}
\sum_{k=1}^M h \left({\bf d}(k) \right) 
& \geq & 
\sum_{k=1}^M h \left( {\bf e}(k) \right) = M h \left( {\bf e}(1) \right). 
\end{eqnarray}
Using Lemma~\ref{lem_poincare} and the increasing property of $h(\cdot)$, we 
have 
\begin{eqnarray}
\label{rhs2}
\sum_{k=1}^M h \left({\bf d}(k) \right) & \geq & 
M h \left( \frac{1}{M} \sum_{k = 1}^M \frac{1} 
{1 + \frac{\rho}{M } \lambda_k(\bH^H \bH)} \right). 
\end{eqnarray}
Since the right-hand side of (\ref{rhs2}) is independent of the choice of ${\bf F}$, 
it serves as a lower bound on the error probability. 

Our goal is to show that the lower bound can be achieved and the choice of ${\bf F}$ 
that leads to the lower bound is ${\bf F}_{\opt}$. For this, let $A$ be defined 
as $A \triangleq \frac{1}{M} \sum_{i=1}^M \frac{1} 
{ 1  + \frac{\rho}{M} \lambda_i(\bH^H \bH) }$. Further, define the two 
$M \times 1$ vectors $\bu$ and $\bv$ such that $\bu(k) = A$ for all $k$ and 
$\bv(k) = \frac{1} { 1  + \frac{\rho}{M} \lambda_k(\bH^H \bH) }$. Since 
$\bu \prec \bv$, 
from Lemma~\ref{lem_ustochastic}, there exists a unitary-stochastic matrix $\bQ$ 
such that 
$\bu = \bv \bQ$ with $\bQ(i,j) = |\bGamma(i,j)|^2$ for some $\bGamma$ unitary. 
Consider the precoder ${\bf F}$ as given in (\ref{fopt2}). The 
$\mse$ across the data-streams with this precoder is given by 
\begin{eqnarray}
\mse_k  & = & \left[ {\bf B}^{-1} \right]_k = 
\left[ \left( \bI_M + \frac{\rho}{M} {\bf F}^H \hsppp {\bf H}^H \hsppp 
{\bf H} \hsppp {\bf F} \right)^{-1} \right]_k  \\ 
& = & \left[ \left( \bI_M + \frac{\rho}{M} \bGamma^H \hsppp \widehat{\bf \Lambda} 
  \hsppp \bGamma \right)^{-1} \right]_k 
= \left[ 
\bGamma^H \hsppp \left( \bI_M + \frac{\rho}{M} \widehat{\bf \Lambda} 
\right)^{-1} \hsppp \bGamma \right]_k 
\end{eqnarray}
with $\widehat{ {\bf \Lambda} }(k)  = \lambda_k( {\bf H}^H {\bf H} )$. 
From the definitions of $\bGamma$, ${\bf v}$ and the relationship ${\bf u} = 
{\bf v} \bQ$, it is easy to check that $\mse_k = A$ for all $k$. Thus, with the 
choice of ${\bf F}$ as in~(\ref{fopt2}), we can achieve the lower bound in 
(\ref{rhs2}).   
\endproof 

\noindent {\bf{\emph{Proof of Prop.~\ref{prop_gem}:}}} 
For the Schur-concave case, from Lemma~\ref{lem_poincare} and~(\ref{eqn_olkin}), it can be 
checked that ${\bf a} \prec_w {\bf b}$, where ${\bf a}(k) = \lambda_k( 
{\bf \Lambda}_{\fixed} \hsppp {\bf V}_{\bf F}^H \hsppp {\bf H}^H \hsppp {\bf H} 
\hsppp {\bf V}_{\bf F} )$ 
and ${\bf b}(k) = {\bf \Lambda}_{\fixed}(k) \hspp \lambda_k({\bf H}^H {\bf H})$. 
Define $g(y) = \frac{1}{1 + \kappa y}$ for some fixed $\kappa > 0$ and note that 
$g(\cdot)$ is convex and decreasing. Thus, from Lemma~\ref{lem_submajorize} we 
have $g(\bb) \prec_w g(\ba)$. Noting that $-f(\cdot)$ is Schur-convex and decreasing, 
from Lemma~\ref{lem_schur_cvx} we have $f \left(g(\ba) \right) \geq 
f\left(g (\bb) \right)$. This universal lower bound is achievable by ${\bf F}_{\opt}$ 
as in~(\ref{foptx}). 

When $f(\cdot)$ is Schur-convex, 
we proceed similar to the semiunitary case. Using $g(y) = \frac{1}{1 + \kappa y}$, 
from Lemma~\ref{lem_submajorize}, we have 
\begin{eqnarray}
\sum_{k=1}^M g( {\bf b}(k)) \leq \sum_{k=1}^M g({\bf a}(k) ). 
\end{eqnarray}
Define ${\bf u}(k) = \frac{1}{M} \sum_{i = 1}^M \frac{1}{ 1 + \frac{\rho}{M} 
{\bf \Lambda}_{\fixed}(i) \hsppp \lambda_i({\bf H}^H {\bf H}) }$ for all $k$ 
and ${\bf w}(k)  = \frac{1}{ 1 + \frac{\rho}{M} 
{\bf \Lambda}_{\fixed}(k) \hsppp \lambda_k({\bf H}^H {\bf H}) }$, and note that 
${\bf u} \prec {\bf w}$. That is, there exists a unitary-stochastic ${\bf Q}$ 
such that ${\bf u} = {\bf v} {\bf Q}$. The result follows as before. 
\endproof 


\subsection{Proof of Proposition~\ref{Delta_I1}} 
\label{term1} 
To characterize the behavior of $\Delta I_1$, recall the structure of the 
optimal semiunitary precoder from Prop.~\ref{prop_opt_precoder} and note 
from Lemma~\ref{lem_unstructured_opt} that the perfect CSI unconstrained 
scheme corresponds to waterfilling along the first $M$ dominant transmit 
singular vectors. Thus, we have 
\begin{eqnarray}
\Delta I_1 \cdot \bEe_{\bH} \left[ I_{\stat, \hspp \semi}(\rho) \right] 
& = & \bEe_{\bH} \left[  \sum_{i= 1}^{n_{\bH}}  
\log \left( 1 + {\bf \Lambda}_{\bH}(i) 
{\bf \Lambda}_{\sf wf}(i)  \right) - \sum_{i=1}^M \log \left( 1 + \frac{ \rho}{M} 
\hspp {\bf \Lambda}_{\bH}(i)  \right) \right],
\end{eqnarray}
where for each realization $\bH$, $n_{\bH}$ modes are excited ($1 \leq n_{\bH} \leq M$) 
with power ${\bf \Lambda}_{\sf wf}(i) \triangleq \left( \mu_{\bH} - 
\frac{1}{ {\bf \Lambda}_{\bH}(i) }  \right)^+$ and the water level $\mu_{\bH}$ 
is chosen such that $\sum_{i=1}^{n_{\bH}} {\bf \Lambda}_{\sf wf}(i) = \rho$. It 
can be easily checked that ${\bf \Lambda}_{\sf wf}(i)$ can be written as 
\begin{eqnarray}
\label{waterfill}
{\bf \Lambda}_{\sf wf}(i) = \frac{\rho}{n_{\bH}} + \frac{1}{n_{\bH}} 
\sum_{j = 1}^ {n_{\bH}} \frac{1}{ {\bf \Lambda}_{\bH}(j) } - 
\frac{1}{ {\bf \Lambda}_{\bH}(i) }, 
\end{eqnarray}
and $n_{\bH}$ is the largest value of $k$ that satisfies: 
\begin{eqnarray}
\sum_{i=1}^k \frac{ {\bf \Lambda}_{\bH}(i) - {\bf \Lambda}_{\bH}(k)  } 
{ {\bf \Lambda}_{\bH}(i) {\bf \Lambda}_{\bH}(k) } \leq \rho. 
\label{nh}
\end{eqnarray}
Hence, we have 
\begin{eqnarray}
\Delta I_1 \cdot \bEe_{\bH} \left[ I_{\stat, \hspp \semi}(\rho) \right] 
& \leq & \bEe_{\bH} \left[ \sum_{i = 1}^{n_{\bH}} \log \left( 1 +  
\frac{ \frac{\rho {\bf \Lambda}_{\bH}(i)(M - n_{\bH} ) } {n_{\bH} M} - 1  + 
\frac{ {\bf \Lambda}_{\bH}(i) } {n_{\bH} } \sum_{j=1}^{n_{\bH} } \frac{1} 
{ {\bf \Lambda}_{\bH}(j) } }
{1 + \frac{\rho {\bf \Lambda}_{\bH}(i)  } {M}  } 
 \right)  \right]. 
\end{eqnarray}
Using the fact that $\log(1 + x) \leq x$ for all $x > -1$, after some 
simplifications we can further upper bound $\Delta I_1$ as 
\begin{eqnarray}
\Delta I_1 \cdot \bEe_{\bH} \left[ I_{\stat, \hspp \semi}(\rho) \right]  
& \leq & \bEe_{\bH} \left[ M - n_{\bH} \right] + \frac{M^2}{\rho^2} \cdot 
\bEe_{\bH} \left[ \sum_{i=1}^M \frac{1}{ {\bf \Lambda}_{\bH}(i)^2 } \right]. 
\label{kl5}
\end{eqnarray}
From~(\ref{nh}), it is easily recognized that if $\rho \geq \frac{k } 
{ {\bf \Lambda}_{\bH}( k ) } - \sum_{i=1}^k \frac{1}{ {\bf \Lambda}_{\bH}(i) }$, 
and in particular, if $\rho \geq \frac{k }  { {\bf \Lambda}_{\bH}( k ) }$, 
then $n_{\bH} \geq k$. Thus, if $\rho > \alpha \bEe_{\bH} \left[ \frac{ M } 
{{\bf \Lambda}_{\bH}(M)} \right]$ for some $\alpha > 1$ as in the statement of the 
theorem, both the terms in~(\ref{kl5}) can be bounded by constants 
that depend only on the channel statistics. For this note that, 
\begin{eqnarray}
\bEe_{\bH} \left[ M - n_{\bH}  \right] & \leq & M \cdot {\rm Pr}( n_{\bH} < M ) 
\leq 
M \cdot {\rm Pr} \left( \frac{M} { {\bf \Lambda}_{\bH}(M) } > \rho \right)
\\ & \leq & M \cdot {\rm Pr} \left( \frac{1}{ {\bf \Lambda}_{\bH}(M)  } > 
\alpha \bEe \left[ \frac{1}{ {\bf \Lambda}_{\bH}(M) } \right]  \right) 
\stackrel{(a)}{\leq} \frac{M}{\alpha^2} \cdot 
\frac{ \bEe \left[ \left( \frac{1} { {\bf \Lambda}_{\bH}(M) } \right)^2  \right] } 
{  \left( \bEe \left[ \frac{1}{ {\bf \Lambda}_{\bH}(M) } \right]  \right)^2 }, 
\label{citing_nh}
\end{eqnarray}
where (a) follows from Chebyshev's inequality. A trivial upper bound for the 
other term gives the desired result. 
\endproof

\subsection{Proof of Theorem~\ref{thm_loss_cap}}
\label{app_thm2} 
It can be checked that $\widetilde{\Delta I}_2$ can be written as 
\begin{eqnarray}
\widetilde{\Delta I}_2 
& = &  \bEe _{{\bf H}} \Bigg[  \frac{ \sum_{k = 1 }^M \log \big( 1 
+ \transnr \hsppp \lambda_k ( {\bf H}^{\sl H} {\bf H}  ) \big)  } 
{\sum_{k = 1 }^M \log \big( 1 
+ \transnr \hsppp \lambda_k ( {\bf F}_{\semi} ^H {\bf H}^{\sl H} 
{\bf H} {\bf F}_{\semi}  ) \big) } - 1  \Bigg] 
\\ 
& \stackrel{(a)}{\leq} & 
\bEe_{ {\bf H} } \left[ \frac{1}{M} \sum_{k = 1 }^M  \frac{  \log \big( 1 
+ \transnr \hsppp \lambda_k ( {\bf H}^{\sl H} {\bf H}  ) \big)  } 
{ \log \big( 1 + \transnr \hsppp \lambda_k ( {\bf F}_{\semi}^H  
{\bf H}^{\sl H} {\bf H} {\bf F}_{\semi}  ) \big) } - 1 
\right]  \\
& \stackrel{(b)}{=} & \frac{1}{M} \sum_{k = 1}^M \bEe_{ {\bf H} } \left[  
\frac{ \log \Big(  1 +  \frac{\rho}{M \hsppp \rho_c} 
\hsppp \lambda_k (  \bfLambda_t \hsppp  {\bf H}_{\iid}^{\sl H} \hsppp 
\bfLambda_r \hsppp {\bf H}_{\iid} ) \Big)   }   
{  \log \Big(  1 +  \frac{\rho}{M \hsppp \rho_c} \hsppp 
\lambda_k (  {\widetilde{\bfLambda} }_t \hsppp {\bf \widetilde{H}}_{\iid}^{\sl H} 
\hsppp \bfLambda_r \hsppp {\bf \widetilde{H}}_{\iid} ) 
 \Big) }  - 1 \right], 
\end{eqnarray} 
where (a) follows from Lemma~\ref{lemma_usable}, and (b) from the notations 
established in Sec.~\ref{nots}. 

Using Lemmas~\ref{lem_eig_bds} and~\ref{yinbailemma}, we have the 
following in the limit of $N_r, \hsppp N_t, \hsppp M$: 
\begin{eqnarray}
\widetilde{\Delta I}_2 & \leq & \frac{1}{M} \sum_{k=1}^M \bEe_{\bH} 
\left[ \frac { \log \left( 1 + \frac{\rho}{M \hsppp \rho_c} 
{\bf \Lambda}_t(k) \lambda_{\max}( \bH_{\iid}^H {\bf \Lambda}_r \bH_{\iid} ) 
\right) } 
{  \log \left( 1 + \frac{\rho}{M \hsppp \rho_c} {\bf \Lambda}_t(k) 
\lambda_{\min}(  \widetilde{\bH}_{\iid}^H {\bf \Lambda}_r 
\widetilde{\bH}_{\iid}  ) \right) } - 1 
\right] \\ 
& \leq & \frac{1}{M} \sum_{k=1}^M \bEe_{\bH} \left[   
\frac{ \log \left( 1 + \frac{\rho }{M} {\bf \Lambda}_t(k) 
\hsppp \left( 1 + \kappa_1 \frac{ \sqrt{ \sum_{i} ( {\bf \Lambda}_r(i) )^2 }  } 
{\rho_c} \right) 
\right)  }
{ \log \left( 1 + \frac{\rho }{M} {\bf \Lambda}_t(k) 
\hsppp \left( 1 - \kappa_1 \frac{ \sqrt{ \sum_{i} ( {\bf \Lambda}_r(i) )^2 }  } 
{\rho_c} \right) \right) } -1 \right] \\ 
& \leq & \frac{2 \kappa_1  
\sqrt{ \sum_i ({\bf \Lambda}_r(i))^2 } }{\rho_c} \cdot 
\frac{1}{M} \sum_{k=1}^M \left[ \frac{1}{  \log\left( 1+ 
\frac{\rho}{M} {\bf \Lambda}_t(k) \right)} \right], 
\end{eqnarray}
where $\kappa_1$ is the constant from an application of Lemma~\ref{yinbailemma} 
in this setting. 
The last inequality follows by using the log-inequality and some trivial 
manipulations. The proof is complete. \endproof

\subsection{Proof of Proposition~\ref{thm_tp_bf}} 
\label{app_tp_bf} 
We have the following well-known facts~\cite{paulraj_book}: 
\begin{eqnarray}
I_{\perf}(\rho) = \log
\big(  1 + \rho \hsppp \lambda_{\max}({\bf H}^H {\bf H}) \big), 
&& 
I_{\stat}(\rho) = \log \bigg( 1 + \rho \sum_{k=1}^{N_t} 
\lambda_k |\bv_k ^{\sl H} \bu_{\stat} |^2 \bigg), 
\end{eqnarray}
where ${\bf u}_{\stat}$ is an eigenvector corresponding to the dominant 
eigenvalue of ${\bf \Sigma}_t = \bEe [ \bH^H \bH ]$, and an eigen-decomposition 
of $\bH^H \bH$ is of the form: $\bH^H \bH = \sum_{k=1}^{N_t } \lambda_k 
{\bf v}_k {\bf v}_k^H$. The following simplifications can then be made: 
\begin{eqnarray}
{\bEe}_{\bH} \left[ I_{\stat}(\rho)  \right] \cdot 
\Delta I_{\sf bf} & = & \bEe_{\bH} \bigg[  \log \bigg(  1 + \rho \hspp 
\frac{ \lambda_1 - \sum_k \lambda_k |\bv_k ^{\sl H} \bu_{\stat} |^2} 
{ 1 + \rho \sum_k \lambda_k |\bv_k ^{\sl H} \bu_{\stat} |^2 }
\bigg) \bigg] 
\\ & \stackrel{(a)}{\leq} & 
\bEe_{\bH} \left[  \log \left(  
1 + \rho \hsppp \lambda_1 \hsppp (1 - |\bv_1^{\sl H} \bu_{\stat} |^2) 
\right) \right] \label{ze1} 
\\ & \stackrel{(b)}{\leq} & \log \left( 1 + \rho \bEe_{\bH} \left[
\lambda_{\max}({\bf H}^H {\bf H}) (1 - |\bv_1^{\sl H} \bu_{\stat} |^2 ) \right] 
\right) 
\\ & \stackrel{(c)}{\leq} & \log \left( 1  + \rho \cdot 
\sqrt{  \bEe_{\bH} \left[ (1 - |\bv_1^{\sl H} \bu_{\stat} |^2 )^2 \right] }  
\cdot \sqrt{ \bEe_{\bH} \left[ \lambda_{\max}^2({\bf H}^H {\bf H}) \right] } 
\right), 
\end{eqnarray} 
where (a) follows trivially by ignoring the contribution of 
$k = 2, \hsppp \cdots, \hsppp N_t$ in the summation, (b) follows 
from Jensen's inequality, and (c) from Cauchy-Schwarz inequality. 
We use the eigenvector perturbation theory developed in~\cite{vasanth_limfb} and 
in particular, the bound in~\cite[Eqn.~(16)]{vasanth_limfb} to 
establish that 
\begin{eqnarray}
\bEe_{\bH} \left[ (1 - |\bv_1^{\sl H} \bu_{\stat} |^2 )^2 \right] \leq 
\kappa_3'  \hsppp \frac{N_t \log(N_r)}{N_r}  
\end{eqnarray} 
for some appropriate constant $\kappa_3'$ that is independent of the channel 
statistics and dimensions. Using Lemma~\ref{lem_eig_bds} and 
Lemma~\ref{yinbailemma}, the conclusion in~(\ref{eqn_kron_bf}) follows for the 
relative asymptotics case. For the proportional growth case, an upper bound needs 
to be established for $\bEe_{\bH} 
\left[ \lambda_{\max}^2({\bf H}^H {\bf H}) \right]$. 
See~\cite{vasanth_allerton04} for an upper bound technique that builds on the 
work by~\cite{yin_bai_and_krishnaiah}, which results in the statement of the 
theorem. 
\endproof

\subsection{Proof of Theorem~\ref{thm_proportional2}}
\label{app_thm_proportional2}
As in App.~\ref{app_thm2}, we can write $\Delta I_2$ as 
\begin{eqnarray}
\Delta I_2 & = & 
\frac{ \bEe _{ {\bf H} } [ I_{\perf, \hsppp \semi}(\rho)] } 
 { \bEe _{ {\bf H} } [ I_{\stat,\hsppp \semi}(\rho)] } - 1 
\\ 
& = &  \frac{ \bEe _{{\bf H}} \left[ \sum_{k = 1 }^M \log \big( 1 
+ \transnr \hsppp \lambda_k ( {\bf H}^{\sl H} {\bf H}  ) \big)  \right] } 
{\bEe _{{\bf H}}  \left[ \sum_{k = 1 }^M \log \big( 1 
+ \frac{\rho}{M \hsppp \rho_c}
\lambda_k ( \widetilde{\bfLambda}_t \widetilde{\bH}_{\iid}^H 
\bfLambda_r \widetilde{\bH}_{\iid}  ) \big) \right] } - 1. 
\label{citer_v1}
\end{eqnarray} 
The denominator of~(\ref{citer_v1}) can be computed following the method 
in~\cite[Theorem 1]{hachem} and equals 
\begin{eqnarray}
\bEe_{\bH}[ I_{\stat, \hsppp \semi}(\rho) ] = 
\sum_{k = 1}^M \log \left(1 + \frac{\rho}{\chanpow} \hsppp \mu_1 \bfLambda_t(k) 
\right)
+ \sum_{k = 1}^M \log \left( 1 + \frac{\rho}{\chanpow} \hsppp \widetilde{\mu}_1 
\bfLambda_r(k) \right) - \frac{\rho M}{\chanpow} \hsppp \mu_1 \widetilde{\mu}_1 , 
\label{istat}
\end{eqnarray}
where $\mu_1$ and $\widetilde{\mu}_1$ satisfy the recursive equations 
\begin{eqnarray}
\mu_1  =  \frac{1}{M} \sum_{k = 1}^M \frac{\bfLambda_r(k)} { 1 + 
\frac{\rho}{\chanpow} \hsppp \widetilde{\mu}_1 \bfLambda_r(k) }, &&
\widetilde{\mu}_1  =  \frac{1}{M} \sum_{k = 1}^M \frac{\bfLambda_t(k)} { 1 + 
\frac{\rho}{\chanpow} \hsppp {\mu}_1 \bfLambda_t(k) }. 
\label{reinforce} 
\end{eqnarray} 

A simple lower bound for $\bEe_{\bH}[ I_{\stat, \hsppp \semi}(\rho) ]$ is 
obtained by using $\log(1 + x) \geq \log(x)$ for $x > 0$: 
\begin{eqnarray}
\label{simple_lb}
\bEe_{\bH}[ I_{\stat, \hsppp \semi}(\rho) ] \geq 
\sum_{k=1}^M \log \left( \frac{ \rho^2 } {\rho_c^2 e} \mu_1 
\widetilde{\mu}_1 {\bf \Lambda}_t(k) {\bf \Lambda}_r(k) \right). 
\end{eqnarray} 
We now establish that the above bound is order-optimal as $\alpha$ increases 
(with $\rho = \alpha \frac{M} { {\bf \Lambda}_t(M) }$), 
by lower bounding $\mu_1 \widetilde{\mu}_1$. We can easily show that 
\begin{eqnarray}
\mu_1  \geq  \frac{\rho_c}{M} \cdot \frac{b_2} { 1 + \alpha b_1 
\frac{ {\bf \Lambda}_r(1) } { {\bf \Lambda}_t(M)}  }, 
{\hspace{0.2in}} 
\widetilde{\mu}_1  \geq  \frac{\rho_c}{M} \cdot \frac{b_1}  
{1 + \alpha b_2 \frac{ {\bf \Lambda}_t(1)  } { {\bf \Lambda}_t(M) } } , 
\end{eqnarray}
and hence, 
\begin{eqnarray}
1 \geq \frac{\rho}{\rho_c} \mu_1 \widetilde{\mu}_1 \geq 
\frac{ \alpha C_1  } {  1 + \alpha (C_1 + C_2)}, 
\end{eqnarray}
where $C_1 = b_1 \frac{ {\bf \Lambda}_r(M) }{ {\bf \Lambda}_t(M) }$ and 
$C_2 = b_2 \frac{ {\bf \Lambda}_t(1) }{ {\bf \Lambda}_t(M) }$. 
Tightness of the bound in~(\ref{simple_lb}) follows from using the fact that 
$\log(1 + x) \leq \log(x) + \frac{1}{x}, \hsppp x > 0$. 

Combining the above relationships, we have 
\begin{eqnarray}
\bEe_{\bH}[ I_{\stat, \hsppp \semi}(\rho) ] 
& \geq & M \log \left( \frac{\rho \alpha C_1} 
{e \left( 1 + \alpha (C_1 + C_2)  \right)} \right) + 
\sum_{k=1}^M \log 
\left( \frac{ \bfLambda_t(k) \bfLambda_r(k) } {\rho_c} \right). 
\label{eqn_stat_ref}
\end{eqnarray}
Proceeding in the same way, one can obtain an upper bound for 
$\bEe_{\bH}[ I_{\perf, \hsppp \semi}(\rho) ]$. Since the main goal here is 
to obtain the trends of $\Delta I_2$, we find it convenient and less cumbersome 
to replace 
the upper bound with an approximation ($\log(1 +x ) \approx \log(x)$) by 
ignoring the term that decays as $\frac{1}{x}$. Thus, we have 
\begin{eqnarray}
\bEe_{\bH}[ I_{\perf, \hsppp \semi}(\rho) ] 
& \approx &  M \log \left(\frac{\rho}{M} \right) + 
\bEe_{\bH} \left[ 
\sum_{k= 1}^M \log \left( \frac{ 
\lambda_k(\bfLambda_t \bH_{\iid}^H \bfLambda_r \bH_{\iid}) 
}{\chanpow} \right) \right] \\ 
& \stackrel{(a)}{\leq} & M \log \left(\frac{\rho}{M} \right) + 
\min (A,B ) \label{eqn_reee} \\ 
A & = &  M \bEe_{\bH} 
\left[\log \left( \frac{ \lambda_{\max} ( \bH_{\iid}^H \bfLambda_r \bH_{\iid}) } 
{\chanpow} \right) \right]
+ \sum_{k=1}^M \log \left( \bfLambda_t(k) \right) \\  
B & = & M \bEe_{\bH} 
\left[\log \left( \frac{ \lambda_{\max} ( \bH_{\iid}\bfLambda_t \bH_{\iid}^H )}{\chanpow} 
\right )  \right]
+ \sum_{k=1}^M \log \left( \bfLambda_r(k) \right), 
\end{eqnarray}
where in (a) we have used Lemma~\ref{lem_poincare}. Combining~(\ref{eqn_stat_ref}) 
and~(\ref{eqn_reee}), we have the statement of the theorem. 
\endproof

\subsection{Proof of Proposition~\ref{prop_prob_err}} 
\label{app_perr_semi}
First, we write $\Delta P_{\sf semi}$ in terms of $\sinr$ of the individual 
data-streams by using $P_{k,\bullet} = \alpha {\cal Q} 
\left( \beta (\sinr_{k, \bullet} )^{1/2} \right)$ and the expression for 
$\sinr_{k,\bullet}$ in~(\ref{sinr_exp}). 
Then, we use the following bound for ${\cal Q}(x)$: 
\begin{eqnarray} 
\label{calq}
\frac{ \exp(-x^2 /2 ) }{x \sqrt{2 \pi}} \left( 1 - \frac{1}{x^2} \right) 
\leq {\cal Q}(x ) \leq \frac{ \exp(-x^2 /2 ) }{x \sqrt{2 \pi}} 
\end{eqnarray} 
to establish the expression in~(\ref{deltap_semi}). 
It is straightforward to check that 
\begin{eqnarray}
\sinr_{k,\hsppp \perf, \hsppp \unstruct} = 
{\bf \Lambda}_{\sf wf}(k) \hsppp \lambda_k({\bf H}^H {\bf H}), 
\end{eqnarray} 
where the waterfilling power allocation $\{ {\bf \Lambda}_{\sf wf}(k) \}$ is 
as in~(\ref{waterfill}) (see App.~\ref{term1}) and normalized to 
\begin{eqnarray}
\sum_{k=1}^M {\bf \Lambda}_{\sf wf}(k) = \rho.  
\end{eqnarray}
Similarly, we have 
\begin{eqnarray}
\sinr_{k, \hsppp \stat, \hsppp \semi}  =  
\frac{1} { \left[ {\bf G}^{-1}\right]_{k} } - 1 
&=& \frac{ \det ({\bf G})} {  \left[{\rm adj}({\bf G}) \right]_k  } -  1, 
\label{stat_snr} \\ 
\label{refg}
{\bf G}  =  {\bI}_M + 
\transnr \hsppp {\bf F}_{\semi}^H {\bf H}^{\sl H} {\bf H} {\bf F}_{\semi} 
& = &{\bI}_M + \frac{\rho}{M \hsppp \rho_c} \cdot 
{\widetilde{\bfLambda} }_t^{1/2} \hsppp {\bf \widetilde{H}}_{\iid}^{\sl H} \hsppp 
\bfLambda_r \hsppp {\bf \widetilde{H}}_{\iid} \hsppp {\widetilde{\bfLambda} }_t^{1/2}. 
\end{eqnarray}
The matrix ${\rm adj}({ \bf G})$ refers to the adjoint of ${\bf G}$, and 
$\left[ {\bf G}^{-1} \right]_k$ and $\left[ {\rm adj}({\bf G}) \right]_k$ refer 
to the $k$-th diagonal entries of ${\bf G}^{-1}$ and ${\rm adj}({\bf G})$, respectively. 
Using the definition of adjoint of a matrix, we have 
\begin{eqnarray} 
\left[ {\rm adj}({\bf G}) \right]_k 
& = & \det \left( {\bI}_{M-1} + \frac{\rho}{M \hsppp \rho_c} \cdot 
{\widehat{\bfLambda} }_t^{1/2} 
\hsppp {\bf \widehat{H}}_{\iid}^{\sl H} \hsppp \bfLambda_r  \hsppp 
{\bf \widehat{H}}_{\iid} \hsppp {\widehat{\bfLambda} }_t^{1/2}  
\right), 
\end{eqnarray}
where ${\widehat{\bfLambda} }_t$ and ${\bf \widehat{H}}_{\iid}$ are as per 
the notations established in~Sec.~\ref{nots}. 
The expression for $\Delta \sinr_k$ in the statement of the proposition follows 
immediately. 
\endproof

\subsection{Proof of Theorem~\ref{thm_prob_loss}} 
\label{app_thm1} 
We have the following upper bound for 
$\sinr_{k, \hsppp \perf, \hsppp \unstruct}$: 
\begin{eqnarray}
\sinr_{k, \hsppp \perf, \hsppp \unstruct} 
& = & {\bf \Lambda}_{\sf wf}(k) \cdot 
\frac{ \lambda_k \left( \bfLambda_t \hsppp {\bH}_{\iid}^{\sl H} \hsppp \bfLambda_r 
\hsppp {\bH}_{\iid}  \right) }{\rho_c} 
\\ 
& \stackrel{(a)}{\leq} & 
{\bf \Lambda}_{\sf wf}(k) \hsppp {\bf \Lambda}_t(k) \cdot 
\frac{ \lambda_{\max}( {\bH}_{\iid}^{\sl H} \hsppp \bfLambda_r \hsppp {\bH}_{\iid}  ) 
}{\rho_c}, 
\label{perf_snr} 
\end{eqnarray} 
where (a) follows from Lemma~\ref{lem_eig_bds}. 
To compute $\sinr_{k,\hsppp \stat, \hsppp \semi}$, note that $\det({\bf G})$, 
where ${\bf G}$ is as in~(\ref{refg}) can be written as 
\begin{eqnarray} 
\det({\bf G}) & = & \prod_{j = 1}^{M} \left(1 + \frac{\rho}{M \hsppp \rho_c} 
\cdot 
\lambda_j ( {\widetilde{\bfLambda} }_t \hsppp {\bf \widetilde{H}}_{\iid}^{\sl H} 
\hsppp \bfLambda_r  \hsppp {\bf \widetilde{H}}_{\iid} ) \right)  \\ 
& \stackrel{(a)}{ \geq} & \prod_{j = 1}^{M} 
\left(1 + \frac{\rho}{M \hsppp \rho_c} \cdot 
\bfLambda_t(j) \hsppp \lambda_{\min}( {\bf \widetilde{H}}_{\iid}^{\sl H} 
\hsppp \bfLambda_r  \hsppp {\bf \widetilde{H}}_{\iid}  ) 
\right), 
\end{eqnarray}
with (a) following from Lemma~\ref{lem_eig_bds}. Similarly, we have 
\begin{eqnarray} 
\left[ {\rm adj}({\bf G})\right]_k 
& = & \prod_{j = 1}^{M-1} \left(1 + \frac{\rho}{M \hsppp \rho_c} 
\cdot \lambda_j ( {\widehat{\bfLambda} }_t 
\hsppp {\bf \widehat{H}}_{\iid}^{\sl H} \hsppp \bfLambda_r  \hsppp 
{\bf \widehat{H}}_{\iid} ) \right) \\ 
& \leq & 
\prod_{j = 1, \hsppp j \neq k}^{M} \left(1 + \frac{\rho}{M \hsppp \rho_c} 
\cdot \bfLambda_t(j) \hsppp \lambda_{\max}( {\bf \widehat{H}}_{\iid}^{\sl H} 
\hsppp \bfLambda_r  \hsppp {\bf \widehat{H}}_{\iid}  ) \right). 
\end{eqnarray} 
Using Lemma~\ref{yinbailemma} from App.~\ref{app_majorize} in 
(\ref{stat_snr}) and (\ref{perf_snr}), 
the following bounds hold with probability $1$ (in 
the limit of $N_r, N_t, M$) for $\sinr_{k, \hsppp \perf, \hsppp \unstruct}$ 
and $\sinr_{k, \hsppp \stat, \hsppp \semi}$: 
\begin{eqnarray} 
\sinr_{k, \hsppp \perf, \hsppp \unstruct} & \leq & 
{\bf \Lambda}_{\sf wf}(k) \hsppp \bfLambda_t(k) \cdot 
\left(1 + \frac{C_1 } {\gamma_r} \sqrt{ \frac{N_t}{N_r} } \hsppp  \right), \\
1 + \sinr_{k, \hsppp \stat, \hsppp \semi} 
& \geq &  \frac{ \prod_{j = 1}^{M} 
\left(1 + \frac{\rho}{M} 
\cdot \bfLambda_t(j) \hsppp \left(1 - \frac{C_1}{\gamma_r} 
\sqrt{ \frac{M}{N_r} } \hsppp \right) \right) } 
{ \prod_{j = 1, \hsppp j \neq k}^{M} \left(1 + \frac{\rho}{M} 
\cdot \bfLambda_t(j) \hsppp \left( 1 + \frac{C_1}{\gamma_r} 
\sqrt{\frac{M-1} { N_r } } \hsppp \right)\right) } 
\end{eqnarray} 
for some universal constant $C_1$ obtained from Lemma~\ref{yinbailemma}. 
If $\rho$ is such that $\rho \geq \alpha \frac{M}{ {\bf \Lambda}_t(M) }$, 
we can trivially lower bound $\sinr_{k,\hsppp \stat, \hsppp \semi}$ as 
\begin{eqnarray}
1 + \sinr_{k, \hsppp \stat, \hsppp \semi}  & \geq &
\left( 1 + \frac{\rho}{M} \hsppp {\bf \Lambda}_t(k) \hsppp \left(1 - 
\frac{C_1}{\gamma_r}  \sqrt{ \frac{M}{N_r} } \right)  \right) 
\left(  \frac{ 1 + \frac{1}{\alpha} - \frac{C_1}{\gamma_r} \sqrt{ \frac{M}{N_r} } } 
{1 + \frac{1}{\alpha} + \frac{C_1}{\gamma_r} \sqrt{ \frac{M-1}{N_r} }  } 
\right)^{M-1} \\ 
& \stackrel{(a)}{\geq} & 
\left( 1 + \frac{\rho}{M} \hsppp {\bf \Lambda}_t(k) \hsppp \left(1 - 
\frac{C_1}{\gamma_r}  \sqrt{ \frac{M}{N_r} } \right)  \right) \cdot 
\left( \frac{ 1 + (M-1) \left( \frac{1}{\alpha}  
- \frac{C_1}{\gamma_r} \sqrt{ \frac{M}{N_r}}  \right) } 
{ 1 + 2(M-1)  \left( \frac{1}{\alpha} + \frac{C_1}{\gamma_r} \sqrt{ \frac{M-1}{N_r} }
\right) }\right), \nonumber \\ 
\label{t1y} 
\end{eqnarray}
where (a) follows from the fact that $1 + ax \leq (1 + x)^a \leq 1 + 2ax$ for 
$x$ sufficiently small and $a > 0$. After some routine manipulations, 
$\Delta \sinr_k$ can be bounded as 
\begin{eqnarray} 
\Delta \sinr_k & \leq & 
M \left( \frac{1}{\alpha} + \frac{3 C_1}{\gamma_r} \sqrt{\frac{M}{N_r}} \right) 
+ {\bf \Lambda}_t(k) \left( {\bf \Lambda}_{\sf wf}(k) - \frac{\rho}{M}  \right) 
\left( 1 + \frac{C_1}{\gamma_r} \sqrt{\frac{N_t}{N_r}}  \right) 
\nonumber \\ 
& & {\hspace{0.2in}} 
+ \frac{\rho}{M} {\bf \Lambda}_t(k) \left( \frac{M}{ \alpha} \left( 1 + 
\frac{C_1}{\gamma_r \hsppp \sqrt{N_r}} (2 \sqrt{N_t } + \sqrt{M}) \right) 
+ \frac{C_1} {\gamma_r \hsppp \sqrt{N_r} } ( 3 M \sqrt{M} + \sqrt{N_t } )
\right) \nonumber \\ 
&=&  \frac{M}{\alpha} \left( 1 + \frac{\rho {\bf \Lambda}_t(k)}{M}  \right) + 
{\bf \Lambda}_t(k) \cdot 
\left( {\bf \Lambda}_{\sf wf}(k) - \frac{\rho}{M} \right) 
+ \frac{\rho { \bf \Lambda}_t(k)} {\gamma_r} \cdot 
 \ord \left( \frac{ \sqrt{M} + \sqrt{N_t} } {\sqrt{N_r}} \right). 
\label{t2} 
\end{eqnarray}

We now use the facts that $\sqrt{1+ x} \leq 1 + \frac{x}{2}$ for any $x$ positive, 
and $\frac{1}{1-x}$ is upper bounded by $1 + 2x$ as long as $x < \frac{1}{2}$ 
for the terms $\sqrt{ 1+ \frac{\Delta \sinr_k} { \sinr_{k, \hsppp \stat, \hsppp \semi}} }$ 
and $\frac{1}{1 - \frac{1}{\beta^2 \sinr_{k, \hsppp \perf, \hsppp \unstruct} } }$, 
respectively. The term $\exp \left( \frac{\beta^2 \Delta \sinr_k }{2} \right)$ 
is bounded by 
using the fact that $e^x$ can be bounded by $1 + a x$ 
for some $a > 1$ in the small $x$ regime. The combination of the above facts 
yields 
\begin{eqnarray}
\Delta P_{\sf semi} 
& \leq & \frac{1}{M \hsppp \beta^2} \bEe_{\bH } \left[ \sum_{k=1}^M \frac{1} 
{{\bf \Lambda}_t(k) {\bf \Lambda}_{\sf wf}(k) } \right] 
+ \beta^2 \frac{M}{\alpha} + 
\frac{\beta^2 }{M} \bEe_{\bH} \left[ \sum_{k=1}^M {\bf \Lambda}_t(k ) 
\left( {\bf \Lambda}_{\sf wf}(k) - \frac{\rho}{M} \right) \right] \nonumber \\ 
& & {\hspace{0.2in}} + \frac{ \rho \beta^2 
\sum_{k=1}^M {\bf \Lambda}_t(k) }{M} \left(  \frac{1}{\alpha} 
+ \frac{1}{\gamma_r} \hspp
\ord \left( \frac{ \sqrt{N_t} + \sqrt{M} }{\sqrt{N_r}} \right) \right) 
\label{eqnciter} 
\end{eqnarray} 
up to a constant scaling multiplicative constant on the right side. For the first 
term, we lower bound ${\bf \Lambda}_{\sf wf}(k)$ from~(\ref{waterfill}) by 
\begin{eqnarray}
{\bf \Lambda}_{\sf wf}(k) & \geq & 
\frac{\rho}{n_{\bH}} - \frac{1}{ {\bf \Lambda}_{\bH}(k) } 
\stackrel{(a)}{\geq} \frac{\rho}{M} - \frac{1} 
{{\bf \Lambda}_t(k) 
\left(1 - \frac{C_1}{\gamma_r} \sqrt{ \frac{N_t}{N_r} } \right)}, 
\end{eqnarray}
where (a) follows from Lemma~\ref{yinbailemma}. 
For the third term, we have 
\begin{eqnarray}
\bEe_{\bH} \left[ \sum_{k=1}^M {\bf \Lambda}_t(k ) 
\left( {\bf \Lambda}_{\sf wf}(k) - \frac{\rho}{M} \right) \right] 
\leq M + \rho \sum_{k=1}^M {\bf \Lambda}_t(k) 
\cdot \left( \bEe_{\bH} \left[ \frac{1}{n_{\bH}} \right] - \frac{1}{M} \right). 
\end{eqnarray}
Finally, we have 
\begin{eqnarray}
\bEe_{\bH} \left[ \frac{1}{n_{\bH}} - \frac{1}{M} \right] 
& \leq & \left( 1 - \frac{1}{M}  \right) {\rm Pr} (n_{\bH} < M)
\leq \frac{1}{\alpha^2} \cdot 
\frac{ \bEe \left[ \left( \frac{1} { {\bf \Lambda}_{\bH}(M) } \right)^2  \right] } 
{  \left( \bEe \left[ \frac{1}{ {\bf \Lambda}_{\bH}(M) } \right]  \right)^2 }, 
\end{eqnarray}
where the second inequality follows from the bound in~(\ref{citing_nh}). 
Combining these facts, we have 
\begin{eqnarray}
\Delta P_{\sf semi} & \leq & \frac{1}{\beta^2 M} \sum_{k=1}^M \frac{1} 
{ \frac{ \rho {\bf \Lambda}_t(k)}{M} - 1 } + 
\beta^2 ( 1 + \frac{M}{\alpha}) \nonumber \\ 
& & {\hspace{0.02in}} 
+  \frac{ \beta^2 \rho \sum_{k=1}^M {\bf \Lambda}_t(k) }{M}  \left( 
\frac{1}{\alpha} + \frac{1}{\alpha^2}\cdot 
\frac{ \bEe \left[ \left( \frac{1} { {\bf \Lambda}_{\bH}(M) } \right)^2  \right] } 
{  \left( \bEe \left[ \frac{1}{ {\bf \Lambda}_{\bH}(M) } \right]  \right)^2 } 
+ \frac{1}{\gamma_r} \hspp \ord \left( \frac{ \sqrt{N_t} + \sqrt{M} } 
{\sqrt{N_r}}   \right) \right) 
\end{eqnarray} 
Thus the proof is complete. 
\endproof 

\bibliographystyle{IEEEbib}
\bibliography{news}

\end{document}